%% file: main.tex
\documentclass[%
 reprint,
 amsmath,
 amssymb,
 aps,
 prl,
 secnumarabic,
]{revtex4-2}
\usepackage{graphicx}
\usepackage{dcolumn}
\usepackage{bm}
\usepackage{bbm}
\usepackage{subcaption}
\usepackage{hyperref}
\usepackage{orcidlink}
\usepackage{algorithm}
\usepackage{algpseudocode}
\usepackage{amsfonts}
\usepackage{float}
\usepackage{caption}
\usepackage{xcolor}
\usepackage{color}
\usepackage{multirow}
\usepackage{tabularx}
\usepackage{makecell} 
\usepackage{multibib}
\usepackage{pgffor}
\usepackage{mathtools}
\usepackage{booktabs}
\usepackage{adjustbox}

\newcommand{\beginSM}{%
        \setcounter{table}{0}
        \renewcommand{\tablename}{Table}
        \renewcommand{\thetable}{S\arabic{table}}%
        \setcounter{figure}{0}
        \renewcommand{\figurename}{FIG.}
        \renewcommand{\thefigure}{S\arabic{figure}}%
        \captionsetup[figure]{labelformat=simple, labelsep=colon}
         \setcounter{algorithm}{0}
         \renewcommand{\thealgorithm}{S\arabic{algorithm}}
        \setcounter{equation}{0}
        \renewcommand{\theequation}{S\arabic{equation}}
     }

\usepackage{titlesec}
\titlespacing{\subsubsection}
  {15pt}
  {13pt}
  {13pt}
\titlespacing{\subsection}
  {15pt}
  {13pt}
  {13pt}

\begin{document}

\makeatletter
\def\frontmatter@title@above{%
  {\centering\color{gray}\small Accepted for publication in \emph{Physical Review Letters}, DOI: \url{https://doi.org/10.1103/b1gc-9c2q}.\par\vspace{1.5em}}%
  \addvspace{6\p@}%
}
\makeatother

\title{AI-Boosted Rare Event Sampling to Characterize Extreme Weather}

\author{Amaury Lancelin\,\orcidlink{0009-0005-3898-3043}$^{1,2,*}$}
\author{Alexander Wikner\,\orcidlink{0000-0001-7575-3953}$^{3,*}$}
\author{Laurent Dubus\,\orcidlink{0000-0002-3987-646X}$^{2,4}$}
\author{Clément Le Priol\,\orcidlink{0000-0002-8214-693X}$^{1,\dagger}$}
\author{Dorian S. Abbot$^{3,\ddagger}$}
\author{Freddy Bouchet\,\orcidlink{0000-0002-1623-0818}$^{1,\S}$}
\author{Pedram Hassanzadeh\,\orcidlink{0000-0001-9425-8085}$^{3,\parallel}$}
\author{Jonathan Weare\,\orcidlink{0000-0001-8745-1821}$^{5,\P}$}

\affiliation{$^1$LMD/IPSL, CNRS, ENS, Université PSL, École Polytechnique, Institut Polytechnique de Paris, Sorbonne Université, Paris, France}
\affiliation{$^2$Réseau de Transport d'Électricité (RTE), Paris, France}
\affiliation{$^3$Department of the Geophysical Sciences, University of Chicago, Chicago, Illinois, USA}
\affiliation{$^4$World Energy $\&$ Meteorology Council (WEMC), Norwich, United Kingdom}
\affiliation{$^5$Courant Institute, New York University, New York, New York, USA}
\affiliation{$^*$\textit{These authors contributed equally to this work.}}
\affiliation{$^\dagger$Now at AXA Climate, Paris, France.}
\affiliation{$^\ddagger$Contact author: abbot@uchicago.edu}
\affiliation{$^\S$Contact author: freddy.bouchet@lmd.ipsl.fr}
\affiliation{$^{\parallel}$Contact author: pedramh@uchicago.edu}
\affiliation{$^\P$Contact author: weare@nyu.edu}

\begin{abstract}

Weather extremes pose major societal risks, especially in a changing climate, but due to their rarity, they are difficult to study using limited observations or complex climate models. We introduce AI+RES, a framework coupling fast AI weather forecasts with a high-fidelity physics model using a rare-event algorithm to efficiently characterize extremes. This approach enables the study of the statistics and physics of very rare events, such as once per millennium heat waves at two orders-of-magnitude lower computational cost. AI+RES can be applied broadly across climate science and other fields concerned with rare events.
  
\end{abstract}

\maketitle

\begin{figure*}[htbp]
  \centering
  \hspace*{-0.7cm}
  \includegraphics[width=0.85\textwidth]{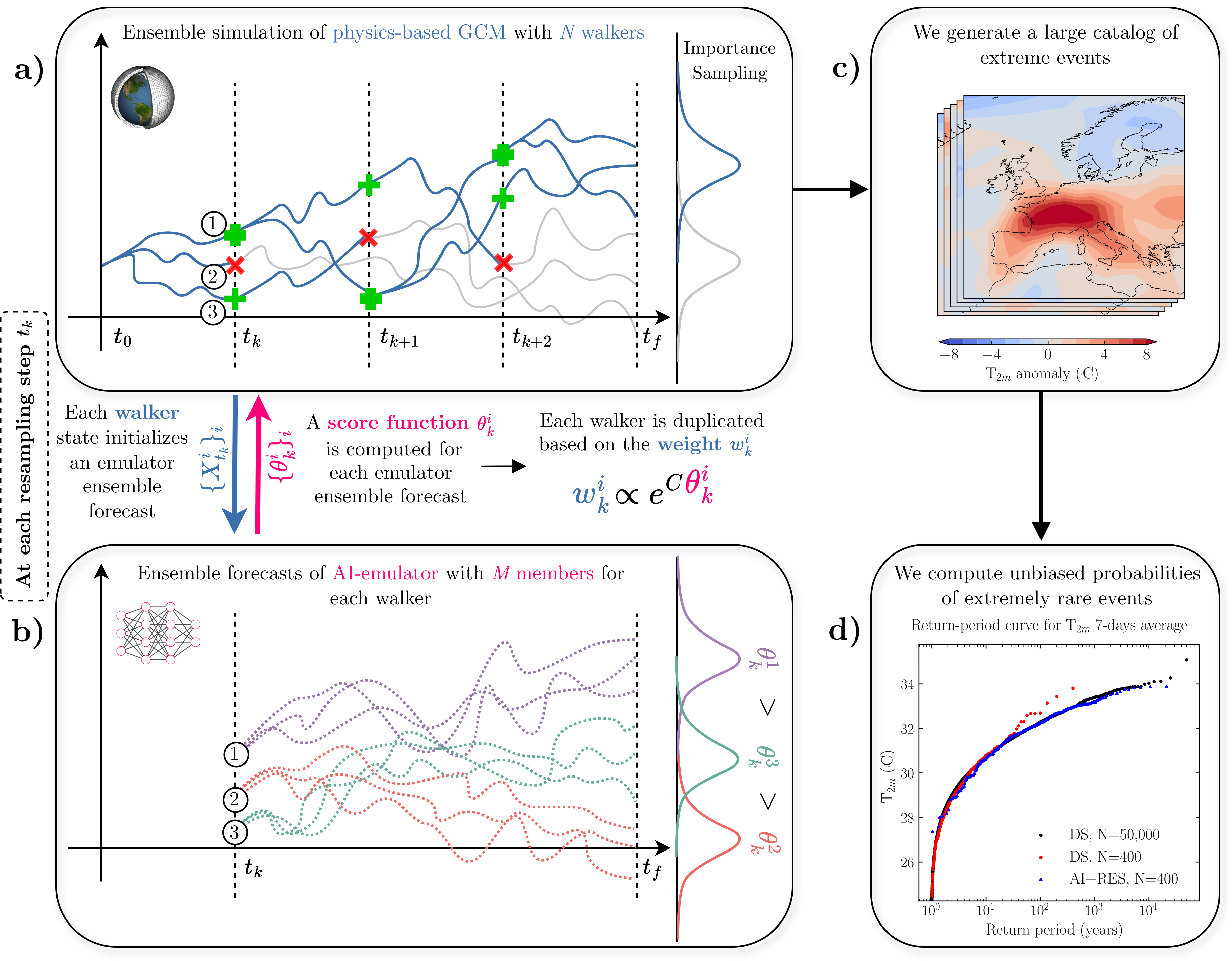}
 \caption{\textbf{Schematic of the AI+RES framework}. 
 We first train an AI weather emulator on 100 years of PlaSim GCM data, and then use it to guide a rare event sampling (RES) algorithm. (a) In AI+RES, we run a PlaSim ensemble simulation with $N$ parallel simulations, called \emph{walkers} and denoted $X_t^i$. (b) At each resampling time $t_k$ in the algorithm (vertical dashed lines), we perform an ensemble forecast with the AI emulator for each PlaSim walker until the target time $t_f$, then duplicate the more promising walkers based on these forecasts (Algorithm~\ref{alg:dmc} and Eq.~\eqref{eq:score_function}). This allows us to both generate a large catalog of extreme events from physics-based simulations (c) and to compute unbiased probabilities for these rare events (Eq.~\eqref{eq:unbiased_estimator}) (d) at a fraction of the cost of direct sampling (DS). An example of schematic (a) with actual data from an AI+RES experiment can be found in Supplemental Material (SM) Fig.~S9.
 }
 \label{fig: schematic1}
\end{figure*} 

Rare, extreme weather and climate events—such as heat waves, floods, and hurricanes—have enormous socioeconomic impacts~\cite{TOL2024,anchen_swiss_2025}. Accurately assessing how their frequency and intensity change in a warming climate is essential for developing effective mitigation and adaptation strategies~\cite{ebi2021extreme, ummenhofer2017extreme, gonccalves2024extreme, ipcc2021wgi, ipcc2022wgii}. However, it is difficult to estimate the frequency and potential impact of the rarest events from observational data~\cite{zeder2023effect}. Extreme value theory \cite{embrechts2013modelling} can be used to extrapolate from short historical records, but it is often overwhelmed by uncertainty \cite{huang_estimating_2016,galfi_convergence_2017,zeder2023effect, lepriol2024using}, and does not provide examples of the event of interest for physical interrogation. High-fidelity physics-based global climate model (GCM) simulations can provide additional data, but accurate sampling requires simulation lengths at least ten times longer than the return time of the event of interest, which is computationally infeasible.  In response to these computational constraints, rare event sampling (RES) techniques and artificial intelligence (AI) emulators have been developed. While promising, both methods have proven insufficient on their own to fully resolve the problem.

RES is a set of importance sampling tools that focus computation on the rare event of interest~\cite{ragone2018computation,webber2019,ragone2021rare,abbot2021rare,finkel_bringing_2024,lepriol2024using, noyelle2025statistical}. RES entails scheduled duplication (``splitting'') or termination (``killing'') of members of an ensemble of independently evolving model simulations to promote efficient sampling of the targeted rare event. The probability that any one ensemble member (a trajectory) is duplicated or terminated is determined by a user-chosen \emph{score function} evaluated at the scheduled resampling times. Choosing a score function typically requires extensive domain knowledge, trial and error, and parameter optimization with costs that may outweigh any advantages of using RES over direct simulation. Recent work has also explored using ensemble boosting, a cousin of RES, to study unprecedented extreme events \cite{gessner2021ensboost, bloin2025estimating}. This approach has recently been extended to allow probability estimation by adopting an RES approach with a one-step splitting scheme \cite{bloin2025estimating, finkel2025boosting}.

RES has proven highly effective for sampling extremes of long-duration weather, e.g., seasonal means \cite{ragone2018computation, webber2019, wouters2023rare, lepriol2024using, noyelle2025statistical}, by using \textit{persistence}, i.e., the \textit{current} value of the index variable defining the event, as a score function. This approach, however, fails catastrophically when the current value of the index variable is a poor predictor of its future value~\cite{webber2019,abbot2021rare}. This is the case for many phenomena in complex atmospheric dynamics, such as blocking-driven heat waves, which have led to some of the most socioeconomically impactful extreme events in recent memory in France, Russia, and the U.S. Pacific Northwest \cite{barriopedro2011hot, stott2004human, white2023unprecedented, galfi2021fingerprinting}. Previous work has therefore established the strong potential of RES, but only if an efficient method for finding an effective and accessible score function can be identified.

One of the most exciting recent developments in climate science and scientific AI has been the introduction of autoregressive AI weather emulators~\cite{pathak2022fourcastnet,lam2023learning,bi2023accurate, price2025probabilistic}. When trained on high-resolution reanalysis data, these models can outperform state-of-the-art physics-based numerical weather prediction systems for short- and medium-range ($\sim$10–15~day) forecasts~\cite{bi2023accurate,lam2023learning,ben2024rise}. Their extreme computational efficiency (up to $10^4$ times faster than traditional models~\cite{pathak2022fourcastnet}) makes them appealing for studying rare extreme weather. More recently, stable AI emulators have been extended to climate timescales, enabling extremely long simulations or large ensembles in which rare events could, in principle, be sampled directly~\cite{watt-meyer_ace2_2024, kochkov2024neural, chapman2025camulator, mahesh2025huge1, mahesh2025huge2}. However, because these models are trained to predict at short lead times, and only on historical data or expensive high-fidelity simulations that contain few examples of the most extreme events, they often struggle to extrapolate reliably and to reproduce the correct frequencies of very rare extremes~\cite{sun_can_2025, sun2025predicting, zhang2025numerical, plasim_long_emulation}. More broadly, beyond having good standard forecast metrics and reproducing stationary statistics, emulators may struggle to capture physically meaningful forced responses under perturbation—arguing for the need for response-theoretic diagnostics \cite{Falasca2025ForcedResponses}.

In this Letter, we present AI+RES, a robust algorithm that leverages the strong forecast capabilities of AI emulators to build an effective score function for RES to characterize rare, extreme events. This algorithm is described in Fig.~\ref{fig: schematic1} and discussed in detail in the End Matter. We demonstrate this new framework by using it to simulate extreme heat waves in an intermediate complexity GCM, PlaSim \cite{fraedrich_planet_2005}, which allows for extensive long-term control runs to compute baseline rare event statistics and rigorously evaluate the performance of our approach. For validation purposes, PlaSim is run in a stationary climate configuration. Heat waves are a well-motivated application area because they are the deadliest extreme weather events and are expected to worsen under climate change~\cite{zhao2021global,newman2023global,thompson2023most}. We show that AI+RES provides accurate rare event statistics at a numerical speedup of up to a factor of $\mathcal{O}(100)$, whereas standard RES fails to even yield examples of the rarest events of interest. This Letter demonstrates a novel methodology that addresses a central difficulty of RES and may broaden its applicability in fields where a fast AI emulator can be constructed.

\subsection*{Midlatitude heat waves}

As mentioned, we focus on sampling rare midlatitude heat waves due to their societal impact and the challenge that they pose to standard RES. Following previous studies \cite{ragone2018computation, miloshevich_probabilistic_2023-1}, we consider heat wave events defined by large values of the spatiotemporal average of the 2-meter air temperature, T$_{2m}$, over a prescribed geographic region and a time interval typically spanning one week during the summer season (see Eq.~\eqref{eq:observable}). We consider two such regions, centered over France and over the U.S. Midwest, respectively (SM Fig.~S7).
In this manuscript, we use \emph{rare} to denote low-probability events and \emph{extreme} to denote very large values of the chosen observable, with substantial overlap between the two. We refer to \cite{kantz2006dynamical, Ghil2011ExtremeEvents} for more formal definitions.

\subsection*{Ground truth and baselines}

To evaluate the performance of AI+RES, we compare it to a ground truth and five baselines. The ground truth, \emph{direct sampling} (DS) with $N=50,000$, is a very large ensemble that is only possible due to the computational efficiency of PlaSim. The first baseline, DS, $N=400$, is a DS with PlaSim using the same ensemble size as the number of walkers used in the AI+RES algorithm. The second baseline, \emph{Standard}-RES, uses the RES algorithm but with the traditional persistence score function used in state-of-the-art weather and climate applications \cite{ragone2018computation, webber2019, wouters2023rare, lepriol2024using, noyelle2025statistical}. The third baseline, AI-DS, is a DS ensemble using the AI emulator started from the same initial conditions as DS and RES, with the same ensemble size as the ground truth ($N=50,000$). This allows us to determine whether the AI emulator has learned enough from its 100-year training period to extrapolate to tail events in PlaSim. The fourth baseline, EVT, applies extreme value theory, specifically the generalized Pareto distribution (GPD) within a peak-over-threshold (POT) framework, to PlaSim datasets of the same size as the AI+RES experiments ($N = 400$). We estimate uncertainty in this method by performing the fits on 100 different datasets of equal size. We discuss a fifth baseline, \emph{Perfect-Forecast-System}+RES (PFS+RES), in SM Sec.~S8.2~\cite{suppmat}. This baseline uses a PlaSim ensemble forecast for the score function and yields an upper bound on algorithm performance given a ``perfect emulator''. PFS denotes ensemble forecasts produced directly with the PlaSim GCM, rather than the AI emulator. PFS provides a near-perfect score function and an upper bound on algorithm performance. This baseline is only possible with GCMs like PlaSim that are computationally inexpensive.

\subsection*{AI+RES accurately estimates long return period events}
\begin{figure}[htbp]
  \vspace{-\intextsep}
  \begin{tabular}{cc}
    \includegraphics[width=1.\linewidth]{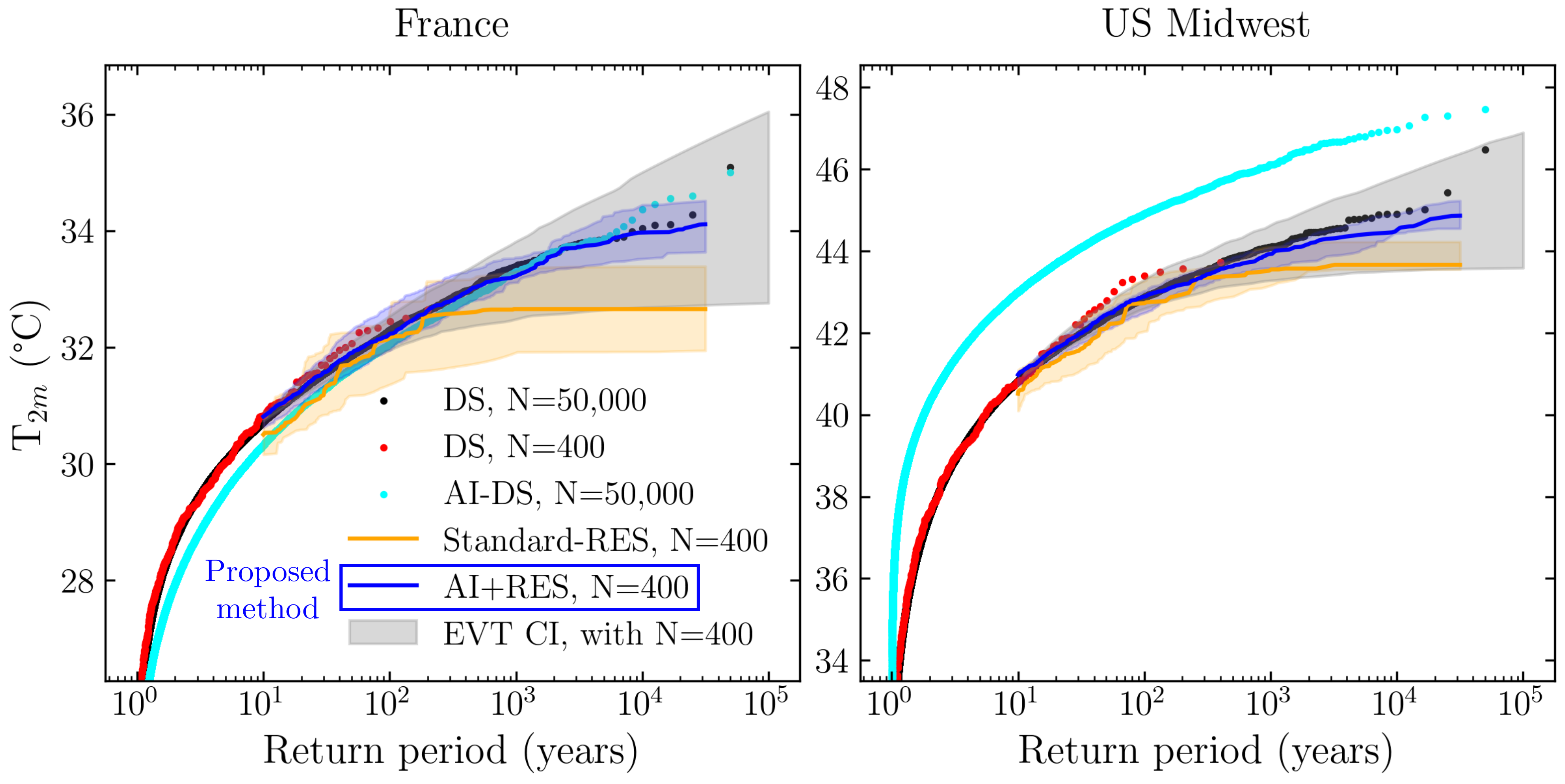}
  \end{tabular}
  \caption{\textbf{Return-period curve for a 7-day heat wave (T$_{2m}$} averaged over 7-days) over France (left panel) and the U.S. Midwest (right panel). The solid blue and yellow lines show the median return-period curve from 10 independent realizations of the AI+RES and standard-RES algorithm, respectively; shaded regions indicate the 10th–90th percentiles. Red and Black dots show DS with N=50,000 and N=400. The gray shaded area (EVT) shows the 10th and 90th percentiles obtained by fitting different GPD distributions with 100 independent training datasets with N=400.}
  \vspace{-\intextsep}
  \vspace{0.6em}
  \label{fig:return_time_curves}
\end{figure}

Figure~\ref{fig:return_time_curves} shows that, for both France and the U.S. Midwest, AI+RES produces accurate, unbiased return period estimates up to our reference ensemble limit of 50,000 years, despite using ensembles with only $N=400$ walkers.  In stark contrast, DS of the same computational cost ($N=400$) can only produce accurate return period estimates up to about 50 years, 1,000 times shorter. This dramatic difference underscores the utility of RES when an {\it effective score function} can be identified. Moreover, Standard-RES saturates for return periods longer than about 100 years, falsely suggesting an upper limit on heat wave intensity. This indicates that persistence misses important trajectories to rare heat waves that the AI-based score function is able to identify. The AI-DS baseline is biased both in the mean and variability of the observable, producing erroneous return period estimates even on a decadal timescale. The correct AI-DS estimate at a return period of about 2,000 years in France is an accident of the curves crossing. It is notable that, despite the errors in AI-DS, the AI emulator's weather forecast skill can still serve as a useful score function by effectively ranking progress of walkers toward strong heat waves. Finally, the
correct return periods do fall within the uncertainty of the EVT baseline even at the longest timescales, but the uncertainty estimate is up to four times larger than that from AI+RES for return periods on the order of 10,000 years. The EVT estimates depend so strongly on the particular years fed into them that they cannot be relied on to produce accurate return period estimates. Also, EVT, unlike AI+RES, does not provide information about the identified extremes and their dynamics.

\subsection*{AI+RES provides a $100\times$ computational speedup for rare events}

\begin{figure}[htbp]
  \centering
  \includegraphics[width=1.\linewidth]{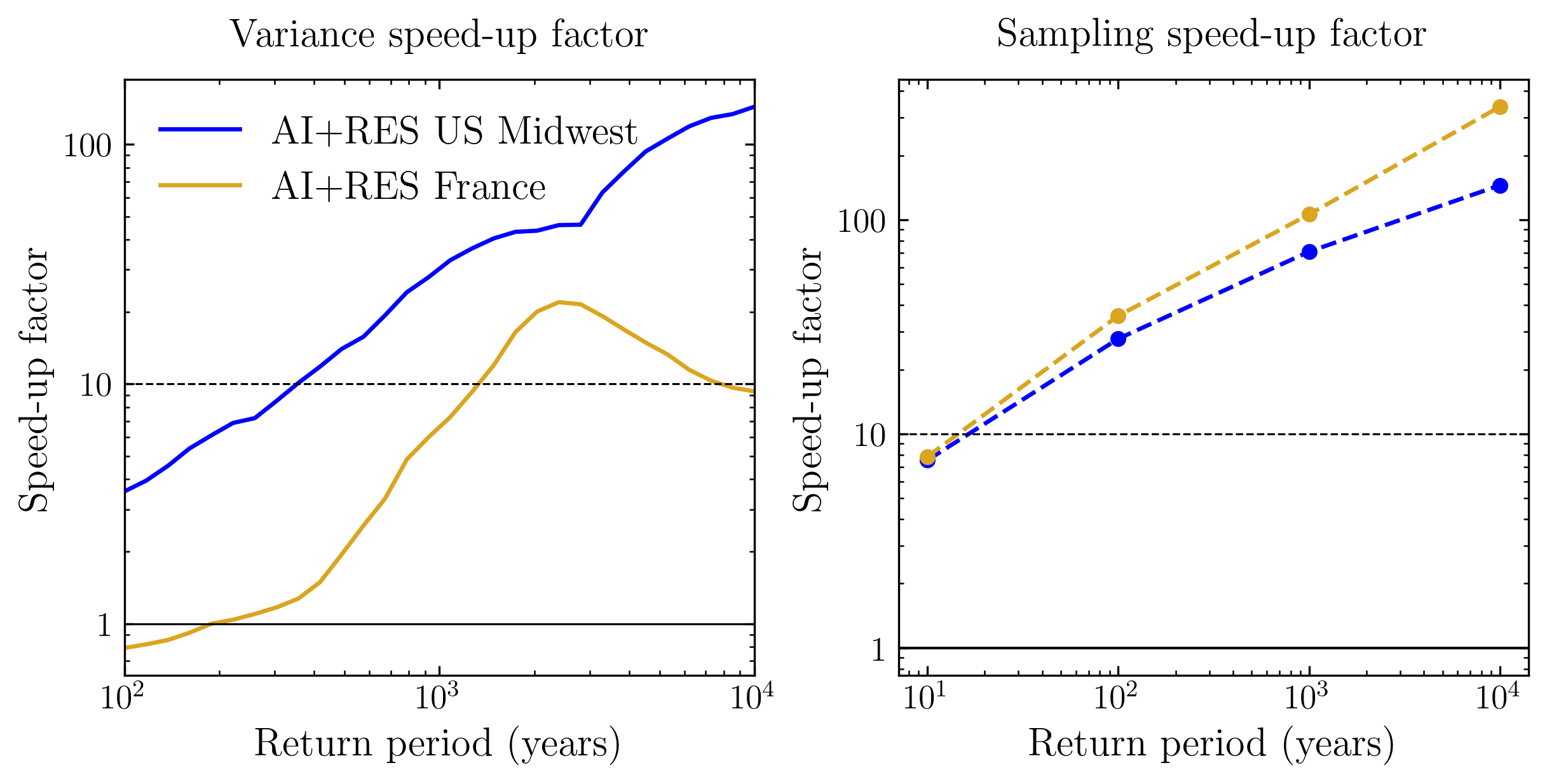}
  \caption{\textbf{Speedup factors for the AI+RES algorithm} over France and the U.S. Midwest as a function of return period for reducing variance in estimated probability and producing additional samples.}\label{fig:speed_up_factors}
\end{figure}

Having established that AI+RES is unbiased, we can quantify its computational speedup compared to DS using the relative efficiency of sampling extreme events (SM Eq.~(S7)) and the relative variance of rare probability estimators (SM Eq.~(S10)). Both metrics tend to yield greater computational gains for rarer events (Fig.~\ref{fig:speed_up_factors}), with speedups of more than 100 for a return period of 10,000 years. The only exception is that the variance speedup estimate for the France region peaks at a factor of 25 around the 2,500-year return period, then gradually decreases to 10 for the rarest events. This nonmonotonic trend is most likely due to a sampling error and is discussed in more detail in SM Sec.~S5.2. Moreover, we hypothesize that the difference in variance speedup for the two regions is due to regional differences in AI emulator weather forecast skill, potentially due to differences in soil moisture variability, as pointed out in SM Fig.~S8.

\subsection*{The rare events generated by AI+RES are physically realistic}

\begin{figure}[htbp]
  \centering
  \includegraphics[width=1.05\linewidth]{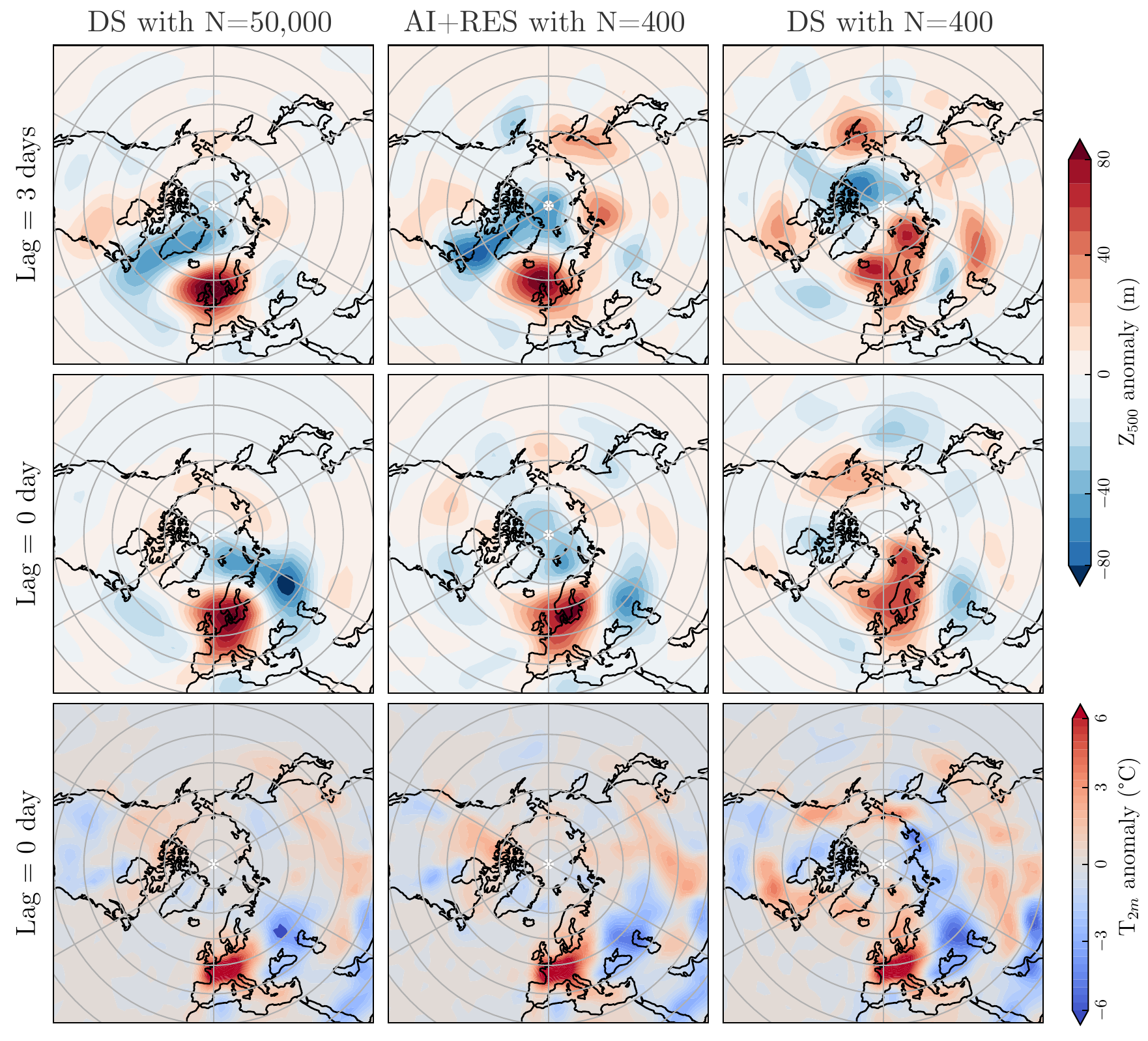}
  \caption{\textbf{Composite maps of heat waves over France with return periods exceeding 100 years.} Events are defined by Eq.~\eqref{eq:observable} with $L=$ 3 days. From top to bottom, the rows show daily mean Z$_{500}$ anomaly composites three days before the heat wave onset, the $L$-day average Z$_{500}$ anomaly composites during the heat wave, and the $L$-day average T$_{2m}$ anomaly composites during the heat wave. 
  } 
  \label{fig: composites}
\end{figure}

A major advantage of AI+RES over the EVT baseline is that it generates full physical trajectories leading to the rare event of interest. This is particularly valuable for investigating the dynamics of the rare event and its precursors \cite{noyelle2025statistical}.

In Fig.~\ref{fig: composites}, we illustrate this capability by presenting composite maps associated with heat waves over France with a return period longer than 100 years from the ground truth DS ($N = 50{,}000$), samples generated with AI+RES ($N = 400$), and samples generated from the DS with the same ensemble ($N = 400$). A strong precursor pattern emerges in the ground truth DS three days before the heat wave onset, marked by a pronounced positive 500-hPa geopotential height anomaly over the British Isles, flanked by a negative anomaly over the Labrador Sea. A weaker negative anomaly off the coast of Portugal suggests potential influence from a cutoff low, while a weak negative anomaly is also visible over Eastern Europe. At the heat wave peak, this synoptic pattern shifts: the high-pressure anomaly migrates eastward, the Labrador Sea anomaly propagates into the North Atlantic, and the Eastern European anomaly intensifies. The AI+RES composite shows similar patterns and captures the correct evolution, although it is slightly noisier than the ground truth due to fewer samples. By contrast, the composite from the DS with the same computational budget as AI+RES is significantly noisier and contains incorrect physical patterns. For example, the 3-day-lagged Z$_{500}$ anomaly north of Scandinavia in this composite is a spurious teleconnection likely due to sampling error. 

Similarly, SM Fig.~S13 shows that AI+RES generates physically realistic samples over the U.S. Midwest that match those from ground truth, while DS with the same computational budget produces samples with spurious features, obscuring the underlying dynamics.

\subsection*{Discussion}

To address the challenge posed by the scarcity of data for studying extreme events, we introduced AI+RES, a novel framework that synergistically combines RES, AI emulators, and GCMs such that they complement each other and alleviate each individual approach's shortcomings. AI+RES provides a practical way to address the long-standing challenge of finding an appropriate score function for RES in a high-dimensional chaotic system. As the first proof-of-concept, we applied this framework on midlatitude heat waves in the PlaSim GCM. We showed that this approach enables the sampling of extremely rare events and yields accurate estimates of their return periods up to 50,000 years with an ensemble size of only 400, with a reduction in computational cost of several hundred compared with direct sampling. We also showed that, even with an imperfect emulator, AI+RES significantly outperforms the current state-of-the-art RES that relies on simple score functions. 

Our work motivates several avenues for further improvement. First, the deterministic nature of the current emulator limits the quality of its ensemble forecasts. Having an emulator trained directly in a probabilistic manner \cite{price2025probabilistic, alet2025skillful, lang2024aifs, couairon2024archesweather,zhou2025reframing} could enhance the ability of the algorithm to explore multiple plausible pathways to extreme events, improving sampling efficiency and reducing estimator variance. Second, while our current AI emulator focuses exclusively on atmospheric variables, surface temperature extremes are also strongly influenced by land surface processes, particularly soil moisture \cite{dandrea_hot_2006, fischer_soil_2007, vautard_summertime_2007, miloshevich_probabilistic_2023-1}. Developing a coupled land-atmosphere emulator remains a challenge, but incorporating soil moisture information, potentially through hybrid strategies that combine emulated forecasts with soil moisture state, could yield a more effective score function.

Beyond the specific case of midlatitude heat waves, the potential applications of AI+RES are broad. It could easily be applied to other types of weather extremes—such as tropical cyclones, blocking events, precipitation, wind, or compound events. AI+RES offers a powerful tool to generate targeted catalogs of rare events in poorly sampled regions of the state space, particularly where direct sampling of computationally expensive GCMs offers insufficient coverage. The framework could also be well suited to characterize the tails of distributions in subseasonal to seasonal ensemble forecasts, where large ensembles are needed. Moreover, by applying AI+RES to climate models running under different climate change scenarios (e.g., CMIP6 or the emerging global high-resolution runs), it would be possible to explore the evolution of the statistics of rare extremes in future climates. Note that the AI+RES can be readily applied to any state-of-the-art GCM; here, we focused the proof-of-concept on PlaSim as its computational efficiency allows for validation against a ground truth dataset. However, this work does not address the question of how faithfully PlaSim rare-event statistics represent the real world.

Regarding the AI emulators themselves, the utility of AI+RES is twofold. As the climate modeling community moves toward building emulators for increasingly high-resolution GCMs, obtaining sufficient physically generated samples to accurately train and validate them with physically generated data becomes a key challenge for rare events. Using AI+RES to generate such catalogs of extremes will constitute precious training and validation datasets for the emulators.

Finally, while this study focused on weather extremes, our
approach should be broadly applicable. AI+RES could be extended beyond climate science, to any high-dimensional dynamical system for which an AI surrogate model can be constructed.

\subsection*{Code and Data availability}
The code implementing AI+RES, along with model configurations, analysis scripts, and the data files used to generate the figures in this Letter, is available at \url{https://github.com/amaurylancelin/AI-RES-public}. For practical reasons, the full long PlaSim runs will not be publicly posted, but they can be made available upon request from the authors.

\subsection*{Acknowledgments}

A.L. was funded in part by RTE France, the French transmission system operator, and benefited from the national agency for research and technology (ANRT) funding under the CIFRE Contract No. 2023/0506. This research received support through Schmidt Sciences, LLC. A.W. and A.L. are also supported by France-Chicago Center FACCTs Award No. 300004248. This work was also supported by the Institut des Mathématiques pour la Planète Terre (IMPT) and funded in part by the U.S. National Science Foundation through Awards No. NSF RISE–2425898, No. NSF-2531264, and No. NSF-2425899. This research is also partially funded by the Institute for Climate and Sustainable Growth at the University of Chicago. In addition, the authors thank Alessandro Lovo for fruitful discussions and insightful ideas on efficiently coupling RES and AI forecasts, which were invaluable to this project; Anna Asch for her work in making the diagnostic plots of the AI emulator; and Justin Finkel for engaging and insightful discussions. Computational resources were provided by NSF ACCESS (allocation ATM170020), NCAR’s CISL (allocation URIC0009), the University of Chicago Research Computing Center, and GENCI-IDRIS (Grant No. 2025-AD010116382).

\subsection*{Author contributions} In alphabetical order, D.S.A., F.B., P.H., A.L., C.L.P., J.W. and A.W. conceptualized the work. D.S.A, F.B., P.H., and J.W. supervised and managed the project. A.L. and A.W. curated the data. A.L. and A.W. developed the codes. A.L. conducted the experiments and did the analysis; A.W. led the development of the AI emulator. A.L. and A.W. wrote the original draft. D.S.A., F.B., L.D., P.H., A.L., C.L.P., J.W. and A.W. reviewed and edited the Letter.

\begin{center}
  \large\textbf{End Matter}\\[16pt]
\end{center}

\subsection*{Definition of the heat wave index}

Following previous studies \cite{ragone2018computation, miloshevich_probabilistic_2023-1}, we define the heat wave index $A_L(t)$ as a spatiotemporal average of 2-meter air temperature, T$_{2m}$, as in the following:
\begin{equation} \label{eq:observable}
A_L(t) := \frac{1}{L}\int_{t}^{t+L} \left( \frac{1}{\mathcal{R}}\int_{\mathcal{R}}\text{T}_{2m}(\vec{r},s)\,d\vec{r} \right)ds.
\end{equation}
We want to estimate the return times of the events $A_L(t_f) >a$ for large thresholds $a$. We typically use $L=7$~days, a window size small enough to capture the peak of midlatitude heat waves, yet large enough to have significant societal impacts.
In this study, we consider two regions, $\mathcal{R}$, each defined as a $3 \times 3$ grid point domain centered over France and the U.S. Midwest (SM Fig.~S7). We initialize simulations at $t_0=$ July 2 and integrate forward the PlaSim GCM until $t_f + L$, with $t_f =$ August 1, chosen to coincide with the climatological peak of $A_L(t_f)$.

\section*{AI+RES with diffusion Monte Carlo}
For the RES component of AI+RES, we use a version of the \emph{diffusion Monte Carlo} (DMC) algorithm that closely follows the formulations proposed in \cite{webber2019} and \cite{abbot2021rare}. It relies on a one-dimensional \emph{score function} (or \emph{reaction coordinate}), $\theta: \mathbb{R}^d \to \mathbb{R}$, which assigns higher values to regions of phase space associated with the rare event of interest. The algorithm enhances sampling in regions where $\theta$ is high (via duplication or ``splitting") and suppresses sampling where it is low (via termination or ``killing").

The mathematical properties of DMC have been rigorously studied---see, for instance, \cite{moral2004feynman}---with results establishing its convergence and asymptotic behavior as the ensemble size $N$ tends to infinity. Under mild integrability assumptions, DMC provides unbiased estimators that converge in the large-$N$ limit. These convergence results are quite general and remain valid even for systems of arbitrarily high dimension $d$. Any statistical quantity that can be computed via direct sampling can, in principle, also be estimated using DMC. This includes observables that depend on the full trajectory of the process from the initial time $t_0$ up to a given time $t_f$. Although those convergence properties are very general, the choice of the score function can greatly influence the variances of the estimators obtained with DMC, highlighting the importance of this work. \\

We denote by $\left(X_t^n\right)_{1 \leq n \leq N}$ the $N$ realizations generated during the simulation and refer to each sample $X_t^n$ as a \emph{walker}.
The DMC-based AI+RES splitting algorithm shown in Fig.~\ref{fig: schematic1}a proceeds as follows: we simulate $N$ walkers in parallel with the physics-based model (the PlaSim GCM, in our case), starting at time $t_0=0$ with small initial condition perturbations. An iterative duplication process is performed at fixed times $t_1 < \cdots < t_K=t_f$, called \textit{resampling times} and marked by dashed vertical lines in Fig.~\ref{fig: schematic1}. These may be spaced either uniformly or nonuniformly; we chose here to perform resampling at uniform time intervals of length $\tau$, where $\tau$ is a tunable parameter. For notational convenience, we set $X_k^i := X_{t_k}^i$. The resampling steps are governed by a sequence of splitting functions $V_k$, which are themselves functions of the score function $\theta$ (itself obtained from the AI emulator forecasts, in the case of AI+RES; see next section). Walkers with large scores are more likely to be duplicated, while walkers with small scores are more likely to be terminated. Following \cite{abbot2021rare}, we choose a splitting function of the form $V_k = C_k \theta(X_k)$, where $C_k > 0$ is a constant that can vary at each resampling step. In practice, the score function is rescaled before computing the splitting function; see SM Sec.~S4.1 for more details. The choice of $C_k$ is critical, as it controls the strength of the selection process; we discuss it in SM Sec.~S4.2. After resampling, we continue to simulate the selected walkers until the next resampling time, when the process is repeated, until the walkers reach the final time horizon $t_f=t_K$. Each walker is accompanied by a weight $w^i_k$ that allows for unbiased estimation of rare event statistics of interest. \\
The procedure is described in Algorithm \ref{alg:dmc}.  \\

\begin{algorithm}[H]
  \caption{Diffusion Monte Carlo (DMC) algorithm}
  \textbf{Input:} $N$ walkers $\left(X_0^n\right)_{1 \leq n \leq N}$ starting from an initial condition.\\
  \textbf{Initialize:} Choose a sequence of resampling times $0 = t_0 < t_1 < \cdots < t_K$ and a family of splitting functions $(V_k)_k$ depending on a score function $\theta$.
  
  \begin{algorithmic}[1]
  \For{$k = 0$ to $K$}
      \State \textbf{(i) Reweighting:} For each walker $i$
          \If{$k = 0$}
              \State Define initial weights: $w_0^i = \exp\left(V_0(X_0^i)\right)$
          \Else
              \State Define weights: \\
               $w_k^i = \bar{w}_{k-1} \exp\left(V_k(X_k^i) - V_{k-1}(\hat{X}_{k-1}^i)\right)$
          \EndIf
          \State Compute average weight: $\bar{w}_k = \frac{1}{N} \sum_{i=1}^N w_k^i$
          
      \State \textbf{(ii) Resampling:} Create an updated ensemble of walkers $\left(\hat{X}_k^i\right)_{1 \leq i \leq N}$ by sampling $N_k^i$ copies of each $X_k^i$ such that $\sum_i N_k^i = N$ and $\mathbb{E}[N_k^i] = \frac{w_k^i}{\bar{w}_k}$.
          
      \State \textbf{(iii) Simulation:} Integrate the model from $t_k$ to $t_{k+1}$: $ \hat{X}_k^i $ $\rightarrow$ $X_{k+1}^i$.
  \EndFor
  \end{algorithmic}
  \textbf{Note:} Walkers for which $N_k^i = 0$ are terminated and replaced by copies of walkers with $N_k^j \geq 2$.
  \label{alg:dmc}
  \end{algorithm}

One should note that the DMC algorithm applies a selection procedure before the first simulation step. In practice, if one does not want to apply this first selection step, it can be avoided by choosing $V_0=0$. We do this in this work. \\

We perform the resampling step $(ii)$ using the \emph{pivotal sampling} scheme \cite{deville1998unequal}. For a stochastic model, duplicated walkers will separate naturally after resampling. For the deterministic chaotic system that we work with, it is necessary to perturb the model immediately after each resampling step. Following the approach of \cite{ragone2018computation}, we apply perturbations to the spherical harmonics of the logarithm of the surface pressure field. In our implementation, we use a perturbation amplitude of $3\times 10^{-3}$. \\

For any function $\psi$ of the state (or history) of the system, the DMC algorithm yields the following unbiased estimator:
\begin{equation}
\mathbb{E}\left[\psi\left(X_{k}\right)\right] \approx \frac{\bar{w}_{k-1}}{N} \sum_{i=1}^N \psi\left(X_{k}^i\right) e^{-V_{k-1}\left(\hat{X}_{k-1}^i\right)}
\label{eq:unbiased_estimator}
\end{equation} 
where the expectation on the left is with respect to the distribution of the original process without RES.  
Using this formula with the choice $\psi=\mathbbm{1}_{A \geq a}$ for a large $a \in \mathbb{R}$ and an observable $A$ of the trajectory of $X$, one then has access to the probability of $A$ taking extreme values $\left(\mathbb{E}\left[\mathbbm{1}_{  A \geq a}\right]=\mathbb{P}\left[A \geq a\right]\right)$. In our case, this observable is $A_L(t_f) $ defined as the $L$-day average of the 2m temperature over the region $\mathcal{R}$ in Eq.~\eqref{eq:observable}. 

\subsection*{Choice of the AI emulator-based score function}
The AI emulator that we use in AI+RES is trained to predict the state of the simulated physical system (the PlaSim GCM) a short time $\Delta t$ in the future given the state at time $t$, and then cycled autoregressively for longer predictions. 
As shown in Fig.~\ref{fig: schematic1}b, at each resampling time and for each walker $i$, we perform an $M$-member ensemble forecast using the AI emulator from the final state of the walker until the target period $[t_f, t_f+L]$.
This yields an ensemble of $M$ forecasts ${\hat{A}^{i,j}_L(t_f | X^i_k)}_{j=1}^M$ of the observable $A^i_L(t_f)$ (see Eq.~\eqref{eq:observable}) starting from the initial condition $X^i_k$. To the extent that the AI emulator faithfully reproduces the GCM's dynamics, the prediction furnished by the AI emulator ensemble is ideally suited for use in constructing the score function. Our results show that the AI emulator's weather forecast skill is sufficient to provide a very effective score function for AI+RES. More details on our AI emulator implementation are given in SM Sec.~S3.

We combine these ensemble forecasts into the score function by taking the ensemble mean of the forecast observable $\hat{A}^{i,j}_L(t_f | X^i_k)$:
\begin{equation} \label{eq:score_function}
\theta (X^i_k) = \frac{1}{M} \sum_{j=1}^M \hat{A}^{i,j}_L(t_f  | X^i_k)
\end{equation}

We set the number of ensemble members per forecast to $M=100$. Since our score function is defined using the ensemble mean, we do not expect substantial performance gains from increasing $M$. This parameter could, however, play a more significant role under alternative formulations of the score function.

We note that $A^{i}_L(t_f)$ could be predicted by other methods. We choose to leverage the AI emulator because of its proven forecasting skill and numerical efficiency, but other methods could be used. In particular, it would be interesting to test simpler methods based on directly learning the observable $A_L(t_f)$ (or its distribution) from an initial condition, such as in \cite{chattopadhyay2020analog,finkel2021learning,miloshevich_probabilistic_2023-1, finkel2023revealing, mascolo2025gaussian, lovo2025tackling}.  As a baseline, we also test the best possible forecast system, using ensembles of the PlaSim model itself to generate the score function as described in the \emph{Ground truth and baselines} section.

We note that one could use a different statistic calculated from the ensemble forecast as the score function. The ideal score function to estimate the probability of achieving the rare event $A_L(t_f)>a$ for large $a$ should anticipate walker paths leading to the rare event. An alternative score function appropriate for this goal is the conditional probability function $\theta (x) = P ( A_L(t_f)>a | X_{t_k} = x)$, which is referred to as the \emph{committor function} \cite{finkel2020path}. This score function has been proven to be optimal for a splitting algorithm similar to quantile DMC \cite{cerou2006genetic}. However, in our case there is no clear threshold, $a$, since we aim to characterize the entire tail of the distribution of $A_L(t_f)$. Our choice of score function, namely, the mean of the ensemble forecasts, favors the duplication of walkers for which the conditional expectation $\mathbb{E}[A_L(t_f) \mid X_{t_k} = x]$ is large. This approach is more conservative than selecting a fixed threshold a priori.

\subsection{Choice of RES parameters}

The RES hyperparameters used in all experiments are summarized in Table~\ref{tab:parameters}.

\begin{table}[H]
  \centering
  \begin{tabular}{|l|c|c|c|c|c|c|}
  \hline
  \textbf{Params} & $N$ & $M$ & $K$ & $\tau$ & $C_k$ & $L$ \\
  \hline
  \textbf{Values} 
  & 400 
  & 100 
  & 6 
  & 5 days 
  & \makecell{(0, 0, 0, 1.6, 1.8, 2.0)} 
  & 7 days \\
  \hline
  \end{tabular}
  \caption{\textbf{Parameters used in the rare-event algorithm} (Algorithm~\ref{alg:dmc}). 
  $N$ is the number of walkers, $M$ the forecast ensemble size, $K$ the number of resampling steps, 
  $\tau$ the resampling interval, $C_k$ the splitting constants, and $L$ the target duration. 
  Simulations start on July~2 and run until $t_f+L$ with $t_f =$ August~1. 
  The $C_k$ were tuned empirically, and steps with $C_k = 0$ ensure that the walkers are well separated in phase space when selection begins.}
  \label{tab:parameters}
  \end{table}

\bibliography{refs_aires}

\newpage
\beginSM  
\include{supmat}

\end{document}

%% file: supmat.tex
\clearpage
\onecolumngrid

\begin{center}
    \Large\textbf{Supplemental Material for:\\ AI-Boosted Rare Event Sampling to Characterize Extreme Weather}\\[6pt]
\end{center}

\begin{center}
Amaury Lancelin, Alexander Wikner, Laurent Dubus, Clément Le Priol, Dorian S.
Abbot, Freddy Bouchet, Pedram Hassanzadeh, and Jonathan Weare
\end{center}

\setcounter{secnumdepth}{3}
\setcounter{section}{0}
\renewcommand\thesection{S\arabic{section}}
\renewcommand\thesubsection{S\arabic{section}.\arabic{subsection}}
\renewcommand\thesubsubsection{S\arabic{section}.\arabic{subsection}.\arabic{subsubsection}}

\section{Why traditional rare event sampling methods fail}
A large part of the applications of trajectory splitting-based RES algorithms in climate science have focused on sampling large deviations of long-time averages of an observable $A$ \cite{ragone2018computation, wouters2023rare, lepriol2024using, noyelle2025statistical}, where the averaging window typically spans several months. In these approaches, the score function is usually constructed from instantaneous values of $A$ (or from its average over the most recent resampling interval) at the resampling time. While this strategy is effective for sampling long-time averages, it is not well suited for sampling extreme events which occur on time scales shorter than the Lyapunov time of the system—about 5–10 days in the atmosphere. To address such events, a score function capable of anticipating extremes weeks in advance is required. This is precisely the approach taken in the present work, where forecasts from an AI-based emulator of the climate model are leveraged to construct the score function.

\section{PlaSim GCM}

PlaSim \cite{fraedrich_planet_2005} is an efficient general circulation model (GCM) of intermediate complexity that includes a spectral dynamical core that solves the primitive equations for vorticity, divergence, temperature, and humidity. We use PlaSim because it allows rigorous validation of our methodology with large, ground truth ensemble simulations and it has previously been employed in various studies as a benchmark model to explore the statistics of persistent heat extremes~\cite{ragone2018computation, lepriol2024using}, as well as for testing deep learning-based forecasting approaches for such extremes~\cite{miloshevich_probabilistic_2023-1, miloshevich_extreme_2024-3}. PlaSim is coupled with simplified representations of land, ocean, and sea ice boundary layers. We use a T42 spectral truncation mapped onto a $64\times128$ Gaussian grid, with 10 vertical levels. We run PlaSim in a stationary climate configuration with fixed CO$_{2}$ concentration of 360 ppm, with sea surface temperature and sea ice thickness fixed to a climatological, annually repeating cycle derived from the AMIP-II boundary dataset (1870–2006) \cite{taylor_sea_2001}, via linear interpolation of monthly climatologies. In the land model, surface temperature is computed through a linear energy balance approach, and soil hydrology is represented using a bucket model with regionally varying water holding capacity \cite{Lunkeit2011PlanetS}.

\section{AI Weather Emulator}
\label{sec:emulator}
\begin{figure}[htbp]
  \centering
  \includegraphics[width=0.9\linewidth]{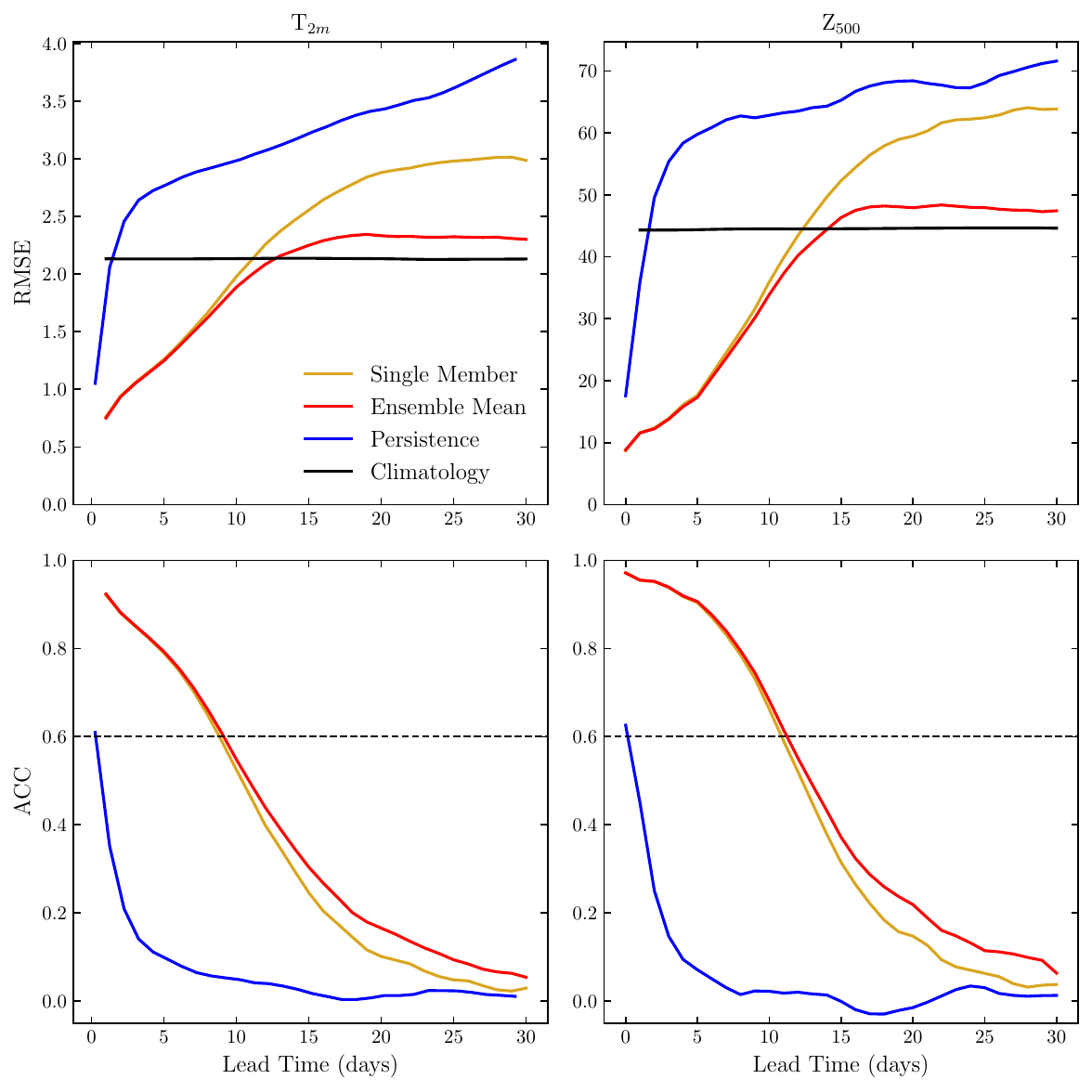}
  \caption{\textbf{Weather forecast skill of the AI emulator}. Global RMSE (top) and ACC (bottom) as a function of lead time for T$_{2m}$ (left) and Z$_{500}$ (right). Gold and red lines show the AI emulator with a single-member forecast and a 100-member ensemble forecast, respectively. The blue line corresponds to a persistence forecast. Horizontal black lines in the RMSE panels indicate the climatological forecast. Dashed horizontal lines in the ACC panels at 0.6 mark the conventional threshold below which forecasts are no longer considered skillful.}
\label{fig: emulator_skill}
\end{figure}

We train a deep neural network-based AI emulator to predict the full atmospheric state of the PlaSim GCM a short time $\Delta t$ in the future given the current atmospheric state $X_{\text{atm}, t}$ and known boundary conditions $X_{\text{bnd}, t}$:
\begin{equation}
    \tilde{X}_{\text{atm},t+\Delta t} = \mathcal{F}_\Theta(X_{\text{atm}, t}, X_{\text{bnd}, t}).
\end{equation}
where $\Theta$ are the trainable parameters of the neural network $\mathcal{F}_\Theta$.
After initialization, the AI emulator may be cycled autoregressively to obtain long-term predictions. While the AI emulator is trained as a deterministic forecast model, we can generate ensemble forecasts by perturbing the initial conditions. See Fig.~\ref{fig: emulator_skill} for an assessment of the forecast skill of the AI emulator.

\subsection{Architecture}\label{sec:est_emulator}

We adopted the PanguWeather 3D Earth-specific transformer (EST) architecture for our AI emulator, which is specifically tailored for global geophysical fields by incorporating Earth-specific inductive biases such as spherical positional encoding and longitudinal periodicity. This architecture has demonstrated strong skill in medium-range forecasting  as well as stability in long-term autoregressive simulations \cite{chattopadhyay_challenges_2024}. To adapt our emulator to the PlaSim GCM, we made a number of modifications to the original EST used in PanguWeather. To account for the decrease from PanguWeather's $0.25^\circ$ horizontal resolution to Pangu-PlaSim's $~2.8^\circ$, we decreased the horizontal patch size from $(4,4)$ to $(2,2)$ and the horizontal attention window size from $(6,12)$ to $(2,4)$. We additionally increased the vertical attention window size to include all tokens to account for the inclusion of the top-of-atmosphere boundary variables. The original patch recovery layer was replaced with a pixel shuffle deconvolution layer, as in the ArchesWeather model of~\cite{couairon2024archesweather}, to mitigate artifacts arising from patch recovery, while an additional convolutional layer for each variable type (3D prognostic, 2D prognostic, and 2D diagnostic) was added following patch recovery to allow for additional processing of the full-resolution recovered variables.

\subsection{Training and Validation}
Prior to input into the AI emulator, the atmospheric fields from the PlaSim GCM are vertically interpolated from the original 10 topography-following sigma levels to 13 equipressure surfaces. The details of these prognostic atmospheric variables, as well as the other variables input and output by the AI emulator, can be found in Table~\ref{tab:emulator_variables}. The AI emulator is trained to minimize a weighted sum of mean absolute errors between the model output and the prognostic and diagnostic variables at a timestep $\Delta t = 6$ hours in the future. The precise weighted loss function is similar to that used in the PanguWeather model:

\begin{equation}
  \mathcal{L}(\tilde{X}_{\text{atm},t}, X_{\text{atm},t}) = \frac{1}{N_a N_z}\sum_{a,z}\left\|\tilde{X}_{a,z,t} - X_{a,z,t}\right\|_{1} + \frac{1}{4 N_s}\sum_{s}\left\|\tilde{X}_{s,t} - X_{s,t}\right\|_{1} + \frac{1}{4 N_d}\sum_{d}\left\|\tilde{X}_{d,t} - X_{d,t}\right\|_{1},
\end{equation}

where, $X_{a,z,t}$ denotes the value of an atmospheric variable $a$ at pressure level $z$ at time $t$,  $X_{s,t}$ and $X_{d,t}$ denote, respectively, the values of surface and diagnostic variables $s$ and $d$ at time $t$, and $\tilde{X}_{\text{atm},t} = \mathcal{F}_\Theta(X_{\text{atm}, t-\Delta t}, X_{\text{bnd}, t-\Delta t})$. Each subscripted $N_{\circ}$ denotes the total number of variables of the corresponding type. The net effect of the addition of the normalized MAE for each variable type and the factor of $1/4$ multiplying the surface and diagnostic variable error is that surface and diagnostic variable error is weighted more highly in the loss relative to each atmospheric variable at a particular pressure level.

The AI emulator is trained on data from a single 111-year PlaSim control run. The first 10 years are discarded for spin-up, year 11 is reserved for validation, and the final 100 years are used for training. Prior to input, this data is standardized using statistics from the 100-year training data set so that each variable at each pressure has a global mean of $0$ and global standard deviation of $1$. We additionally add a Gaussian noise vector independently sampled from $\mathcal{N}(0, 2.5\times10^{-3})$ to each standardized atmospheric variable input to the emulator during training. This noise addition mitigates an instability that can arise during autoregressive prediction with the AI emulator due to the reduced spectral resolution of the PlaSim GCM's dynamical core relative to its spatial resolution. All other AI emulator training parameters and methods are taken from standard methods for training large-scale vision transformer-based models (see Table~\ref{tab:hyperparams}).

\subsection{Training hyperparameters}
\begin{table}[ht]
\centering
\caption{Hyperparameters used to train the Pangu-PlaSim emulator.}
\label{tab:hyperparams}
\begin{tabular}{l l}
\toprule
\textbf{Hyperparameter} & \textbf{Value} \\
\midrule
Optimizer             & AdamW \\
Learning rate scheduler & OneCycleLR \\
Total epochs          & 100 \\
Annealing epochs      & 10 \\
Min. learning rate    & 1e-6 \\
Max. learning rate    & 1e-4 \\
Batch size            & 64 \\
Weight decay          & 3e-6 \\
Drop path rate        & 0.2 \\
Epoch selection       & Lowest loss\\
\bottomrule
\end{tabular}
\end{table}

\subsection{Ensemble Forecasting}
Ensemble forecasts are generated by adding independent initial condition perturbations to all atmospheric variables at the beginning of the forecast. Perturbations are sampled from a Gaussian distribution with mean $0$ and a standard deviation such that the resulting ensemble spread over time is similar to that of the PlaSim GCM with our selected magnitude of perturbations to the initial surface pressure (see Section~\ref{sec:perturbation_method}).

\begin{table}[htbp]
  \renewcommand\cellalign{tl}
  \renewcommand\theadalign{tl}
  \centering
  \begin{tabularx}{\textwidth}{|l||X|}
  \hline
  \makecell[l]{Boundary\\Variables} & \makecell[l]{Land-sea mask, surface roughness, and surface geopotential height (constant); \\ Sea surface temperature, sea ice cover, and total incoming solar radiation \\(yearly repeating)} \\ \hline
  \makecell[l]{Prognostic \\ Atmospheric \\ Variables} & \makecell[l]{Specific humidity, temperature, U and V wind, and geopotential height at \\ 50, 100, 150, 200, 250, 300, 400, 500, 600, 700, 850, 925, and 1000 hPa} \\ \hline
  \makecell[l]{Prognostic \\ Surface \\ Variables} & $\log(\text{Surface pressure})$ and air temperature at 2 m \\ \hline
  \makecell[l]{Diagnostic \\ Variables} & Precipitation accumulated over 6 Hours \\ \hline
  \end{tabularx}
  \caption{\textbf{Description of the variables input to and predicted by the AI emulator}. Boundary variables are prescribed variables that are only input to the AI emulator. Prognostic variables refer to the variables both input to and predicted by the AI emulator. Diagnostic variables are only predicted by the AI emulator.}
  \label{tab:emulator_variables}
\end{table}

\section{Rare event algorithm: Diffusion Monte Carlo} 
\label{sec:choosing_hyperparameters_algo}

We used a rare event algorithm called \emph{Diffusion Monte Carlo} (DMC), in a version that closely follows the formulations proposed in \cite{webber2019} and \cite{abbot2021rare}. The procedure is described in Algorithm \ref{alg:dmc}.

\subsection{\emph{Quantile} Diffusion Monte Carlo (QDMC)} \label{sec:QDMC}

As mentioned, the splitting functions $\left(V_k\right)_k$ can critically influence the behavior of the algorithm. A frequently used choice is $V_k(X) = C_k \theta(X)$, where $C_k > 0$ controls the duplication rate of walkers. While simple, this formulation has drawbacks. In nonlinear systems, walkers with large $\theta$ values may produce an excessive number of offspring, eventually leading to the so-called \emph{extinction} phenomenon in which the population becomes overly concentrated around a few trajectories. Although such a scheme remains unbiased, the variance of the resulting estimator can be prohibitively large. On the opposite end, if selection is too weak, particle weights are nearly equal, and the method essentially reduces to naive Monte Carlo sampling—again yielding high variance. Proper tuning is therefore essential to balance these two extremes.\\

To alleviate these issues, \cite{webber2019} proposed a modification of DMC designed to improve robustness. The idea is to rescale the score function $\theta$ dynamically at each resampling step so that its distribution matches a prescribed target distribution $\nu_k$, most often taken as Gaussian. The transformed score, denoted $\theta_k^{\prime}$, is then used in the splitting and pruning procedure. This additional step gives the algorithm its full name: \emph{Quantile} Diffusion Monte Carlo (QDMC).\\

The distinguishing feature of QDMC is this rescaling step. After estimating the empirical distribution of $\theta(X_{t_k})$, one constructs a transformation $\theta_k^{\prime} = \gamma_k(\theta)$ such that the distribution of $\theta_k^{\prime}(X_{t_k})$ is close to $\nu_k$. More precisely, QDMC introduces a transport map from the empirical law of $\theta_k$ to the target distribution $\nu_k$ via

\begin{equation}
\gamma_k(y) = F_{\nu_k}^{-1}(F_{\theta_k}(y)),
\end{equation}

where $F_{\theta_k}$ is the cumulative distribution function (CDF) of $\theta_k$, and $F_{\nu_k}^{-1}$ is the quantile function (inverse CDF) of $\nu_k$, typically $\mathcal{N}(0,1)$.\\

An attractive property for rare-event simulation algorithms is invariance with respect to monotone bijective transformations of the score function $\theta$. Unlike standard DMC, QDMC satisfies this invariance, which provides greater flexibility. Nevertheless, some elements of the method, such as the choice of target distribution or the splitting constants $C_k$, remain somewhat arbitrary and require further discussion.\\

In our experiments, we observed that when the number of walkers $N$ is too small, the quantile-mapping step—based on the empirical estimate of $F_{\theta_k}$, which only relies on the ranking between walkers—can have an undesired effect, sometimes assigning very different weights to walkers with very close scores. To avoid this issue, we opted for a simpler alternative by rescaling the score function $\theta$ at each resampling step using the following transformation

\begin{equation} \label{eq:rescaling_theta}
\tilde{\theta}_k(X^i_k) = \frac{\theta(X^i_k) - \mu_{\theta_k}}{\sigma_{\theta_k}},
\end{equation}

where $\mu_{\theta_k}$ and $\sigma_{\theta_k}$ denote the mean and standard deviation of $\theta$ across all walkers at time $t_k$.
The rescaled score function $\tilde{\theta}_k$ is then used to compute the splitting function via 
\begin{equation} \label{eq:splitting_function}
V_k = C_k \tilde{\theta}_k.
\end{equation}
Using this approach, we obtained results that were very similar to those achieved with the quantile-mapping step for large values of $N$ and better at small $N$. 

\subsection{Scaling the constants $C_k$} \label{sec:scaling_constants_Ck}

With our choice of splitting function in Eq.~\eqref{eq:splitting_function}, some key parameters left to be tuned are the splitting constants $C_k$, which determine the rate at which walkers are duplicated at each resampling step.

A higher value of $C_k$ results in a larger number of walker clones, making the algorithm more selective. However, if $C_k$ is too large, it may lead to the \emph{extinction} phenomenon described in Section \ref{sec:QDMC}, whereas if it is too small, the algorithm behaves similarly to direct sampling by not sampling enough extreme events.
It is often desirable not to use the same constant $C_k$ for all resampling steps. In our case, we are forecasting the future state of the trajectory to compute the score function, and the uncertainty in the forecast decreases as we approach the final time $t_f$. Therefore, we chose to set $C_k = C e^{-\alpha (t_f - t_k)}$, where $C > 0$ is a constant and $\alpha > 0$ is a decay factor. A similar approach is used in \cite{webber2019}. Both $C$ and $\alpha$ are hyperparameters of the algorithm. Using a non-exhaustive grid search, we found that setting $C = 2.0$ and $\alpha = 0.02$ yielded good results in terms of variance of the final estimators of rare event probabilities. Furthermore, we turned off resampling at the first three resampling times, $t_1$, $t_2$, and $t_3$, by setting $C_1 = C_2 = C_3 = 0$ to ensure that the walkers were sufficiently separated in phase space before actually applying a resampling. \\

While we expect that further tuning of the hyperparameters could improve performance, determining the optimal values for all hyperparameters involved in the algorithm is computationally expensive and, although mathematically interesting, lies beyond the scope of this work.

\subsection{Choice of the resampling scheme}

\begin{algorithm}[H]
  \caption{Pivotal Resampling (adapted from \cite{deville1998unequal})}
  \textbf{Input:} Normalized weights $\{\tilde{w}_k^1, \dots, \tilde{w}_k^N\}$ with $\sum_{i=1}^N \tilde{w}_k^i = N$.\\
  \textbf{Output:} Number of clones $N_k^i$ for each particle $i=1,\dots,N$ at resampling step $k$.
  
  \begin{algorithmic}[1]
    \State \textbf{Step 1: Decomposition.} For each $i$, decompose
    \[
    \tilde{w}_k^i = \lfloor \tilde{w}_k^i \rfloor + \delta_k^i, \qquad \delta_k^i \in [0,1).
    \]
  
    \State \textbf{Step 2: Pivotal sampling on fractional parts.}
    \While{there exist at least two indices $i,j$ with $\delta_k^i,\delta_k^j \in (0,1)$}
        \State Let $s = \delta_k^i + \delta_k^j$.
        \If{$s \leq 1$}
            \State With probability $\frac{\delta_k^i}{s}$: set $\delta_k^i \gets s, \; \delta_k^j \gets 0$.
            \State Otherwise: set $\delta_k^j \gets s, \; \delta_k^i \gets 0$.
        \Else
            \State With probability $\frac{1-\delta_k^j}{2-s}$: set $\delta_k^i \gets 1, \; \delta_k^j \gets s-1$.
            \State Otherwise: set $\delta_k^j \gets 1, \; \delta_k^i \gets s-1$.
        \EndIf
    \EndWhile
  
    \State \textbf{Step 3: Final number of clones.}
    \For{$i = 1$ to $N$}
        \State $N_k^i \gets \lfloor \tilde{w}_k^i \rfloor + \delta_k^i$
    \EndFor
  
    \State \Return $\{N_k^i\}_{i=1}^N$
  \end{algorithmic}
  \label{alg:pivotal}
\end{algorithm}

As described in Algorithm~\ref{alg:dmc} in the main text, at each resampling step $k$, we duplicate each walker a random number of times $N_k^i$, subject to the constraints
\begin{equation}
\mathbb{E}[N_k^i] = \frac{w_k^i}{\bar{w}_k} \coloneq \tilde{w}_k^i, \quad \sum_{i=1}^N N_k^i = N.
\end{equation}
Several resampling strategies can be used to generate the integers $N_k^i$ while preserving these properties. In this work, we employ the \emph{pivotal} resampling scheme \cite{deville1998unequal}, detailed in Algorithm \ref{alg:pivotal}, which we found yields lower variance than alternative approaches.

\subsection{Choice of the initial condition}

Motivated by sampling the most extreme events possible, we chose for the France region an initial condition at $t_0$ corresponding to the largest value of the observable $A_L(t_f)$ (see Eq.~\eqref{eq:observable} in the main text) in the 100-year training set of the emulator. For the U.S. Midwest region, in order to test the robustness of the algorithm to the choice of initial condition, we instead selected a random initial condition at $t_0$ from the same training set.

\subsection{How we generate PlaSim ensembles} \label{sec:perturbation_method}

\begin{figure}[htbp]
  \centering
  \includegraphics[width=0.9\linewidth]{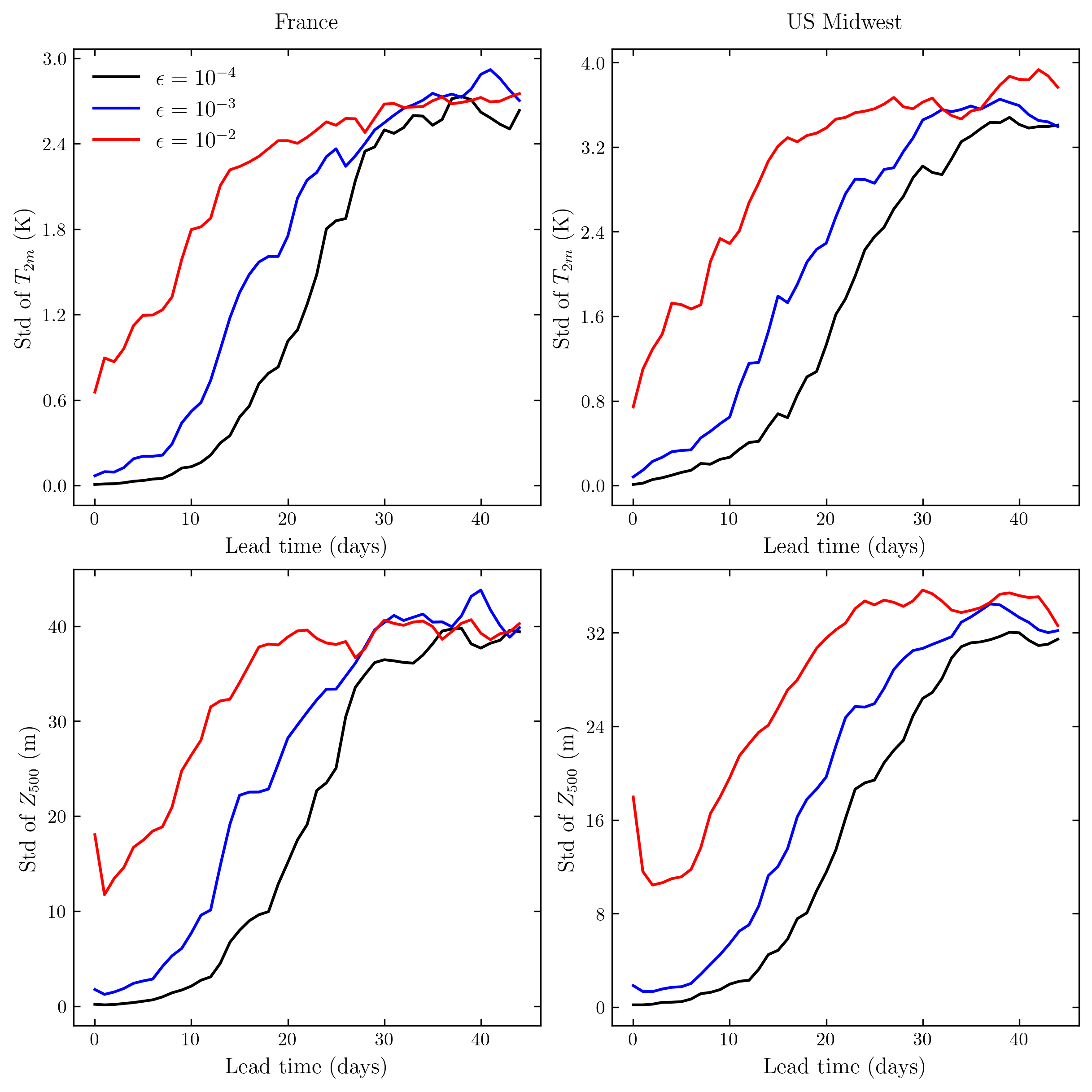}
  \caption{\textbf{Time evolution of the standard deviation of PlaSim ensembles for different values of perturbation amplitude $\epsilon$}. The perturbation is applied to the spherical harmonic coefficients of the logarithm of the surface pressure field of the initial condition. The ensemble size is 100 and the results are averaged over 10 independent initial conditions. Top panels: Standard deviation of the spatially-averaged daily T$_{2m}$ for the France (left) and the U.S. Midwest (right) regions.  Bottom panels: Standard deviation of the spatially-averaged daily $Z_{500}$ for the France (left) and the U.S. Midwest (right) regions. The black, blue, and red curves show results for $\epsilon = 10^{-4}$, $\epsilon = 10^{-3}$, and $\epsilon = 10^{-2}$, respectively.}
  \label{fig:perturbation_amplitude}
\end{figure}

As described in the \emph{End Matter} in the main text, for the duplication of walkers at each resampling step in the DMC algorithm to have an effect, it is necessary to perturb their initial conditions. Following \cite{ragone2018computation}, we introduce perturbations by adding a small noise to the coefficients of the spherical harmonics of the logarithm of the surface pressure field. Fig.~\ref{fig:perturbation_amplitude} shows the time evolution of the standard deviation of 100-member PlaSim ensembles generated with different values of $\epsilon$. In our DMC implementation, we use a perturbation amplitude of $\epsilon = 3 \times 10^{-3}$. We tested several values of $\epsilon$ and found that this choice provides a good balance: it enhances the ability of the PlaSim ensembles to explore the phase space, while remaining small enough to avoid altering the final statistics of the target observable or introducing nonphysical discontinuities in the trajectories. We anticipate that more sophisticated perturbation strategies, such as Bred vectors \citep{toth1993ensemble} or \emph{early splitting} \citep{finkel_bringing_2024}, could further improve the performance of the algorithm.

\section{Derivation of the computational speedup factors}

In the following, we derive two criteria to evaluate the computational gains provided by the AI+RES algorithm compared to DS.
We make the key assumption that the computational cost of running the AI emulator ensemble forecasts is negligible compared to that of running the physics-based model (in this case, the GCM). For reference, \cite{pathak2022fourcastnet} report speedups of up to $\mathcal{O}(10^4\!-\!10^5)$ and energy-consumption reductions of $\mathcal{O}(10^4)$ for FourCastNet (AI emulator) relative to the IFS model (numerical weather prediction model). In our case, the cost comparison between the AI emulator and PlaSim is less favorable, as PlaSim is specifically designed to be computationally inexpensive. This is precisely what allows us to run a 50,000-member PlaSim ensemble within reasonable computational limits, enabling a robust validation of the methodology. Nevertheless, our analysis indicates that our emulator runs approximately 10 times faster on a single A100 GPU than PlaSim using 64 Intel Xeon Platinum 8160 CPU cores, even without any emulator variable or weight pruning or inference-specific optimizations. This does not undermine our assumption of negligible cost for the AI emulator, as the present study serves as a proof of concept. 

\subsection{Sampling speedup factor}
\label{sec:sampling_speed_up}
One of the goals of RES is to efficiently generate a large catalog of extreme events. To quantify the improvement achieved by RES in this task, we count the average number of events, $n_a$, that the algorithm samples above a given threshold $a$ (corresponding to a return period $T_a$), using a fixed budget of $N_{RES}$ walkers.

To obtain the same number of events with DS, one would need to have on average a budget of $N_{DS}(a) = T_a n_a$ members by definition of the return period. We then introduce the \emph{Sampling Speedup Factor} (SSUF) as the ratio between the DS and RES budgets needed to sample the same number of extreme events:
\begin{equation}
  \text{SSUF}(a) = \frac{N_{DS}(a)}{N_{RES}} = \frac{T_a n_a}{N_{RES}}
  \label{eq:sampling_speed_up_ratio}
\end{equation}

We also introduced a metric to quantify the diversity of the sampled trajectories (Section~\ref{sec:diversity_of_sampled_trajectories}), demonstrating that the generated samples are not overly correlated.

\subsection{Variance speedup factor}
\label{sec:variance_speed_up}
The other main objective of RES is to reduce the uncertainty in rare probability estimates. Following \cite{webber2019, abbot2021rare}, we also quantify speedup based on the reduction in the variance of estimated rare event probabilities. Suppose we have used RES to obtain an estimate of the probability of exceedance $\hat{p}_a$ above the value $a$ with a variance $\sigma_{RES}(a)^2$ using a budget of $N_{RES}$ walkers. We can compare this with the variance of the DS estimator (Eq.~\eqref{eq:empirical_probability_of_exceedance_DNS}) as a function of $N$ and $a$, which is given by
\begin{equation}
  \sigma_{DS}(a,N)^2 = \frac{p_a (1 - p_a)}{N}
  \label{eq:variance_DNS}
\end{equation}
We can then estimate the number of DS samples that would be necessary to produce a similarly accurate estimate as RES as follows
\begin{equation}
N_{DS}(a) \approx \frac{\hat{p}_a (1 - \hat{p}_a)}{\hat{\sigma}_{RES}(a)^2}.
\end{equation}
We then define the \emph{Variance Speedup Factor} (VSUF) as 
\begin{equation}
  \text{VSUF}(a) = \frac{N_{DS}(a)}{N_{RES}} = \frac{p_a (1 - p_a)}{\hat{\sigma}_{RES}(a)^2 N_{RES}} = \frac{T_a-1}{T_a^2 \hat{\sigma}_{RES}(a)^2 N_{RES}}
  \label{eq:variance_speed_up_ratio}
\end{equation}
In practice, we estimate the variance of RES, $\hat{\sigma}_{RES}(a)^2$, empirically from only 10 independent realizations of the algorithm, as running a larger number of experiments is computationally expensive. Consequently, the estimate of $\hat{\sigma}_{RES}(a)^2$ itself is subject to sampling error. In Section~\ref{sec:error_bars_DMC}, we also present alternative estimates of the variance using only a single realization of the algorithm. 

\subsubsection{Computing error bars for the DMC algorithm} \label{sec:error_bars_DMC}

To compute the empirical variance estimates with the DMC algorithm, we performed 10 independent runs of the algorithm with the same hyperparameters and the same initial conditions. The same methodology is used in \cite{lepriol2024using,ragone2018computation,ragone2021rare}.

However, it is possible to compute variance estimates for the DMC algorithm using a single realization of the algorithm \cite{lee2018variance, webber2019, abbot2021rare}. As discussed in \cite{abbot2021rare}, two different approaches can be used to estimate the variance in DMC: one that systematically overestimates it, and another that underestimates it.
For a general DMC estimator $\hat{f_\psi}$ of the quantity $\mathbb{E}[\psi(X_k)]$, defined via Eq.~\eqref{eq:unbiased_estimator} in the main text, the \emph{pessimistic} variance estimate, which leans toward overestimation, is given by

\begin{equation}
\hat{\sigma}^2_{\mathrm{pess}} = \frac{1}{N^2} \sum_{i=1}^N \left| \sum_{\mathrm{anc}(X_k^j) = i}  \psi\left(X_{k}^j\right) \bar{w}_{k-1} e^{-V_{k-1}\left(\hat{X}_{k-1}^j\right)} \right|^2 - \hat{f_\psi}^2.
\label{eq:variance_pessimistic}
\end{equation}

Here, $\mathrm{anc}(X_k^j)$ identifies the index of the original ancestor at $t=t_0$ from which the trajectory $X_k^j$ descends.
Conversely, an \emph{optimistic} estimator, which tends to underestimate the variance, is expressed as

\begin{equation}
\hat{\sigma}^2_{\mathrm{opt}} = \frac{1}{N^2} \sum_{i=1}^N \left| \psi\left(X_k^i\right) \bar{w}_{k-1} e^{-V_{k-1} \left(\hat{X}_{k-1}^i\right)} \right|^2 - \hat{f_\psi}^2.
\label{eq:variance_optimistic}
\end{equation}

\begin{figure}[htbp]
  \vspace{-0.\intextsep}
  \centering
  \begin{tabular}{cc}
      \includegraphics[width=0.45\linewidth]{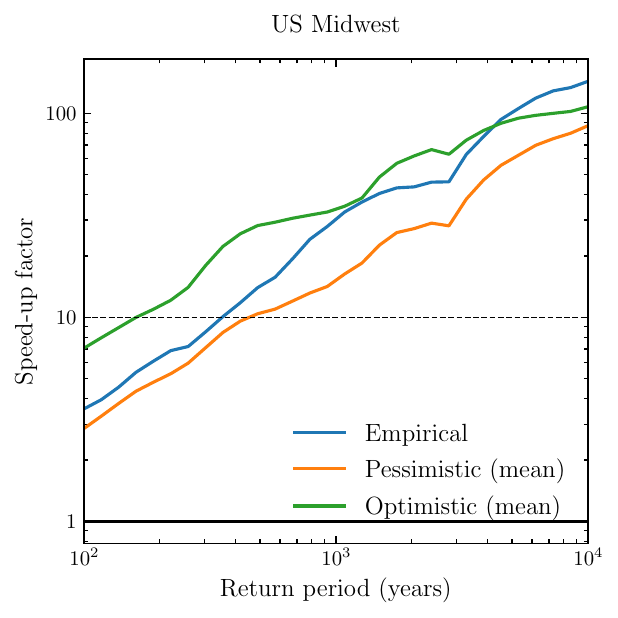} &
      \includegraphics[width=0.45\linewidth]{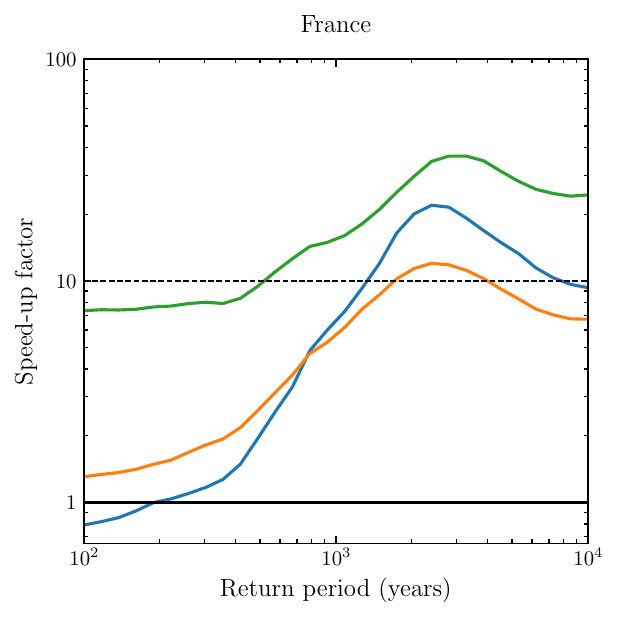} \\
  \end{tabular}
  \caption{\textbf{Comparison of speedup factors} obtained via the empirical, the pessimistic (Eq.~\eqref{eq:variance_pessimistic}) and optimistic (Eq.~\eqref{eq:variance_optimistic}) variance estimates. The experiments shown here are described in Table~\ref{tab:parameters} in the main text, for the U.S. Midwest (left panel) and France (right panel). The variance speedup factor measures the reduction in the variance of return time estimates compared to direct sampling (Eq.~\eqref{eq:variance_speed_up_ratio}). The blue curve shows the empirical speedup ratio obtained by running the AI+RES algorithm 10 times and computing the variance of the return time estimates across the 10 runs. The green curve shows the mean over the 10 optimistic variance estimates, each computed from a single realization of the algorithm. The orange curve shows the mean over the 10 pessimistic variance estimates, each computed from a single realization of the algorithm.}
  \vspace{-\intextsep}
  \label{fig:speed_up_emp_pess_opt}
\end{figure}

The intuition behind these two forms is the following: the optimistic estimator considers each trajectory as an independent contribution, similar to importance sampling \cite{liu2001monte}, while the pessimistic estimator aggregates all descendants of a given ancestor into a single effective data point. In practice, the effective number of independent samples lies between these two limiting cases.
As there is no theoretical guarantee that one of these estimators is better than the other, we preferred to use the empirical variance computed from the 10 independent runs of the algorithm to compute variance speedup factors (see Eq.~\eqref{eq:variance_speed_up_ratio} and Fig.~\ref{fig:speed_up_factors} in the main text). In Fig.~\ref{fig:speed_up_emp_pess_opt}, we verify that the empirical variance lies between the pessimistic and optimistic estimators, within statistical error.

\section{Return period curves}

To assess the ability of our model to accurately estimate the probability of very rare events, a classical diagnostic is the return-period curve. Since we have constrained the target observable to a specific period in summer, we only observe a single event per year (i.e., per summer simulated). In this setting, the return period $T_a$ associated with a return value $a$ is defined as the inverse of the probability of exceedance $p_a = \mathbb{P}(A_L(t_f) > a)$. A return-period curve is then defined as the plot of the return values $a$ as a function of their corresponding return periods $T_a$.

Let $(a_1, a_2, \ldots, a_N)$ be all the values of the observable $A_L(t_f)$ sampled by DS or RES. To draw the return-period curve, we first need to estimate $p_{a_j}$ for each value $a_j$. For DS, we estimate the probability of exceedance with the following empirical estimator:
\begin{equation}
  \mathbb{P}(A_L(t_f) > a_j) \approx \frac{1}{N} \sum_{i=1}^N \mathbbm{1}_{A_L(t_f) > a_j}\left(Y_{t_f+L}^i\right)
  \label{eq:empirical_probability_of_exceedance_DNS}
\end{equation}
where, in this expression, $Y_t^i$ are independent PlaSim simulations without RES. 

For the RES algorithm, we use the formula \eqref{eq:unbiased_estimator} with the observable $\psi = \mathbbm{1}_{A_L(t_f) > a_j}$ to compute the probability of exceedance. Namely,
\begin{equation}
 \mathbb{P}(A_L(t_f) > a_j) \approx \frac{\bar{w}_K}{N} \sum_{i=1}^N \mathbbm{1}_{A_L(t_f) > a_j}\left(X_{t_f+L}^i\right) e^{-V_K\left(\hat{X}_{K}^i\right)}
  \label{eq:unbiased_estimator_probability_of_exceedance}
\end{equation} 
The return period for each value is then obtained by taking the inverse of its probability of exceedance. 

\subsection{Interpretation of the return times}
Since we have constrained the target period to a specific moment in summer, we only observe a single event per year (i.e., per summer simulated). In this setting, the return time associated with a return value $a$ is defined as the inverse of the probability of the event $A_L(t_f) > a$. \\

In our framework, these return times can be interpreted as the average number of years one would have to wait before observing an event of magnitude greater than $a$, assuming we repeatedly simulate the same summer, each time conditioned on the same initial state at the beginning of the season. In that sense, they correspond to \emph{conditional return times}. Given our finite computational budget, we chose this approach to allow for a larger number of experiments, thereby enabling a robust validation of the algorithm. However, extending this to the unconditional case is straightforward. This can be achieved either by initiating the simulations earlier in spring—so that the influence of initial conditions no longer persists into the target period $[t_f, t_f+L]$—or by relying on appropriate heuristics, such as those proposed in \cite{bloin2025estimating}. 

\section{Characterizing the diversity of trajectories sampled with the Rare event algorithm} \label{sec:diversity_of_sampled_trajectories}
\begin{figure}[htbp]
  \centering
  \includegraphics[width=0.45\linewidth]{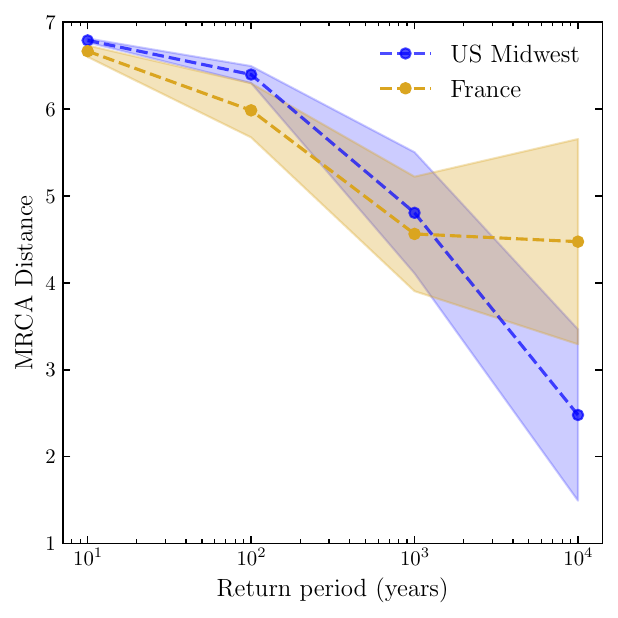}
  \caption{
    \textbf{Most Recent Common Ancestor Distance (MRCAD) for AI+RES sampled trajectories}. The figure shows results for France and the U.S. Midwest with the setup given in Table~\ref{tab:parameters} in the main text. The MRCAD is the average number of resampling steps since the current population last shared a common ancestor (max = 7, min = 1, with $K=6$ resampling steps). The solid lines show the mean MRCAD over 10 independent realizations of the AI+RES algorithm, with the shading indicating the $95\%$ confidence interval. A larger MRCAD indicates greater trajectory diversity. The AI+RES generated trajectories maintain substantial diversity, yielding a representative catalog of rare events.}
  \label{fig: MRCA_distance}
\end{figure}

The DMC algorithm duplicates walkers at each resampling step to generate a large catalog of extreme events. When two walkers share the same parent, their trajectories are highly correlated, which reduces the diversity of the sampled events. To quantify this diversity, we introduce the \emph{Most Recent Common Ancestor Distance} (MRCAD). The MRCAD is defined as the average number of resampling steps since a given population $\mathcal{P}$ last shared a common ancestor, with a maximum of 7 and a minimum of 1 for $K = 6$ resampling steps. A higher MRCAD indicates greater diversity in the sampled trajectories.

To assess how diversity is affected when focusing on rarer events, we computed the MRCAD for different populations $\{\mathcal{P}_p\}_p$, where $\mathcal{P}_p$ consists of walkers whose associated values of $A_L(t_f)$ have a return time larger than $1/p$. Figure~\ref{fig: MRCA_distance} shows the MRCAD as a function of the return period $1/p$ for AI+RES experiments. We observe that trajectory diversity remains relatively high (between 4 and 5) up to return periods of approximately $10^3$ years for both regions studied. Beyond this threshold, diversity decreases rapidly for the U.S. Midwest, whereas it remains high for France. These results indicate that the AI+RES algorithm successfully samples a large catalog of extreme events with sufficient diversity, rather than merely replicating a single trajectory with an extreme value of $A_L(t_f)$.

\section{Baseline methods}
\subsection{Extreme value theory fits}
To compare the uncertainty of our return period estimates from RES with those obtained via extreme value theory (EVT), we use the standard peak-over-threshold (POT) method, fitting a generalized Pareto distribution (GPD) to the data.

The family of distributions $GPD(\sigma, \xi)$ is suitable to characterize the tail of random variables. If $V_1, \dots, V_n \in \mathbb{R}$ is a sequence of independent and identically distributed random variables, then, under suitable assumptions,  the asymptotic distribution of the excesses $Z = V - u \ \vert\ V > u$ over a sufficiently high threshold $u$ is given by:

\begin{equation}
F_{(\sigma, \xi)}(z) = 1 - \left[ 1 + \xi \frac{(z - u)}{\sigma} \right]^{-1/\xi}_{+},
\end{equation}
where $[a]_+ = \max\{0, a\}$, $\sigma > 0$ is the scale parameter, $-\infty < \xi < \infty$ is the shape parameter and $F_{(\sigma, \xi)}$ is the cumulative distribution function. The parameters $(\sigma, \xi)$ are then fitted on the exceedances $Z$ and used to determine the GPD return values.

To ensure a fair comparison in terms of sample size, the GPD distributions were fitted on datasets of the same size as the AI+RES experiments (namely $N = 400$ in Fig.~\ref{fig:return_time_curves} of the main text), on 100 independent training datasets. The gray shading represents the 10th–90th percentile range of return-period curves obtained from the GPD fits across these datasets.

In the peak-over-threshold method, the key hyperparameter to tune is the threshold $u$ above which the data is modeled. We determine $u$ using the heuristic of the \emph{Mean Residual Life} plot (\cite{coles2001introduction}), selecting a percentile that can be applied consistently across all datasets. Based on this criterion, we retain the 90th percentile of the data as the threshold.
Parameters were estimated using a maximum likelihood approach with the \texttt{scipy} Python package (\cite{2020SciPy-NMeth}).

\subsection{Using PlaSim itself to produce forecasts} \label{sec: PFS+RES}

\begin{figure}[htbp]
  \centering
  \includegraphics[width=0.45\linewidth]{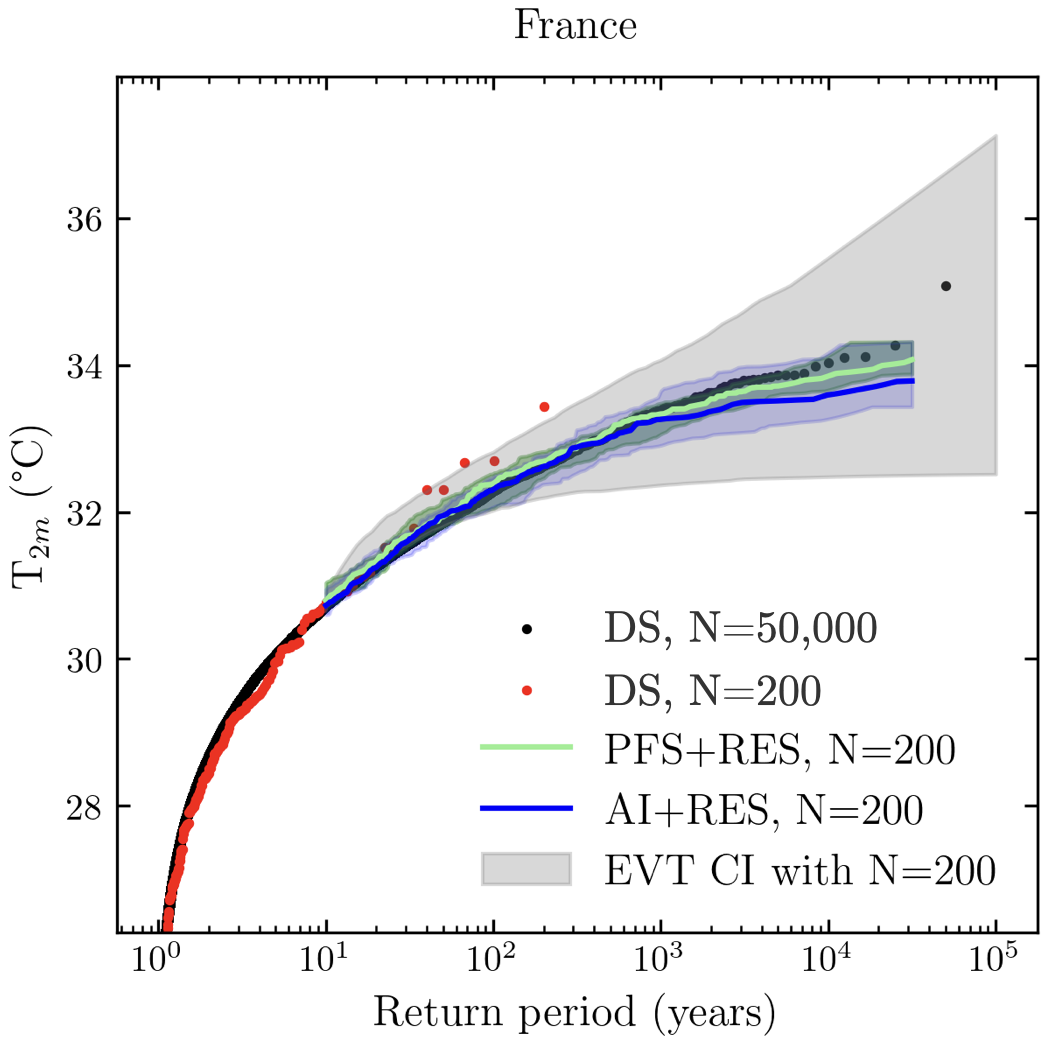}
  \caption{\textbf{Return-time curve comparison between AI+RES and PFS+RES.} The setup matches that described in Table~\ref{tab:parameters} in the main text, except that it uses $N=200$ walkers (instead of $N=400$), since larger values are computationally prohibitive for PFS+RES. Black dots are the empirical return periods obtained with a $50,000$-member control simulation. The red dots are from this control simulation but using the same computational budget as the AI+RES algorithm ($N=200$). The solid blue line shows the median return time curve produced by 10 independent realizations of the AI+RES algorithm with $N=200$ walkers and the blue-shaded area represents the 10th to 90th percentile range. The solid green line shows the median return time curve produced by 10 independent realizations of the PFS+RES algorithm with $N=200$ walkers and the green shaded area represents the 10th to 90th percentile range. The gray shaded area is obtained by fitting different GPD distributions with independent $N=200$ training datasets, and showing the 10th to 90th percentile range.}
  \label{fig: return_time_curve_PFS}
\end{figure}

\begin{figure}[htbp]
  \includegraphics[width=0.9\linewidth]{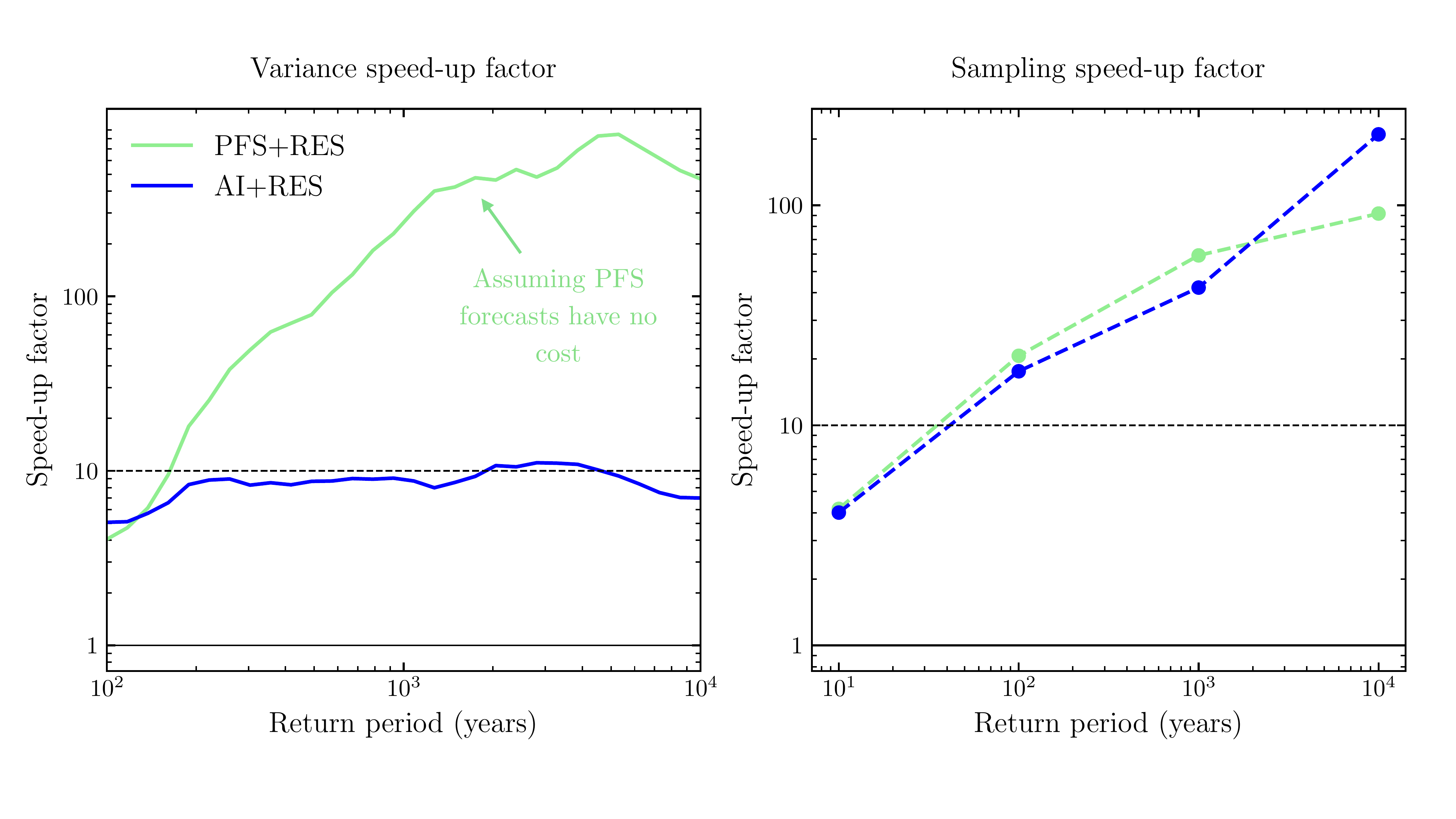}
    \caption{\textbf{Speedup ratio for AI+RES and PFS+RES.} 
    The setup matches that described in Table~\ref{tab:parameters}, with France as the region of interest, except that it uses $N=200$ walkers (instead of $N=400$ in the main text), since larger values are computationally prohibitive for PFS+RES. Left: variance speedup factor (Eq.~\eqref{eq:variance_speed_up_ratio}). Right: sampling speedup factor (Eq.~\eqref{eq:sampling_speed_up_ratio}), averaged over 10 independent algorithm realizations. Importantly, the speedup factors are computed assuming that the cost of running the ensemble forecasts is negligible. This is certainly not the case for the PFS+RES algorithm. Instead, PFS+RES shows the speedup factor that could be obtained if we had a perfect emulator.}
  \label{fig:speed_up_ratio_comparison}
\end{figure}

The \emph{Perfect-Forecast-System}+RES (PFS+RES) approach is designed to provide an upper bound on algorithmic performance. As noted earlier, using the AI emulator introduces an imperfect approximation of the score function due to the emulator’s limited weather forecast skill, particularly beyond 10–15 days of lead time. Here, we propose performing the ensemble forecast directly with the PlaSim GCM, thereby obtaining a near-perfect forecast to use in the score function.

We conducted experiments using the same setup as the AI+RES experiments, with parameters listed in Table~\ref{tab:parameters} in the main text. The only difference is that the number of walkers was set to $N=200$ instead of $N=400$, as the latter is computationally prohibitive for PFS+RES.

Fig.~\ref{fig: return_time_curve_PFS} shows the return-time curves for AI+RES and PFS+RES, demonstrating that PFS+RES can produce highly accurate return-period estimates with smaller variance than AI+RES. To quantify precisely how much we lose in terms of variance when moving from a perfect forecast system (PFS+RES) to an imperfect one (AI+RES), we compared the speedup factors relative to DS for both methods in Fig.~\ref{fig:speed_up_ratio_comparison}. Importantly, these speedup factors assume the cost of running ensemble forecasts is negligible. While this is not the case for PFS+RES, the comparison provides an indication of the potential performance achievable with near-perfect forecasts, under the assumption that forecast computations are inexpensive relative to the cost of running the physical model (here, PlaSim).

\newpage
\section{Additional material}

\begin{figure}[htbp]
  \centering
  \includegraphics[width=0.9\linewidth]{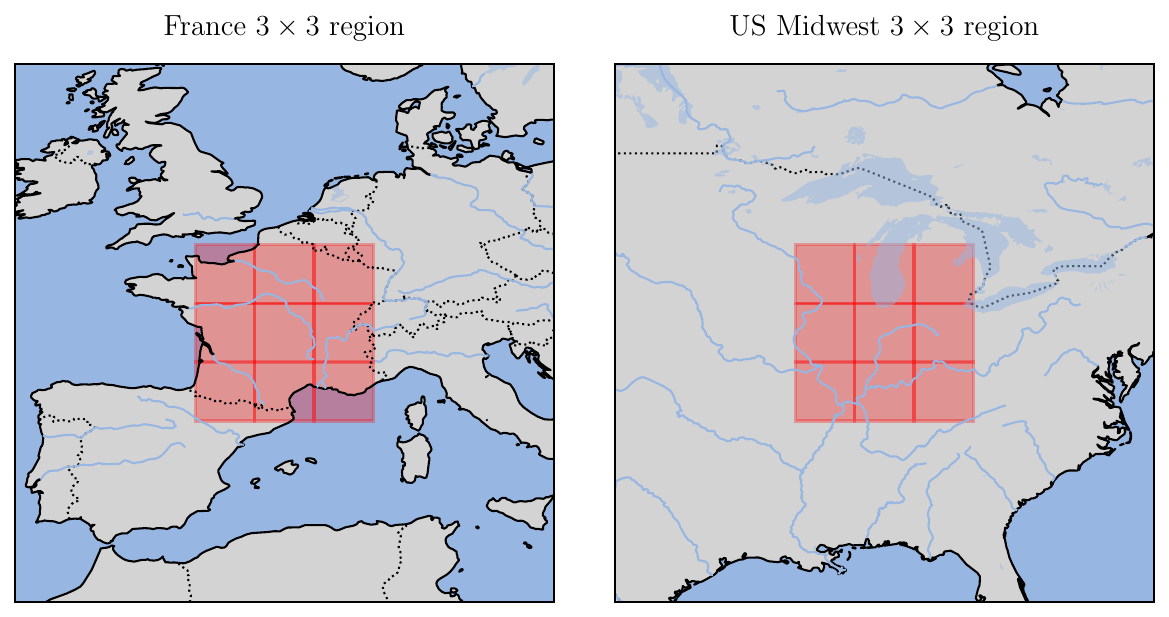}
  \caption{
    \textbf{Map of the regions of interest in this study.} The left panel shows the region of France, and the right panel shows the region of the U.S. Midwest. Each region is a box of $3 \times 3$ pixels (for PlaSim resolution, each pixel is approximately 2.8 degrees). We chose to study midlatitude heat waves over France because previous studies using PlaSim or RES focused on this region \cite{ragone2018computation, miloshevich_probabilistic_2023-1}. We also selected the U.S. Midwest region, as it is one of the hottest areas in summer in the PlaSim world, with the aim of sampling the most extreme events possible.
  }
  \label{fig:map_regions}
\end{figure}

\begin{figure}[htbp]  
  \centering  
  \includegraphics[width=0.9\linewidth]{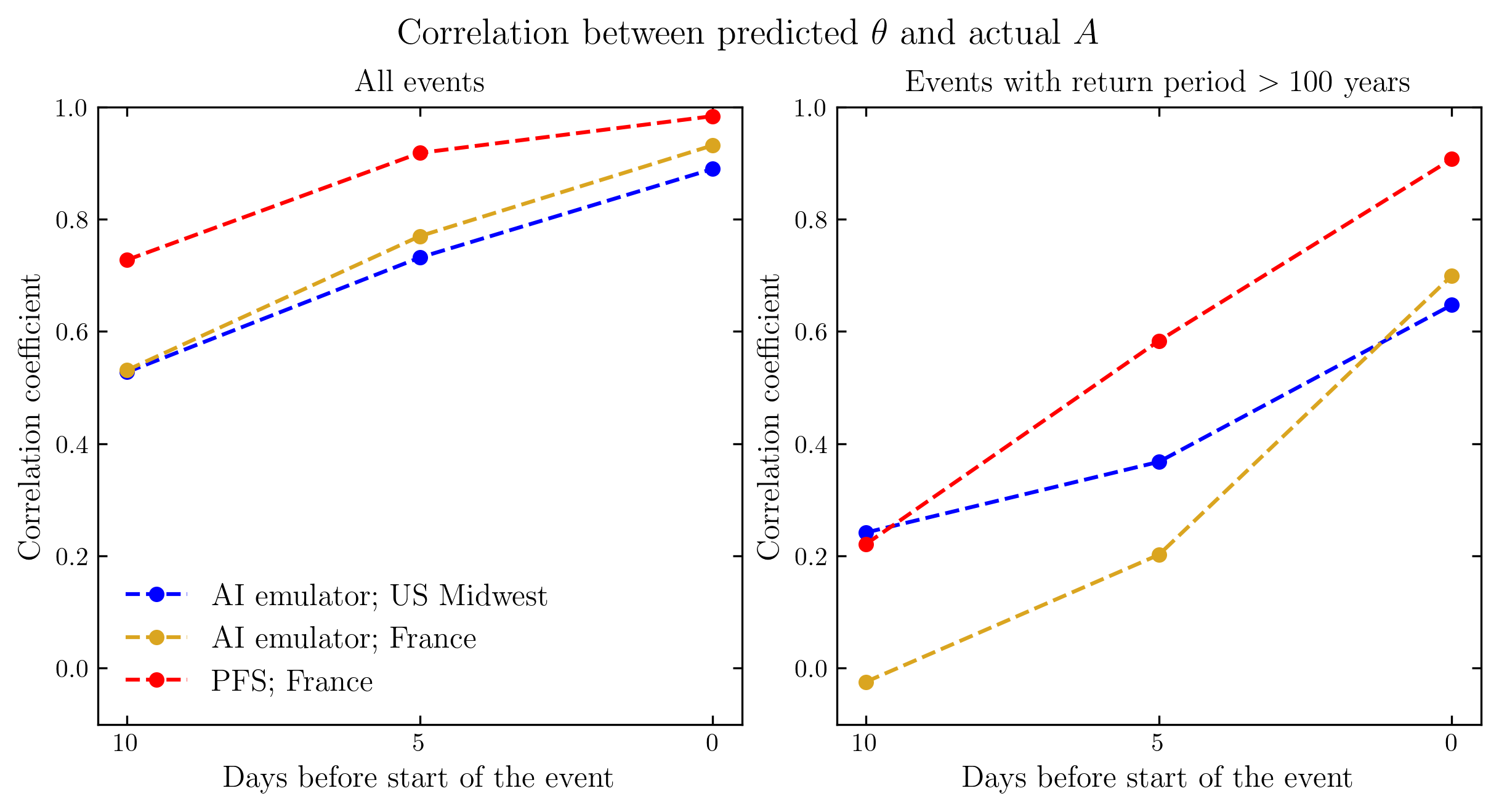}
  \caption{\textbf{Correlation between predicted and actual $A_L(t_f)$} across forecast systems and regions. The blue curves correspond to forecasts made with the AI emulator within AI+RES for the U.S. Midwest, while gold shows AI emulator results for France. Red curves represent forecasts made with the PlaSim GCM itself—referred to as the Perfect Forecast System (PFS) for the France region. Results are based on 10 independent realizations of the experiments in Table~\ref{tab:parameters} in the main text. For PFS, the number of walkers is $N=200$ instead of $N=400$ for the AI+RES experiments, and the analysis is limited to France, for computational reasons. At each resampling step in AI+RES, score functions $\theta_k^i$ are computed from ensemble forecasts (with $M=100$ members) with the AI emulator for each PlaSim walker (Eq.~\eqref{eq:score_function} in the main text). Here, three resampling steps are used (at $t_f-t_k=10$, $5$, and $0$ days before the event). We show the correlation between predicted (ensemble mean from the emulator) and actual (PlaSim realizations) $A_L(t_f)$ for all events (left) and for events with return periods $>100$ years (right). For rare events, the emulator forecasts are notably less skillful at the earliest resampling time in France compared to U.S. Midwest, likely reflecting the stronger role of soil moisture in the former region (since soil moisture is absent from emulator inputs but evolved by PlaSim). This result explains the higher variance of rare-probability estimators in France compared to U.S. Midwest (left panel of Fig.~\ref{fig:speed_up_factors} in the main text). Similarly, the smaller variances obtained for PFS-RES compared to AI-RES (Fig.~\ref{fig:speed_up_ratio_comparison}), can also be attributed to an overall higher forecast skill of PFS both for typical and extreme events.}
  \label{fig: correlation_between_predicted_and_actual_A_EXP15_AIRES}
\end{figure}

\begin{figure}[htbp]
  \centering
  \includegraphics[width=0.9\linewidth]{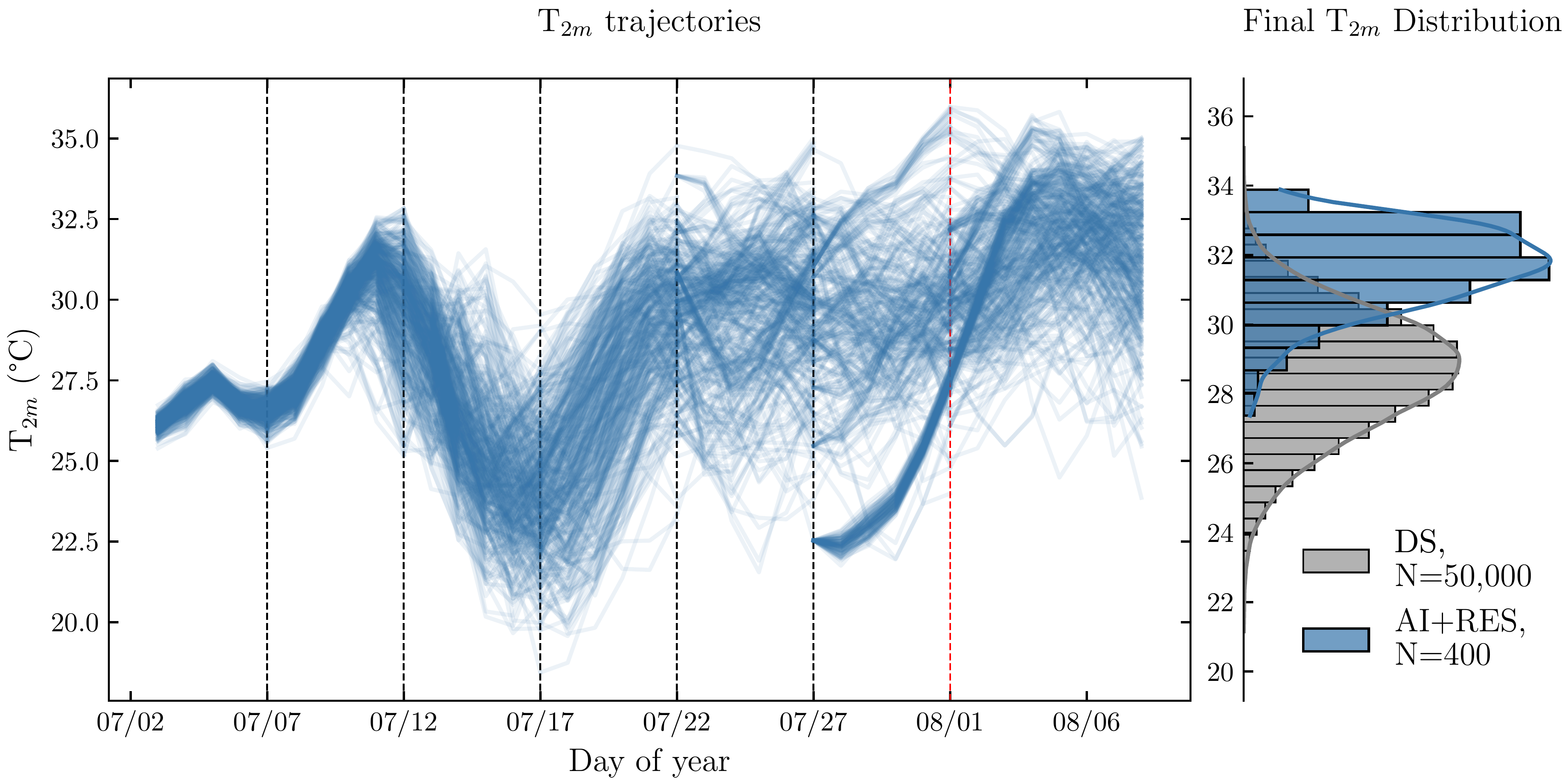}
  \caption{\textbf{T$_{2m}$ trajectories and histogram of the $A_L(t_f)$ observable} from a single AI+RES experiment over France. The setup matches the one described in Table~\ref{tab:parameters}. Left: solid blue lines show the evolution of daily T$_{2m}$ averaged over the region of interest for all $N=400$ walkers. At each resampling time (vertical dashed lines), ensemble forecasts with the emulator are run for each PlaSim walker until the end of the simulation, and the most promising are duplicated based on these forecasts (Eq.~\eqref{eq:score_function}). In particular, at the very bottom of the figure at the resampling time of 07/27, we observe that a large number of clones are drawn for a walker that exhibited a low T$_{2m}$ value at the time of resampling, but whose temperature rapidly increased thereafter, illustrating the emulator’s ability to anticipate this event. The red dashed line at $t_f$ marks the start of the event of interest on 08/01. Right: histograms of the distribution of $A_L(t_f)$ (Eq.~\eqref{eq:observable} with $L=7$ days) for direct sampling (DS, gray) and AI+RES (blue). The shift between them illustrates the algorithm’s \emph{importance sampling} effect. 
}
  \label{fig:traj_and_hist}
\end{figure}

\begin{figure}[htbp]
  \centering
  \includegraphics[width=0.45\linewidth]{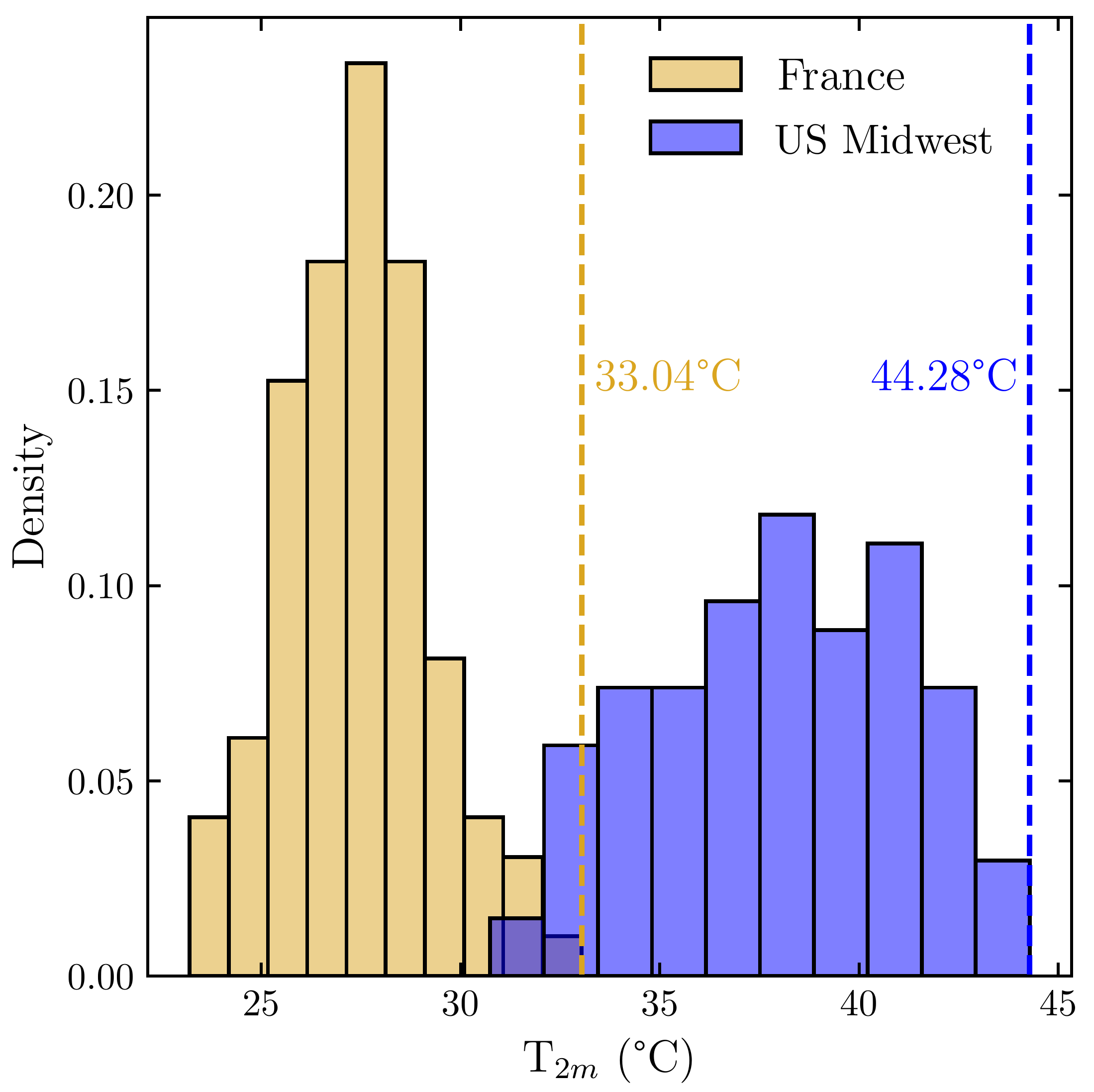}
  \caption{\textbf{Histogram of the $A_L(t_f)$ observable in the training dataset of the AI emulator} for the France (gold) and the U.S. Midwest (blue) regions with $L=7$ days and $t_f$ = August 1. The maximum value seen during training by the AI emulator is $33.04 \text{°C}$ in France and $44.28 \text{°C}$ in the U.S. Midwest. As seen in the AI-DS results (cyan dots) in Fig.~\ref{fig:return_time_curves}, the AI emulator is able to simulate events much more extreme than what was seen during the training.
}
\label{fig: histogram_A_train_France_and_Chicago_T7}
\end{figure}

\begin{figure}[htbp]
  \centering
  \includegraphics[width=0.45\linewidth]{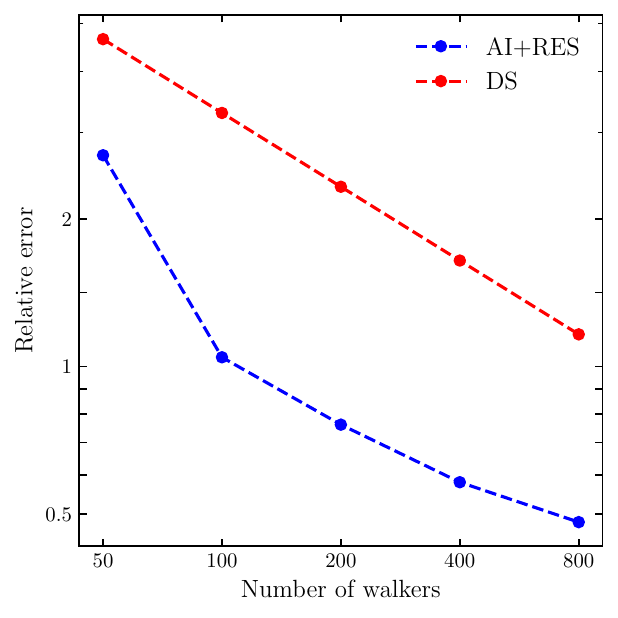}
  \caption{\textbf{Scaling of the relative error on the estimation of the probability $p=0.001$ as a function of the number of walkers $N$}. The setup matches the one described in Table~\ref{tab:parameters} (except the varying $N$), with France as the region of interest. The relative error for the probability $p$ is defined as $\mathbb{R} \mathbb{E}:=\frac{\sqrt{\mathbb{V}\left(\hat{p}\right)}}{\mathbb{E}\left(\hat{p}\right)}$. The slope of the DS curve (red line) is $-1/2$, as can be derived from Eq.~\eqref{eq:variance_DNS}. For AI+RES, we observe that increasing $N$ yields a greater reduction in relative standard deviation when the number of walkers is low (between 50 and 200), while for larger $N$, the improvement becomes comparable to that of DS.
}
\label{fig: scaling_with_N_of_relative_std_p0.001}
\end{figure}

\begin{figure}[htbp]
  \centering
  \includegraphics[width=0.45\linewidth]{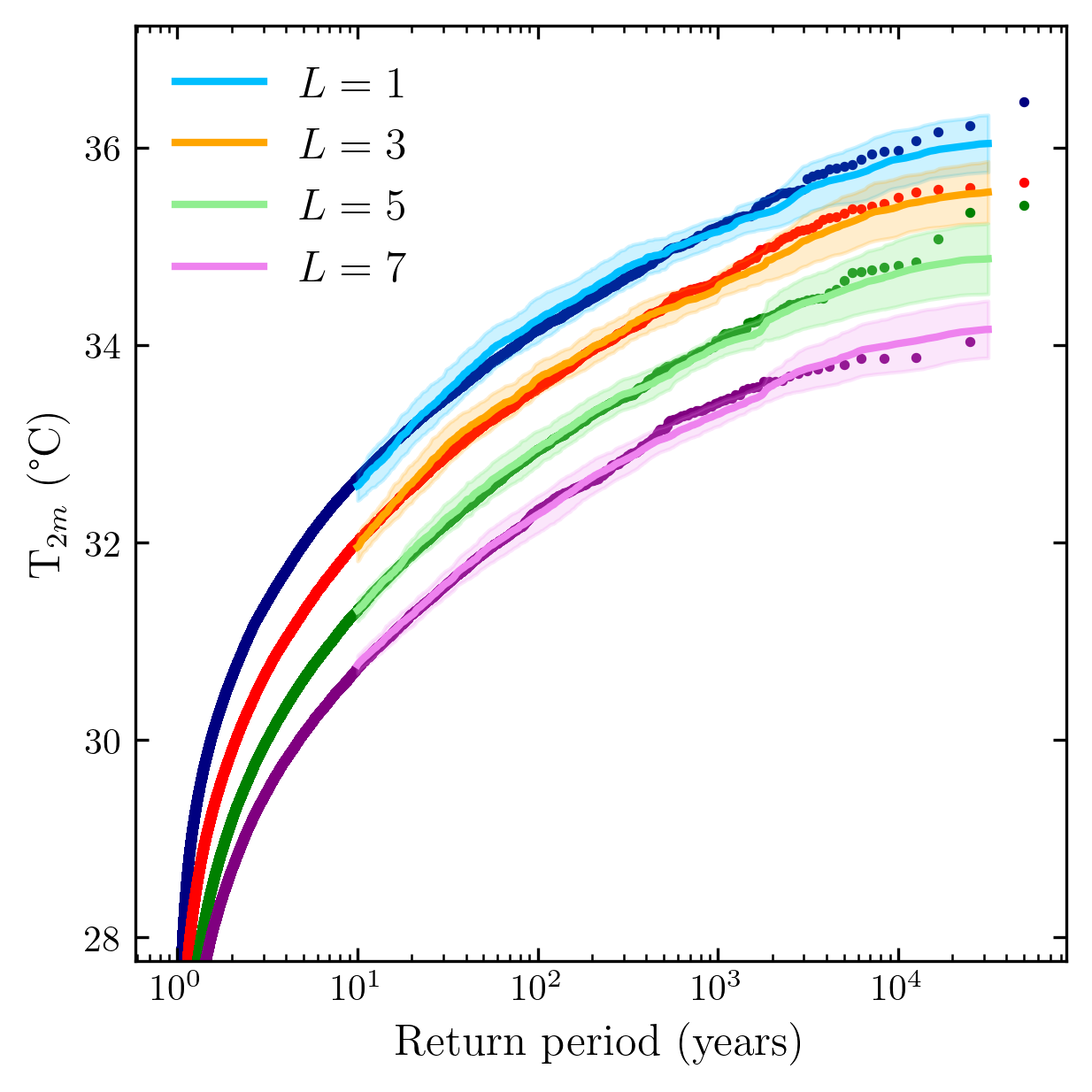}
  \caption{\textbf{Return-time curves of $A_L(t_f)$ for $L = 1, 3, 5$ and $7$ days} (Eq.~\eqref{eq:observable}) and France as the region of interest. The results shown here are from the same experiments presented in Table~\ref{tab:parameters}. In particular, while the score function used here is the 7-day average surface temperature over the region of interest, the trajectories sampled by the AI+RES algorithm allow us to accurately estimate the return period of shorter-duration events. The darker dots are the empirical return periods obtained with a $50,000$-member control simulation. The solid lighter lines show the mean of return time curves produced by 10 independent realizations of the AI+RES algorithm with $N=400$ walkers. The light shaded areas represent $95\%$ confidence intervals. 
}
\label{fig: varying_T_return_time_plot}
\end{figure}

\begin{figure}[htbp]
  \centering
  \includegraphics[width=0.9\linewidth]{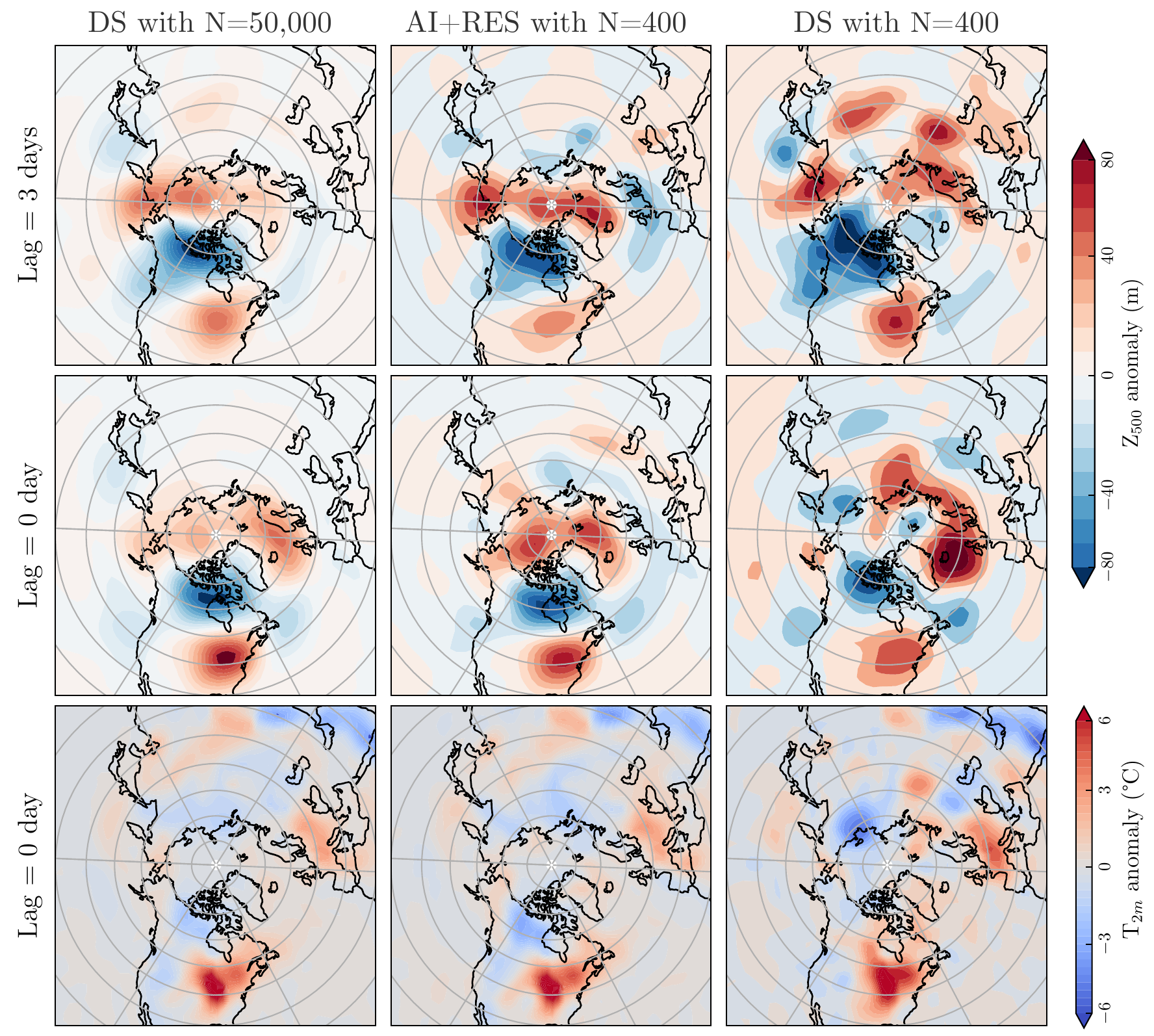}
  \caption{\textbf{Composite maps of heat waves over the U.S. Midwest with return periods exceeding 100 years.} Events are defined by Eq.~\eqref{eq:observable} of the main text with $L=$ 3 days. The first row shows daily mean Z$_{500}$ anomaly composites three days before the heat wave onset; the second row shows the $L$-day average Z$_{500}$ anomaly composites during the heat wave; the third row shows the $L$-day average T$_{2m}$ anomaly composites during the heat wave. First column: DS with $N = 50{,}000$ (ground truth). Second column: results from the AI+RES algorithm with $N = 400$. Third column: DS with $N = 400$.}
  \label{fig: composites_Chicago}
\end{figure}

%% file: main.bbl
\begin{thebibliography}{74}%
\makeatletter
\providecommand \@ifxundefined [1]{%
 \@ifx{#1\undefined}
}%
\providecommand \@ifnum [1]{%
 \ifnum #1\expandafter \@firstoftwo
 \else \expandafter \@secondoftwo
 \fi
}%
\providecommand \@ifx [1]{%
 \ifx #1\expandafter \@firstoftwo
 \else \expandafter \@secondoftwo
 \fi
}%
\providecommand \natexlab [1]{#1}%
\providecommand \enquote  [1]{``#1''}%
\providecommand \bibnamefont  [1]{#1}%
\providecommand \bibfnamefont [1]{#1}%
\providecommand \citenamefont [1]{#1}%
\providecommand \href@noop [0]{\@secondoftwo}%
\providecommand \href [0]{\begingroup \@sanitize@url \@href}%
\providecommand \@href[1]{\@@startlink{#1}\@@href}%
\providecommand \@@href[1]{\endgroup#1\@@endlink}%
\providecommand \@sanitize@url [0]{\catcode `\\12\catcode `\$12\catcode
  `\&12\catcode `\#12\catcode `\^12\catcode `\_12\catcode `\%12\relax}%
\providecommand \@@startlink[1]{}%
\providecommand \@@endlink[0]{}%
\providecommand \url  [0]{\begingroup\@sanitize@url \@url }%
\providecommand \@url [1]{\endgroup\@href {#1}{\urlprefix }}%
\providecommand \urlprefix  [0]{URL }%
\providecommand \Eprint [0]{\href }%
\providecommand \doibase [0]{https://doi.org/}%
\providecommand \selectlanguage [0]{\@gobble}%
\providecommand \bibinfo  [0]{\@secondoftwo}%
\providecommand \bibfield  [0]{\@secondoftwo}%
\providecommand \translation [1]{[#1]}%
\providecommand \BibitemOpen [0]{}%
\providecommand \bibitemStop [0]{}%
\providecommand \bibitemNoStop [0]{.\EOS\space}%
\providecommand \EOS [0]{\spacefactor3000\relax}%
\providecommand \BibitemShut  [1]{\csname bibitem#1\endcsname}%
\let\auto@bib@innerbib\@empty
\bibitem [{\citenamefont {Tol}(2024)}]{TOL2024}%
  \BibitemOpen
  \bibfield  {author} {\bibinfo {author} {\bibfnamefont {R.~S.}\ \bibnamefont
  {Tol}},\ }\bibfield  {title} {\bibinfo {title} {{A meta-analysis of the total
  economic impact of climate change}},\ }\href
  {https://doi.org/https://doi.org/10.1016/j.enpol.2023.113922} {\bibfield
  {journal} {\bibinfo  {journal} {Energy Policy}\ }\textbf {\bibinfo {volume}
  {185}},\ \bibinfo {pages} {113922} (\bibinfo {year} {2024})}\BibitemShut
  {NoStop}%
\bibitem [{\citenamefont {Anchen}\ \emph {et~al.}(2025)\citenamefont {Anchen},
  \citenamefont {Gonzalez}, \citenamefont {Chatterjee}, \citenamefont {Egloff},
  \citenamefont {Felderer}, \citenamefont {Mejler{\"o}}, \citenamefont
  {Vischer},\ and\ \citenamefont {Wilke}}]{anchen_swiss_2025}%
  \BibitemOpen
  \bibfield  {author} {\bibinfo {author} {\bibfnamefont {J.}~\bibnamefont
  {Anchen}}, \bibinfo {author} {\bibfnamefont {V.~B.}\ \bibnamefont
  {Gonzalez}}, \bibinfo {author} {\bibfnamefont {M.}~\bibnamefont
  {Chatterjee}}, \bibinfo {author} {\bibfnamefont {R.}~\bibnamefont {Egloff}},
  \bibinfo {author} {\bibfnamefont {A.}~\bibnamefont {Felderer}}, \bibinfo
  {author} {\bibfnamefont {A.}~\bibnamefont {Mejler{\"o}}}, \bibinfo {author}
  {\bibfnamefont {A.}~\bibnamefont {Vischer}},\ and\ \bibinfo {author}
  {\bibfnamefont {B.}~\bibnamefont {Wilke}},\ }\href@noop {} {\emph {\bibinfo
  {title} {{Swiss {{Re SONAR}}: {{New}} Emerging Risk Insights}}}},\ \bibinfo
  {type} {Tech. Rep.}\ (\bibinfo  {institution} {Swiss Re Institute},\ \bibinfo
  {address} {Zurich, Switzerland},\ \bibinfo {year} {2025})\BibitemShut
  {NoStop}%
\bibitem [{\citenamefont {Ebi}\ \emph {et~al.}(2021)\citenamefont {Ebi},
  \citenamefont {Vanos}, \citenamefont {Baldwin}, \citenamefont {Bell},
  \citenamefont {Hondula}, \citenamefont {Errett}, \citenamefont {Hayes},
  \citenamefont {Reid}, \citenamefont {Saha}, \citenamefont {Spector} \emph
  {et~al.}}]{ebi2021extreme}%
  \BibitemOpen
  \bibfield  {author} {\bibinfo {author} {\bibfnamefont {K.~L.}\ \bibnamefont
  {Ebi}}, \bibinfo {author} {\bibfnamefont {J.}~\bibnamefont {Vanos}}, \bibinfo
  {author} {\bibfnamefont {J.~W.}\ \bibnamefont {Baldwin}}, \bibinfo {author}
  {\bibfnamefont {J.~E.}\ \bibnamefont {Bell}}, \bibinfo {author}
  {\bibfnamefont {D.~M.}\ \bibnamefont {Hondula}}, \bibinfo {author}
  {\bibfnamefont {N.~A.}\ \bibnamefont {Errett}}, \bibinfo {author}
  {\bibfnamefont {K.}~\bibnamefont {Hayes}}, \bibinfo {author} {\bibfnamefont
  {C.~E.}\ \bibnamefont {Reid}}, \bibinfo {author} {\bibfnamefont
  {S.}~\bibnamefont {Saha}}, \bibinfo {author} {\bibfnamefont {J.}~\bibnamefont
  {Spector}}, \emph {et~al.},\ }\bibfield  {title} {\bibinfo {title} {{Extreme
  weather and climate change: population health and health system
  implications}},\ }\href@noop {} {\bibfield  {journal} {\bibinfo  {journal}
  {Annual review of public health}\ }\textbf {\bibinfo {volume} {42}},\
  \bibinfo {pages} {293} (\bibinfo {year} {2021})}\BibitemShut {NoStop}%
\bibitem [{\citenamefont {Ummenhofer}\ and\ \citenamefont
  {Meehl}(2017)}]{ummenhofer2017extreme}%
  \BibitemOpen
  \bibfield  {author} {\bibinfo {author} {\bibfnamefont {C.~C.}\ \bibnamefont
  {Ummenhofer}}\ and\ \bibinfo {author} {\bibfnamefont {G.~A.}\ \bibnamefont
  {Meehl}},\ }\bibfield  {title} {\bibinfo {title} {{Extreme weather and
  climate events with ecological relevance: a review}},\ }\href@noop {}
  {\bibfield  {journal} {\bibinfo  {journal} {Philosophical Transactions of the
  Royal Society B: Biological Sciences}\ }\textbf {\bibinfo {volume} {372}},\
  \bibinfo {pages} {20160135} (\bibinfo {year} {2017})}\BibitemShut {NoStop}%
\bibitem [{\citenamefont {Gon{\c{c}}alves}\ \emph {et~al.}(2024)\citenamefont
  {Gon{\c{c}}alves}, \citenamefont {Costoya}, \citenamefont {Nieto},\ and\
  \citenamefont {Liberato}}]{gonccalves2024extreme}%
  \BibitemOpen
  \bibfield  {author} {\bibinfo {author} {\bibfnamefont {A.~C.}\ \bibnamefont
  {Gon{\c{c}}alves}}, \bibinfo {author} {\bibfnamefont {X.}~\bibnamefont
  {Costoya}}, \bibinfo {author} {\bibfnamefont {R.}~\bibnamefont {Nieto}},\
  and\ \bibinfo {author} {\bibfnamefont {M.~L.}\ \bibnamefont {Liberato}},\
  }\bibfield  {title} {\bibinfo {title} {{Extreme weather events on energy
  systems: a comprehensive review on impacts, mitigation, and adaptation
  measures}},\ }\href@noop {} {\bibfield  {journal} {\bibinfo  {journal}
  {Sustainable Energy Research}\ }\textbf {\bibinfo {volume} {11}},\ \bibinfo
  {pages} {4} (\bibinfo {year} {2024})}\BibitemShut {NoStop}%
\bibitem [{\citenamefont {IPCC}(2021)}]{ipcc2021wgi}%
  \BibitemOpen
  \bibfield  {author} {\bibinfo {author} {\bibnamefont {IPCC}},\ }\href
  {https://doi.org/10.1017/9781009157896} {\emph {\bibinfo {title} {{Climate
  Change 2021: The Physical Science Basis. Contribution of Working Group I to
  the Sixth Assessment Report of the Intergovernmental Panel on Climate
  Change}}}},\ edited by\ \bibinfo {editor} {\bibfnamefont {V.}~\bibnamefont
  {Masson-Delmotte}}, \bibinfo {editor} {\bibfnamefont {P.}~\bibnamefont
  {Zhai}}, \bibinfo {editor} {\bibfnamefont {A.}~\bibnamefont {Pirani}},
  \bibinfo {editor} {\bibfnamefont {S.}~\bibnamefont {Connors}}, \bibinfo
  {editor} {\bibfnamefont {C.}~\bibnamefont {Péan}}, \bibinfo {editor}
  {\bibfnamefont {S.}~\bibnamefont {Berger}}, \bibinfo {editor} {\bibfnamefont
  {N.}~\bibnamefont {Caud}}, \bibinfo {editor} {\bibfnamefont {Y.}~\bibnamefont
  {Chen}}, \bibinfo {editor} {\bibfnamefont {L.}~\bibnamefont {Goldfarb}},
  \bibinfo {editor} {\bibfnamefont {M.}~\bibnamefont {Gomis}}, \bibinfo
  {editor} {\bibfnamefont {M.}~\bibnamefont {Huang}}, \bibinfo {editor}
  {\bibfnamefont {K.}~\bibnamefont {Leitzell}}, \bibinfo {editor}
  {\bibfnamefont {E.}~\bibnamefont {Lonnoy}}, \bibinfo {editor} {\bibfnamefont
  {J.}~\bibnamefont {Matthews}}, \bibinfo {editor} {\bibfnamefont
  {T.}~\bibnamefont {Maycock}}, \bibinfo {editor} {\bibfnamefont
  {T.}~\bibnamefont {Waterfield}}, \bibinfo {editor} {\bibfnamefont
  {O.}~\bibnamefont {Yelekçi}}, \bibinfo {editor} {\bibfnamefont
  {R.}~\bibnamefont {Yu}},\ and\ \bibinfo {editor} {\bibfnamefont
  {B.}~\bibnamefont {Zhou}}\ (\bibinfo  {publisher} {Cambridge University
  Press},\ \bibinfo {year} {2021})\BibitemShut {NoStop}%
\bibitem [{\citenamefont {IPCC}(2022)}]{ipcc2022wgii}%
  \BibitemOpen
  \bibfield  {author} {\bibinfo {author} {\bibnamefont {IPCC}},\ }\href
  {https://doi.org/10.1017/9781009325844} {\emph {\bibinfo {title} {{Climate
  Change 2022: Impacts, Adaptation and Vulnerability. Contribution of Working
  Group II to the Sixth Assessment Report of the Intergovernmental Panel on
  Climate Change}}}},\ edited by\ \bibinfo {editor} {\bibfnamefont {H.-O.}\
  \bibnamefont {Pörtner}}, \bibinfo {editor} {\bibfnamefont {D.}~\bibnamefont
  {Roberts}}, \bibinfo {editor} {\bibfnamefont {M.}~\bibnamefont {Tignor}},
  \bibinfo {editor} {\bibfnamefont {E.}~\bibnamefont {Poloczanska}}, \bibinfo
  {editor} {\bibfnamefont {K.}~\bibnamefont {Mintenbeck}}, \bibinfo {editor}
  {\bibfnamefont {A.}~\bibnamefont {Alegría}}, \bibinfo {editor}
  {\bibfnamefont {M.}~\bibnamefont {Craig}}, \bibinfo {editor} {\bibfnamefont
  {S.}~\bibnamefont {Langsdorf}}, \bibinfo {editor} {\bibfnamefont
  {S.}~\bibnamefont {Löschke}}, \bibinfo {editor} {\bibfnamefont
  {V.}~\bibnamefont {Möller}}, \bibinfo {editor} {\bibfnamefont
  {A.}~\bibnamefont {Okem}},\ and\ \bibinfo {editor} {\bibfnamefont
  {B.}~\bibnamefont {Rama}}\ (\bibinfo  {publisher} {Cambridge University
  Press},\ \bibinfo {year} {2022})\BibitemShut {NoStop}%
\bibitem [{\citenamefont {Zeder}\ \emph {et~al.}(2023)\citenamefont {Zeder},
  \citenamefont {Sippel}, \citenamefont {Pasche}, \citenamefont {Engelke},\
  and\ \citenamefont {Fischer}}]{zeder2023effect}%
  \BibitemOpen
  \bibfield  {author} {\bibinfo {author} {\bibfnamefont {J.}~\bibnamefont
  {Zeder}}, \bibinfo {author} {\bibfnamefont {S.}~\bibnamefont {Sippel}},
  \bibinfo {author} {\bibfnamefont {O.~C.}\ \bibnamefont {Pasche}}, \bibinfo
  {author} {\bibfnamefont {S.}~\bibnamefont {Engelke}},\ and\ \bibinfo {author}
  {\bibfnamefont {E.~M.}\ \bibnamefont {Fischer}},\ }\bibfield  {title}
  {\bibinfo {title} {{The effect of a short observational record on the
  statistics of temperature extremes}},\ }\href@noop {} {\bibfield  {journal}
  {\bibinfo  {journal} {Geophysical Research Letters}\ }\textbf {\bibinfo
  {volume} {50}},\ \bibinfo {pages} {e2023GL104090} (\bibinfo {year}
  {2023})}\BibitemShut {NoStop}%
\bibitem [{\citenamefont {Embrechts}\ \emph {et~al.}(2013)\citenamefont
  {Embrechts}, \citenamefont {Kl{\"u}ppelberg},\ and\ \citenamefont
  {Mikosch}}]{embrechts2013modelling}%
  \BibitemOpen
  \bibfield  {author} {\bibinfo {author} {\bibfnamefont {P.}~\bibnamefont
  {Embrechts}}, \bibinfo {author} {\bibfnamefont {C.}~\bibnamefont
  {Kl{\"u}ppelberg}},\ and\ \bibinfo {author} {\bibfnamefont {T.}~\bibnamefont
  {Mikosch}},\ }\href {https://books.google.com/books?id=BXOI2pICfJUC} {\emph
  {\bibinfo {title} {{Modelling Extremal Events: for Insurance and
  Finance}}}},\ Stochastic Modelling and Applied Probability\ (\bibinfo
  {publisher} {Springer Berlin Heidelberg},\ \bibinfo {year}
  {2013})\BibitemShut {NoStop}%
\bibitem [{\citenamefont {Huang}\ \emph {et~al.}(2016)\citenamefont {Huang},
  \citenamefont {Stein}, \citenamefont {McInerney}, \citenamefont {Sun},\ and\
  \citenamefont {Moyer}}]{huang_estimating_2016}%
  \BibitemOpen
  \bibfield  {author} {\bibinfo {author} {\bibfnamefont {W.~K.}\ \bibnamefont
  {Huang}}, \bibinfo {author} {\bibfnamefont {M.~L.}\ \bibnamefont {Stein}},
  \bibinfo {author} {\bibfnamefont {D.~J.}\ \bibnamefont {McInerney}}, \bibinfo
  {author} {\bibfnamefont {S.}~\bibnamefont {Sun}},\ and\ \bibinfo {author}
  {\bibfnamefont {E.~J.}\ \bibnamefont {Moyer}},\ }\bibfield  {title} {\bibinfo
  {title} {{Estimating Changes in Temperature Extremes from Millennial-Scale
  Climate Simulations Using Generalized Extreme Value ({{GEV}})
  Distributions}},\ }\href {https://doi.org/10.5194/ascmo-2-79-2016} {\bibfield
   {journal} {\bibinfo  {journal} {Advances in Statistical Climatology,
  Meteorology and Oceanography}\ }\textbf {\bibinfo {volume} {2}},\ \bibinfo
  {pages} {79} (\bibinfo {year} {2016})}\BibitemShut {NoStop}%
\bibitem [{\citenamefont {G{\'a}lfi}\ \emph {et~al.}(2017)\citenamefont
  {G{\'a}lfi}, \citenamefont {B{\'o}dai},\ and\ \citenamefont
  {Lucarini}}]{galfi_convergence_2017}%
  \BibitemOpen
  \bibfield  {author} {\bibinfo {author} {\bibfnamefont {V.~M.}\ \bibnamefont
  {G{\'a}lfi}}, \bibinfo {author} {\bibfnamefont {T.}~\bibnamefont
  {B{\'o}dai}},\ and\ \bibinfo {author} {\bibfnamefont {V.}~\bibnamefont
  {Lucarini}},\ }\bibfield  {title} {\bibinfo {title} {{Convergence of
  {{Extreme Value Statistics}} in a {{Two-Layer Quasi-Geostrophic Atmospheric
  Model}}}},\ }\href {https://doi.org/10.1155/2017/5340858} {\bibfield
  {journal} {\bibinfo  {journal} {Complexity}\ }\textbf {\bibinfo {volume}
  {2017}},\ \bibinfo {pages} {5340858} (\bibinfo {year} {2017})}\BibitemShut
  {NoStop}%
\bibitem [{\citenamefont {Le~Priol}\ \emph {et~al.}(2024)\citenamefont
  {Le~Priol}, \citenamefont {Monteiro},\ and\ \citenamefont
  {Bouchet}}]{lepriol2024using}%
  \BibitemOpen
  \bibfield  {author} {\bibinfo {author} {\bibfnamefont {C.}~\bibnamefont
  {Le~Priol}}, \bibinfo {author} {\bibfnamefont {J.~M.}\ \bibnamefont
  {Monteiro}},\ and\ \bibinfo {author} {\bibfnamefont {F.}~\bibnamefont
  {Bouchet}},\ }\bibfield  {title} {\bibinfo {title} {{Using rare event
  algorithms to understand the statistics and dynamics of extreme heatwave
  seasons in South Asia}},\ }\href@noop {} {\bibfield  {journal} {\bibinfo
  {journal} {Environmental Research: Climate}\ }\textbf {\bibinfo {volume}
  {3}},\ \bibinfo {pages} {045016} (\bibinfo {year} {2024})}\BibitemShut
  {NoStop}%
\bibitem [{\citenamefont {Ragone}\ \emph {et~al.}(2018)\citenamefont {Ragone},
  \citenamefont {Wouters},\ and\ \citenamefont
  {Bouchet}}]{ragone2018computation}%
  \BibitemOpen
  \bibfield  {author} {\bibinfo {author} {\bibfnamefont {F.}~\bibnamefont
  {Ragone}}, \bibinfo {author} {\bibfnamefont {J.}~\bibnamefont {Wouters}},\
  and\ \bibinfo {author} {\bibfnamefont {F.}~\bibnamefont {Bouchet}},\
  }\bibfield  {title} {\bibinfo {title} {{Computation of extreme heat waves in
  climate models using a large deviation algorithm}},\ }\href@noop {}
  {\bibfield  {journal} {\bibinfo  {journal} {Proceedings of the National
  Academy of Sciences}\ }\textbf {\bibinfo {volume} {115}},\ \bibinfo {pages}
  {24} (\bibinfo {year} {2018})}\BibitemShut {NoStop}%
\bibitem [{\citenamefont {Webber}\ \emph {et~al.}(2019)\citenamefont {Webber},
  \citenamefont {Plotkin}, \citenamefont {O’Neill}, \citenamefont {Abbot},\
  and\ \citenamefont {Weare}}]{webber2019}%
  \BibitemOpen
  \bibfield  {author} {\bibinfo {author} {\bibfnamefont {R.}~\bibnamefont
  {Webber}}, \bibinfo {author} {\bibfnamefont {D.}~\bibnamefont {Plotkin}},
  \bibinfo {author} {\bibfnamefont {M.}~\bibnamefont {O’Neill}}, \bibinfo
  {author} {\bibfnamefont {D.}~\bibnamefont {Abbot}},\ and\ \bibinfo {author}
  {\bibfnamefont {J.}~\bibnamefont {Weare}},\ }\bibfield  {title} {\bibinfo
  {title} {{Practical rare event sampling for extreme mesoscale weather}},\
  }\href {https://doi.org/10.1063/1.5081461} {\bibfield  {journal} {\bibinfo
  {journal} {Chaos: An Interdisciplinary Journal of Nonlinear Science}\
  }\textbf {\bibinfo {volume} {29}},\ \bibinfo {pages} {053109} (\bibinfo
  {year} {2019})}\BibitemShut {NoStop}%
\bibitem [{\citenamefont {Ragone}\ and\ \citenamefont
  {Bouchet}(2021)}]{ragone2021rare}%
  \BibitemOpen
  \bibfield  {author} {\bibinfo {author} {\bibfnamefont {F.}~\bibnamefont
  {Ragone}}\ and\ \bibinfo {author} {\bibfnamefont {F.}~\bibnamefont
  {Bouchet}},\ }\bibfield  {title} {\bibinfo {title} {{Rare event algorithm
  study of extreme warm summers and heatwaves over Europe}},\ }\href@noop {}
  {\bibfield  {journal} {\bibinfo  {journal} {Geophysical Research Letters}\
  }\textbf {\bibinfo {volume} {48}},\ \bibinfo {pages} {e2020GL091197}
  (\bibinfo {year} {2021})}\BibitemShut {NoStop}%
\bibitem [{\citenamefont {Abbot}\ \emph {et~al.}(2021)\citenamefont {Abbot},
  \citenamefont {Webber}, \citenamefont {Hadden}, \citenamefont {Seligman},\
  and\ \citenamefont {Weare}}]{abbot2021rare}%
  \BibitemOpen
  \bibfield  {author} {\bibinfo {author} {\bibfnamefont {D.~S.}\ \bibnamefont
  {Abbot}}, \bibinfo {author} {\bibfnamefont {R.~J.}\ \bibnamefont {Webber}},
  \bibinfo {author} {\bibfnamefont {S.}~\bibnamefont {Hadden}}, \bibinfo
  {author} {\bibfnamefont {D.}~\bibnamefont {Seligman}},\ and\ \bibinfo
  {author} {\bibfnamefont {J.}~\bibnamefont {Weare}},\ }\bibfield  {title}
  {\bibinfo {title} {{Rare event sampling improves Mercury instability
  statistics}},\ }\href@noop {} {\bibfield  {journal} {\bibinfo  {journal} {The
  Astrophysical Journal}\ }\textbf {\bibinfo {volume} {923}},\ \bibinfo {pages}
  {236} (\bibinfo {year} {2021})}\BibitemShut {NoStop}%
\bibitem [{\citenamefont {Finkel}\ and\ \citenamefont
  {O'Gorman}(2024)}]{finkel_bringing_2024}%
  \BibitemOpen
  \bibfield  {author} {\bibinfo {author} {\bibfnamefont {J.}~\bibnamefont
  {Finkel}}\ and\ \bibinfo {author} {\bibfnamefont {P.~A.}\ \bibnamefont
  {O'Gorman}},\ }\bibfield  {title} {\bibinfo {title} {{Bringing {{Statistics}}
  to {{Storylines}}: {{Rare Event Sampling}} for {{Sudden}}, {{Transient
  Extreme Events}}}},\ }\href {https://doi.org/10.1029/2024MS004264} {\bibfield
   {journal} {\bibinfo  {journal} {Journal of Advances in Modeling Earth
  Systems}\ }\textbf {\bibinfo {volume} {16}},\ \bibinfo {pages}
  {e2024MS004264} (\bibinfo {year} {2024})}\BibitemShut {NoStop}%
\bibitem [{\citenamefont {Noyelle}\ \emph {et~al.}(2025)\citenamefont
  {Noyelle}, \citenamefont {Caubel}, \citenamefont {Meurdesoif}, \citenamefont
  {Yiou},\ and\ \citenamefont {Faranda}}]{noyelle2025statistical}%
  \BibitemOpen
  \bibfield  {author} {\bibinfo {author} {\bibfnamefont {R.}~\bibnamefont
  {Noyelle}}, \bibinfo {author} {\bibfnamefont {A.}~\bibnamefont {Caubel}},
  \bibinfo {author} {\bibfnamefont {Y.}~\bibnamefont {Meurdesoif}}, \bibinfo
  {author} {\bibfnamefont {P.}~\bibnamefont {Yiou}},\ and\ \bibinfo {author}
  {\bibfnamefont {D.}~\bibnamefont {Faranda}},\ }\bibfield  {title} {\bibinfo
  {title} {{Statistical and dynamical aspects of extremely hot summers in
  Western Europe sampled with a rare events algorithm}},\ }\href
  {https://doi.org/10.1175/JCLI-D-24-0635.1} {\bibfield  {journal} {\bibinfo
  {journal} {Journal of Climate}\ }\textbf {\bibinfo {volume} {38}},\ \bibinfo
  {pages} {4763} (\bibinfo {year} {2025})}\BibitemShut {NoStop}%
\bibitem [{\citenamefont {Gessner}\ \emph {et~al.}(2021)\citenamefont
  {Gessner}, \citenamefont {Fischer}, \citenamefont {Beyerle},\ and\
  \citenamefont {Knutti}}]{gessner2021ensboost}%
  \BibitemOpen
  \bibfield  {author} {\bibinfo {author} {\bibfnamefont {C.}~\bibnamefont
  {Gessner}}, \bibinfo {author} {\bibfnamefont {E.~M.}\ \bibnamefont
  {Fischer}}, \bibinfo {author} {\bibfnamefont {U.}~\bibnamefont {Beyerle}},\
  and\ \bibinfo {author} {\bibfnamefont {R.}~\bibnamefont {Knutti}},\
  }\bibfield  {title} {\bibinfo {title} {{Very Rare Heat Extremes: Quantifying
  and Understanding Using Ensemble Reinitialization}},\ }\href
  {https://doi.org/10.1175/JCLI-D-20-0916.1} {\bibfield  {journal} {\bibinfo
  {journal} {Journal of Climate}\ }\textbf {\bibinfo {volume} {34}},\ \bibinfo
  {pages} {6619 } (\bibinfo {year} {2021})}\BibitemShut {NoStop}%
\bibitem [{\citenamefont {Bloin-Wibe}\ \emph {et~al.}(2025)\citenamefont
  {Bloin-Wibe}, \citenamefont {Noyelle}, \citenamefont {Humphrey},
  \citenamefont {Beyerle}, \citenamefont {Knutti},\ and\ \citenamefont
  {Fischer}}]{bloin2025estimating}%
  \BibitemOpen
  \bibfield  {author} {\bibinfo {author} {\bibfnamefont {L.}~\bibnamefont
  {Bloin-Wibe}}, \bibinfo {author} {\bibfnamefont {R.}~\bibnamefont {Noyelle}},
  \bibinfo {author} {\bibfnamefont {V.}~\bibnamefont {Humphrey}}, \bibinfo
  {author} {\bibfnamefont {U.}~\bibnamefont {Beyerle}}, \bibinfo {author}
  {\bibfnamefont {R.}~\bibnamefont {Knutti}},\ and\ \bibinfo {author}
  {\bibfnamefont {E.}~\bibnamefont {Fischer}},\ }\bibfield  {title} {\bibinfo
  {title} {{Estimating return periods for extreme events in climate models
  through Ensemble Boosting}},\ }\href@noop {} {\bibfield  {journal} {\bibinfo
  {journal} {EGUsphere}\ }\textbf {\bibinfo {volume} {2025}},\ \bibinfo {pages}
  {1} (\bibinfo {year} {2025})}\BibitemShut {NoStop}%
\bibitem [{\citenamefont {Finkel}\ and\ \citenamefont
  {O'Gorman}(2025)}]{finkel2025boosting}%
  \BibitemOpen
  \bibfield  {author} {\bibinfo {author} {\bibfnamefont {J.}~\bibnamefont
  {Finkel}}\ and\ \bibinfo {author} {\bibfnamefont {P.~A.}\ \bibnamefont
  {O'Gorman}},\ }\bibfield  {title} {\bibinfo {title} {{Boosting Ensembles for
  Statistics of Tails at Conditionally Optimal Advance Split Times}},\
  }\href@noop {} {\bibfield  {journal} {\bibinfo  {journal} {arXiv preprint
  arXiv:2507.22310}\ } (\bibinfo {year} {2025})}\BibitemShut {NoStop}%
\bibitem [{\citenamefont {Wouters}\ \emph {et~al.}(2023)\citenamefont
  {Wouters}, \citenamefont {Schiemann},\ and\ \citenamefont
  {Shaffrey}}]{wouters2023rare}%
  \BibitemOpen
  \bibfield  {author} {\bibinfo {author} {\bibfnamefont {J.}~\bibnamefont
  {Wouters}}, \bibinfo {author} {\bibfnamefont {R.~K.}\ \bibnamefont
  {Schiemann}},\ and\ \bibinfo {author} {\bibfnamefont {L.~C.}\ \bibnamefont
  {Shaffrey}},\ }\bibfield  {title} {\bibinfo {title} {{Rare event simulation
  of extreme European winter rainfall in an intermediate complexity climate
  model}},\ }\href@noop {} {\bibfield  {journal} {\bibinfo  {journal} {Journal
  of Advances in Modeling Earth Systems}\ }\textbf {\bibinfo {volume} {15}},\
  \bibinfo {pages} {e2022MS003537} (\bibinfo {year} {2023})}\BibitemShut
  {NoStop}%
\bibitem [{\citenamefont {Barriopedro}\ \emph {et~al.}(2011)\citenamefont
  {Barriopedro}, \citenamefont {Fischer}, \citenamefont {Luterbacher},
  \citenamefont {Trigo},\ and\ \citenamefont
  {Garc{\'\i}a-Herrera}}]{barriopedro2011hot}%
  \BibitemOpen
  \bibfield  {author} {\bibinfo {author} {\bibfnamefont {D.}~\bibnamefont
  {Barriopedro}}, \bibinfo {author} {\bibfnamefont {E.~M.}\ \bibnamefont
  {Fischer}}, \bibinfo {author} {\bibfnamefont {J.}~\bibnamefont
  {Luterbacher}}, \bibinfo {author} {\bibfnamefont {R.~M.}\ \bibnamefont
  {Trigo}},\ and\ \bibinfo {author} {\bibfnamefont {R.}~\bibnamefont
  {Garc{\'\i}a-Herrera}},\ }\bibfield  {title} {\bibinfo {title} {{The hot
  summer of 2010: redrawing the temperature record map of Europe}},\
  }\href@noop {} {\bibfield  {journal} {\bibinfo  {journal} {Science}\ }\textbf
  {\bibinfo {volume} {332}},\ \bibinfo {pages} {220} (\bibinfo {year}
  {2011})}\BibitemShut {NoStop}%
\bibitem [{\citenamefont {Stott}\ \emph {et~al.}(2004)\citenamefont {Stott},
  \citenamefont {Stone},\ and\ \citenamefont {Allen}}]{stott2004human}%
  \BibitemOpen
  \bibfield  {author} {\bibinfo {author} {\bibfnamefont {P.~A.}\ \bibnamefont
  {Stott}}, \bibinfo {author} {\bibfnamefont {D.~A.}\ \bibnamefont {Stone}},\
  and\ \bibinfo {author} {\bibfnamefont {M.~R.}\ \bibnamefont {Allen}},\
  }\bibfield  {title} {\bibinfo {title} {{Human contribution to the European
  heatwave of 2003}},\ }\href@noop {} {\bibfield  {journal} {\bibinfo
  {journal} {Nature}\ }\textbf {\bibinfo {volume} {432}},\ \bibinfo {pages}
  {610} (\bibinfo {year} {2004})}\BibitemShut {NoStop}%
\bibitem [{\citenamefont {White}\ \emph {et~al.}(2023)\citenamefont {White},
  \citenamefont {Anderson}, \citenamefont {Booth}, \citenamefont {Braich},
  \citenamefont {Draeger}, \citenamefont {Fei}, \citenamefont {Harley},
  \citenamefont {Henderson}, \citenamefont {Jakob}, \citenamefont {Lau} \emph
  {et~al.}}]{white2023unprecedented}%
  \BibitemOpen
  \bibfield  {author} {\bibinfo {author} {\bibfnamefont {R.~H.}\ \bibnamefont
  {White}}, \bibinfo {author} {\bibfnamefont {S.}~\bibnamefont {Anderson}},
  \bibinfo {author} {\bibfnamefont {J.~F.}\ \bibnamefont {Booth}}, \bibinfo
  {author} {\bibfnamefont {G.}~\bibnamefont {Braich}}, \bibinfo {author}
  {\bibfnamefont {C.}~\bibnamefont {Draeger}}, \bibinfo {author} {\bibfnamefont
  {C.}~\bibnamefont {Fei}}, \bibinfo {author} {\bibfnamefont {C.~D.}\
  \bibnamefont {Harley}}, \bibinfo {author} {\bibfnamefont {S.~B.}\
  \bibnamefont {Henderson}}, \bibinfo {author} {\bibfnamefont {M.}~\bibnamefont
  {Jakob}}, \bibinfo {author} {\bibfnamefont {C.-A.}\ \bibnamefont {Lau}},
  \emph {et~al.},\ }\bibfield  {title} {\bibinfo {title} {{The unprecedented
  Pacific Northwest heatwave of June 2021}},\ }\href@noop {} {\bibfield
  {journal} {\bibinfo  {journal} {Nature communications}\ }\textbf {\bibinfo
  {volume} {14}},\ \bibinfo {pages} {727} (\bibinfo {year} {2023})}\BibitemShut
  {NoStop}%
\bibitem [{\citenamefont {Galfi}\ and\ \citenamefont
  {Lucarini}(2021)}]{galfi2021fingerprinting}%
  \BibitemOpen
  \bibfield  {author} {\bibinfo {author} {\bibfnamefont {V.~M.}\ \bibnamefont
  {Galfi}}\ and\ \bibinfo {author} {\bibfnamefont {V.}~\bibnamefont
  {Lucarini}},\ }\bibfield  {title} {\bibinfo {title} {{Fingerprinting
  heatwaves and cold spells and assessing their response to climate change
  using large deviation theory}},\ }\href@noop {} {\bibfield  {journal}
  {\bibinfo  {journal} {Physical review letters}\ }\textbf {\bibinfo {volume}
  {127}},\ \bibinfo {pages} {058701} (\bibinfo {year} {2021})}\BibitemShut
  {NoStop}%
\bibitem [{\citenamefont {Pathak}\ \emph {et~al.}(2022)\citenamefont {Pathak},
  \citenamefont {Subramanian}, \citenamefont {Harrington}, \citenamefont
  {Raja}, \citenamefont {Chattopadhyay}, \citenamefont {Mardani}, \citenamefont
  {Kurth}, \citenamefont {Hall}, \citenamefont {Li}, \citenamefont
  {Azizzadenesheli} \emph {et~al.}}]{pathak2022fourcastnet}%
  \BibitemOpen
  \bibfield  {author} {\bibinfo {author} {\bibfnamefont {J.}~\bibnamefont
  {Pathak}}, \bibinfo {author} {\bibfnamefont {S.}~\bibnamefont {Subramanian}},
  \bibinfo {author} {\bibfnamefont {P.}~\bibnamefont {Harrington}}, \bibinfo
  {author} {\bibfnamefont {S.}~\bibnamefont {Raja}}, \bibinfo {author}
  {\bibfnamefont {A.}~\bibnamefont {Chattopadhyay}}, \bibinfo {author}
  {\bibfnamefont {M.}~\bibnamefont {Mardani}}, \bibinfo {author} {\bibfnamefont
  {T.}~\bibnamefont {Kurth}}, \bibinfo {author} {\bibfnamefont
  {D.}~\bibnamefont {Hall}}, \bibinfo {author} {\bibfnamefont {Z.}~\bibnamefont
  {Li}}, \bibinfo {author} {\bibfnamefont {K.}~\bibnamefont {Azizzadenesheli}},
  \emph {et~al.},\ }\bibfield  {title} {\bibinfo {title} {{FourCastNet: A
  global data-driven high-resolution weather model using adaptive Fourier
  neural operators}},\ }\href@noop {} {\bibfield  {journal} {\bibinfo
  {journal} {arXiv preprint arXiv:2202.11214}\ } (\bibinfo {year}
  {2022})}\BibitemShut {NoStop}%
\bibitem [{\citenamefont {Lam}\ \emph {et~al.}(2023)\citenamefont {Lam},
  \citenamefont {Sanchez-Gonzalez}, \citenamefont {Willson}, \citenamefont
  {Wirnsberger}, \citenamefont {Fortunato}, \citenamefont {Alet}, \citenamefont
  {Ravuri}, \citenamefont {Ewalds}, \citenamefont {Eaton-Rosen}, \citenamefont
  {Hu} \emph {et~al.}}]{lam2023learning}%
  \BibitemOpen
  \bibfield  {author} {\bibinfo {author} {\bibfnamefont {R.}~\bibnamefont
  {Lam}}, \bibinfo {author} {\bibfnamefont {A.}~\bibnamefont
  {Sanchez-Gonzalez}}, \bibinfo {author} {\bibfnamefont {M.}~\bibnamefont
  {Willson}}, \bibinfo {author} {\bibfnamefont {P.}~\bibnamefont
  {Wirnsberger}}, \bibinfo {author} {\bibfnamefont {M.}~\bibnamefont
  {Fortunato}}, \bibinfo {author} {\bibfnamefont {F.}~\bibnamefont {Alet}},
  \bibinfo {author} {\bibfnamefont {S.}~\bibnamefont {Ravuri}}, \bibinfo
  {author} {\bibfnamefont {T.}~\bibnamefont {Ewalds}}, \bibinfo {author}
  {\bibfnamefont {Z.}~\bibnamefont {Eaton-Rosen}}, \bibinfo {author}
  {\bibfnamefont {W.}~\bibnamefont {Hu}}, \emph {et~al.},\ }\bibfield  {title}
  {\bibinfo {title} {{Learning skillful medium-range global weather
  forecasting}},\ }\href@noop {} {\bibfield  {journal} {\bibinfo  {journal}
  {Science}\ }\textbf {\bibinfo {volume} {382}},\ \bibinfo {pages} {1416}
  (\bibinfo {year} {2023})}\BibitemShut {NoStop}%
\bibitem [{\citenamefont {Bi}\ \emph {et~al.}(2023)\citenamefont {Bi},
  \citenamefont {Xie}, \citenamefont {Zhang}, \citenamefont {Chen},
  \citenamefont {Gu},\ and\ \citenamefont {Tian}}]{bi2023accurate}%
  \BibitemOpen
  \bibfield  {author} {\bibinfo {author} {\bibfnamefont {K.}~\bibnamefont
  {Bi}}, \bibinfo {author} {\bibfnamefont {L.}~\bibnamefont {Xie}}, \bibinfo
  {author} {\bibfnamefont {H.}~\bibnamefont {Zhang}}, \bibinfo {author}
  {\bibfnamefont {X.}~\bibnamefont {Chen}}, \bibinfo {author} {\bibfnamefont
  {X.}~\bibnamefont {Gu}},\ and\ \bibinfo {author} {\bibfnamefont
  {Q.}~\bibnamefont {Tian}},\ }\bibfield  {title} {\bibinfo {title} {{Accurate
  medium-range global weather forecasting with {3D} neural networks}},\
  }\href@noop {} {\bibfield  {journal} {\bibinfo  {journal} {Nature}\ }\textbf
  {\bibinfo {volume} {619}},\ \bibinfo {pages} {533} (\bibinfo {year}
  {2023})}\BibitemShut {NoStop}%
\bibitem [{\citenamefont {Price}\ \emph {et~al.}(2025)\citenamefont {Price},
  \citenamefont {Sanchez-Gonzalez}, \citenamefont {Alet}, \citenamefont
  {Andersson}, \citenamefont {El-Kadi}, \citenamefont {Masters}, \citenamefont
  {Ewalds}, \citenamefont {Stott}, \citenamefont {Mohamed}, \citenamefont
  {Battaglia} \emph {et~al.}}]{price2025probabilistic}%
  \BibitemOpen
  \bibfield  {author} {\bibinfo {author} {\bibfnamefont {I.}~\bibnamefont
  {Price}}, \bibinfo {author} {\bibfnamefont {A.}~\bibnamefont
  {Sanchez-Gonzalez}}, \bibinfo {author} {\bibfnamefont {F.}~\bibnamefont
  {Alet}}, \bibinfo {author} {\bibfnamefont {T.~R.}\ \bibnamefont {Andersson}},
  \bibinfo {author} {\bibfnamefont {A.}~\bibnamefont {El-Kadi}}, \bibinfo
  {author} {\bibfnamefont {D.}~\bibnamefont {Masters}}, \bibinfo {author}
  {\bibfnamefont {T.}~\bibnamefont {Ewalds}}, \bibinfo {author} {\bibfnamefont
  {J.}~\bibnamefont {Stott}}, \bibinfo {author} {\bibfnamefont
  {S.}~\bibnamefont {Mohamed}}, \bibinfo {author} {\bibfnamefont
  {P.}~\bibnamefont {Battaglia}}, \emph {et~al.},\ }\bibfield  {title}
  {\bibinfo {title} {{Probabilistic weather forecasting with machine
  learning}},\ }\href@noop {} {\bibfield  {journal} {\bibinfo  {journal}
  {Nature}\ }\textbf {\bibinfo {volume} {637}},\ \bibinfo {pages} {84}
  (\bibinfo {year} {2025})}\BibitemShut {NoStop}%
\bibitem [{\citenamefont {Ben~Bouallegue}\ \emph {et~al.}(2024)\citenamefont
  {Ben~Bouallegue}, \citenamefont {Clare}, \citenamefont {Magnusson},
  \citenamefont {Gascon}, \citenamefont {Maier-Gerber}, \citenamefont
  {Janou{\v{s}}ek}, \citenamefont {Rodwell}, \citenamefont {Pinault},
  \citenamefont {Dramsch}, \citenamefont {Lang} \emph {et~al.}}]{ben2024rise}%
  \BibitemOpen
  \bibfield  {author} {\bibinfo {author} {\bibfnamefont {Z.}~\bibnamefont
  {Ben~Bouallegue}}, \bibinfo {author} {\bibfnamefont {M.~C.}\ \bibnamefont
  {Clare}}, \bibinfo {author} {\bibfnamefont {L.}~\bibnamefont {Magnusson}},
  \bibinfo {author} {\bibfnamefont {E.}~\bibnamefont {Gascon}}, \bibinfo
  {author} {\bibfnamefont {M.}~\bibnamefont {Maier-Gerber}}, \bibinfo {author}
  {\bibfnamefont {M.}~\bibnamefont {Janou{\v{s}}ek}}, \bibinfo {author}
  {\bibfnamefont {M.}~\bibnamefont {Rodwell}}, \bibinfo {author} {\bibfnamefont
  {F.}~\bibnamefont {Pinault}}, \bibinfo {author} {\bibfnamefont {J.~S.}\
  \bibnamefont {Dramsch}}, \bibinfo {author} {\bibfnamefont {S.~T.}\
  \bibnamefont {Lang}}, \emph {et~al.},\ }\bibfield  {title} {\bibinfo {title}
  {{The rise of data-driven weather forecasting: A first statistical assessment
  of machine learning--based weather forecasts in an operational-like
  context}},\ }\href@noop {} {\bibfield  {journal} {\bibinfo  {journal}
  {Bulletin of the American Meteorological Society}\ }\textbf {\bibinfo
  {volume} {105}},\ \bibinfo {pages} {E864} (\bibinfo {year}
  {2024})}\BibitemShut {NoStop}%
\bibitem [{\citenamefont {{Watt-Meyer}}\ \emph {et~al.}(2024)\citenamefont
  {{Watt-Meyer}}, \citenamefont {Henn}, \citenamefont {McGibbon}, \citenamefont
  {Clark}, \citenamefont {Kwa}, \citenamefont {Perkins}, \citenamefont {Wu},
  \citenamefont {Harris},\ and\ \citenamefont
  {Bretherton}}]{watt-meyer_ace2_2024}%
  \BibitemOpen
  \bibfield  {author} {\bibinfo {author} {\bibfnamefont {O.}~\bibnamefont
  {{Watt-Meyer}}}, \bibinfo {author} {\bibfnamefont {B.}~\bibnamefont {Henn}},
  \bibinfo {author} {\bibfnamefont {J.}~\bibnamefont {McGibbon}}, \bibinfo
  {author} {\bibfnamefont {S.~K.}\ \bibnamefont {Clark}}, \bibinfo {author}
  {\bibfnamefont {A.}~\bibnamefont {Kwa}}, \bibinfo {author} {\bibfnamefont
  {W.~A.}\ \bibnamefont {Perkins}}, \bibinfo {author} {\bibfnamefont
  {E.}~\bibnamefont {Wu}}, \bibinfo {author} {\bibfnamefont {L.}~\bibnamefont
  {Harris}},\ and\ \bibinfo {author} {\bibfnamefont {C.~S.}\ \bibnamefont
  {Bretherton}},\ }\href {https://doi.org/10.48550/arXiv.2411.11268} {\bibinfo
  {title} {{{{ACE2}}: {{Accurately}} Learning Subseasonal to Decadal
  Atmospheric Variability and Forced Responses}}} (\bibinfo {year}
  {2024})\BibitemShut {NoStop}%
\bibitem [{\citenamefont {Kochkov}\ \emph {et~al.}(2024)\citenamefont
  {Kochkov}, \citenamefont {Yuval}, \citenamefont {Langmore}, \citenamefont
  {Norgaard}, \citenamefont {Smith}, \citenamefont {Mooers}, \citenamefont
  {Kl{\"o}wer}, \citenamefont {Lottes}, \citenamefont {Rasp}, \citenamefont
  {D{\"u}ben} \emph {et~al.}}]{kochkov2024neural}%
  \BibitemOpen
  \bibfield  {author} {\bibinfo {author} {\bibfnamefont {D.}~\bibnamefont
  {Kochkov}}, \bibinfo {author} {\bibfnamefont {J.}~\bibnamefont {Yuval}},
  \bibinfo {author} {\bibfnamefont {I.}~\bibnamefont {Langmore}}, \bibinfo
  {author} {\bibfnamefont {P.}~\bibnamefont {Norgaard}}, \bibinfo {author}
  {\bibfnamefont {J.}~\bibnamefont {Smith}}, \bibinfo {author} {\bibfnamefont
  {G.}~\bibnamefont {Mooers}}, \bibinfo {author} {\bibfnamefont
  {M.}~\bibnamefont {Kl{\"o}wer}}, \bibinfo {author} {\bibfnamefont
  {J.}~\bibnamefont {Lottes}}, \bibinfo {author} {\bibfnamefont
  {S.}~\bibnamefont {Rasp}}, \bibinfo {author} {\bibfnamefont {P.}~\bibnamefont
  {D{\"u}ben}}, \emph {et~al.},\ }\bibfield  {title} {\bibinfo {title} {{Neural
  general circulation models for weather and climate}},\ }\href@noop {}
  {\bibfield  {journal} {\bibinfo  {journal} {Nature}\ }\textbf {\bibinfo
  {volume} {632}},\ \bibinfo {pages} {1060} (\bibinfo {year}
  {2024})}\BibitemShut {NoStop}%
\bibitem [{\citenamefont {Chapman}\ \emph {et~al.}(2025)\citenamefont
  {Chapman}, \citenamefont {Schreck}, \citenamefont {Sha}, \citenamefont
  {Gagne~II}, \citenamefont {Kimpara}, \citenamefont {Zanna}, \citenamefont
  {Mayer},\ and\ \citenamefont {Berner}}]{chapman2025camulator}%
  \BibitemOpen
  \bibfield  {author} {\bibinfo {author} {\bibfnamefont {W.~E.}\ \bibnamefont
  {Chapman}}, \bibinfo {author} {\bibfnamefont {J.~S.}\ \bibnamefont
  {Schreck}}, \bibinfo {author} {\bibfnamefont {Y.}~\bibnamefont {Sha}},
  \bibinfo {author} {\bibfnamefont {D.~J.}\ \bibnamefont {Gagne~II}}, \bibinfo
  {author} {\bibfnamefont {D.}~\bibnamefont {Kimpara}}, \bibinfo {author}
  {\bibfnamefont {L.}~\bibnamefont {Zanna}}, \bibinfo {author} {\bibfnamefont
  {K.~J.}\ \bibnamefont {Mayer}},\ and\ \bibinfo {author} {\bibfnamefont
  {J.}~\bibnamefont {Berner}},\ }\bibfield  {title} {\bibinfo {title}
  {{CAMulator: Fast emulation of the community atmosphere model}},\ }\href@noop
  {} {\bibfield  {journal} {\bibinfo  {journal} {arXiv preprint
  arXiv:2504.06007}\ } (\bibinfo {year} {2025})}\BibitemShut {NoStop}%
\bibitem [{\citenamefont {Mahesh}\ \emph
  {et~al.}(2025{\natexlab{a}})\citenamefont {Mahesh}, \citenamefont {Collins},
  \citenamefont {Bonev}, \citenamefont {Brenowitz}, \citenamefont {Cohen},
  \citenamefont {Elms}, \citenamefont {Harrington}, \citenamefont {Kashinath},
  \citenamefont {Kurth}, \citenamefont {North} \emph
  {et~al.}}]{mahesh2025huge1}%
  \BibitemOpen
  \bibfield  {author} {\bibinfo {author} {\bibfnamefont {A.}~\bibnamefont
  {Mahesh}}, \bibinfo {author} {\bibfnamefont {W.~D.}\ \bibnamefont {Collins}},
  \bibinfo {author} {\bibfnamefont {B.}~\bibnamefont {Bonev}}, \bibinfo
  {author} {\bibfnamefont {N.}~\bibnamefont {Brenowitz}}, \bibinfo {author}
  {\bibfnamefont {Y.}~\bibnamefont {Cohen}}, \bibinfo {author} {\bibfnamefont
  {J.}~\bibnamefont {Elms}}, \bibinfo {author} {\bibfnamefont {P.}~\bibnamefont
  {Harrington}}, \bibinfo {author} {\bibfnamefont {K.}~\bibnamefont
  {Kashinath}}, \bibinfo {author} {\bibfnamefont {T.}~\bibnamefont {Kurth}},
  \bibinfo {author} {\bibfnamefont {J.}~\bibnamefont {North}}, \emph {et~al.},\
  }\bibfield  {title} {\bibinfo {title} {{Huge ensembles--Part 1: Design of
  ensemble weather forecasts using spherical Fourier neural operators}},\
  }\href@noop {} {\bibfield  {journal} {\bibinfo  {journal} {Geoscientific
  Model Development}\ }\textbf {\bibinfo {volume} {18}},\ \bibinfo {pages}
  {5575} (\bibinfo {year} {2025}{\natexlab{a}})}\BibitemShut {NoStop}%
\bibitem [{\citenamefont {Mahesh}\ \emph
  {et~al.}(2025{\natexlab{b}})\citenamefont {Mahesh}, \citenamefont
  {D~Collins}, \citenamefont {Bonev}, \citenamefont {Brenowitz}, \citenamefont
  {Cohen}, \citenamefont {Harrington}, \citenamefont {Kashinath}, \citenamefont
  {Kurth}, \citenamefont {North}, \citenamefont {O'Brien} \emph
  {et~al.}}]{mahesh2025huge2}%
  \BibitemOpen
  \bibfield  {author} {\bibinfo {author} {\bibfnamefont {A.}~\bibnamefont
  {Mahesh}}, \bibinfo {author} {\bibfnamefont {W.}~\bibnamefont {D~Collins}},
  \bibinfo {author} {\bibfnamefont {B.}~\bibnamefont {Bonev}}, \bibinfo
  {author} {\bibfnamefont {N.}~\bibnamefont {Brenowitz}}, \bibinfo {author}
  {\bibfnamefont {Y.}~\bibnamefont {Cohen}}, \bibinfo {author} {\bibfnamefont
  {P.}~\bibnamefont {Harrington}}, \bibinfo {author} {\bibfnamefont
  {K.}~\bibnamefont {Kashinath}}, \bibinfo {author} {\bibfnamefont
  {T.}~\bibnamefont {Kurth}}, \bibinfo {author} {\bibfnamefont
  {J.}~\bibnamefont {North}}, \bibinfo {author} {\bibfnamefont {T.~A.}\
  \bibnamefont {O'Brien}}, \emph {et~al.},\ }\bibfield  {title} {\bibinfo
  {title} {{Huge ensembles--Part 2: Properties of a huge ensemble of hindcasts
  generated with spherical Fourier neural operators}},\ }\href@noop {}
  {\bibfield  {journal} {\bibinfo  {journal} {Geoscientific Model Development}\
  }\textbf {\bibinfo {volume} {18}},\ \bibinfo {pages} {5605} (\bibinfo {year}
  {2025}{\natexlab{b}})}\BibitemShut {NoStop}%
\bibitem [{\citenamefont {Sun}\ \emph {et~al.}(2025{\natexlab{a}})\citenamefont
  {Sun}, \citenamefont {Hassanzadeh}, \citenamefont {Zand}, \citenamefont
  {Chattopadhyay}, \citenamefont {Weare},\ and\ \citenamefont
  {Abbot}}]{sun_can_2025}%
  \BibitemOpen
  \bibfield  {author} {\bibinfo {author} {\bibfnamefont {Y.~Q.}\ \bibnamefont
  {Sun}}, \bibinfo {author} {\bibfnamefont {P.}~\bibnamefont {Hassanzadeh}},
  \bibinfo {author} {\bibfnamefont {M.}~\bibnamefont {Zand}}, \bibinfo {author}
  {\bibfnamefont {A.}~\bibnamefont {Chattopadhyay}}, \bibinfo {author}
  {\bibfnamefont {J.}~\bibnamefont {Weare}},\ and\ \bibinfo {author}
  {\bibfnamefont {D.~S.}\ \bibnamefont {Abbot}},\ }\bibfield  {title} {\bibinfo
  {title} {{Can {{AI}} Weather Models Predict Out-of-Distribution Gray Swan
  Tropical Cyclones?}},\ }\href {https://doi.org/10.1073/pnas.2420914122}
  {\bibfield  {journal} {\bibinfo  {journal} {Proceedings of the National
  Academy of Sciences}\ }\textbf {\bibinfo {volume} {122}},\ \bibinfo {pages}
  {e2420914122} (\bibinfo {year} {2025}{\natexlab{a}})}\BibitemShut {NoStop}%
\bibitem [{\citenamefont {Sun}\ \emph {et~al.}(2025{\natexlab{b}})\citenamefont
  {Sun}, \citenamefont {Hassanzadeh}, \citenamefont {Shaw},\ and\ \citenamefont
  {Pahlavan}}]{sun2025predicting}%
  \BibitemOpen
  \bibfield  {author} {\bibinfo {author} {\bibfnamefont {Y.~Q.}\ \bibnamefont
  {Sun}}, \bibinfo {author} {\bibfnamefont {P.}~\bibnamefont {Hassanzadeh}},
  \bibinfo {author} {\bibfnamefont {T.}~\bibnamefont {Shaw}},\ and\ \bibinfo
  {author} {\bibfnamefont {H.~A.}\ \bibnamefont {Pahlavan}},\ }\bibfield
  {title} {\bibinfo {title} {{Predicting Beyond Training Data via Extrapolation
  versus Translocation: AI Weather Models and Dubai's Unprecedented 2024
  Rainfall}},\ }\href@noop {} {\bibfield  {journal} {\bibinfo  {journal} {arXiv
  preprint arXiv:2505.10241}\ } (\bibinfo {year} {2025}{\natexlab{b}})},\
  \Eprint {https://arxiv.org/abs/2505.10241} {arXiv:2505.10241} \BibitemShut
  {NoStop}%
\bibitem [{\citenamefont {Zhang}\ \emph {et~al.}(2025)\citenamefont {Zhang},
  \citenamefont {Fischer}, \citenamefont {Zscheischler},\ and\ \citenamefont
  {Engelke}}]{zhang2025numerical}%
  \BibitemOpen
  \bibfield  {author} {\bibinfo {author} {\bibfnamefont {Z.}~\bibnamefont
  {Zhang}}, \bibinfo {author} {\bibfnamefont {E.}~\bibnamefont {Fischer}},
  \bibinfo {author} {\bibfnamefont {J.}~\bibnamefont {Zscheischler}},\ and\
  \bibinfo {author} {\bibfnamefont {S.}~\bibnamefont {Engelke}},\ }\bibfield
  {title} {\bibinfo {title} {{Numerical models outperform AI weather forecasts
  of record-breaking extremes}},\ }\href@noop {} {\bibfield  {journal}
  {\bibinfo  {journal} {arXiv preprint arXiv:2508.15724}\ } (\bibinfo {year}
  {2025})}\BibitemShut {NoStop}%
\bibitem [{\citenamefont {Wikner}\ \emph {et~al.}(2025)\citenamefont {Wikner},
  \citenamefont {Lancelin}, \citenamefont {Arcomano}, \citenamefont {Jakhar},
  \citenamefont {Patel}, \citenamefont {Bouchet},\ and\ \citenamefont
  {Hassanzadeh}}]{plasim_long_emulation}%
  \BibitemOpen
  \bibfield  {author} {\bibinfo {author} {\bibfnamefont {A.}~\bibnamefont
  {Wikner}}, \bibinfo {author} {\bibfnamefont {A.}~\bibnamefont {Lancelin}},
  \bibinfo {author} {\bibfnamefont {T.}~\bibnamefont {Arcomano}}, \bibinfo
  {author} {\bibfnamefont {K.}~\bibnamefont {Jakhar}}, \bibinfo {author}
  {\bibfnamefont {D.}~\bibnamefont {Patel}}, \bibinfo {author} {\bibfnamefont
  {F.}~\bibnamefont {Bouchet}},\ and\ \bibinfo {author} {\bibfnamefont
  {P.}~\bibnamefont {Hassanzadeh}},\ }\href@noop {} {\bibinfo {title} {{Can AI
  climate emulators quantify the statistics of unseen weather extremes?}}},\
  \bibinfo {howpublished} {Presented at AGU Annual Meeting 2025, New Orleans,
  LA, December 16, 2025, Abstract NG24A-06} (\bibinfo {year}
  {2025})\BibitemShut {NoStop}%
\bibitem [{\citenamefont {Falasca}(2025)}]{Falasca2025ForcedResponses}%
  \BibitemOpen
  \bibfield  {author} {\bibinfo {author} {\bibfnamefont {F.}~\bibnamefont
  {Falasca}},\ }\bibfield  {title} {\bibinfo {title} {Probing forced responses
  and causality in data-driven climate emulators: Conceptual limitations and
  the role of reduced-order models},\ }\href
  {https://doi.org/10.1103/2f4r-k8lr} {\bibfield  {journal} {\bibinfo
  {journal} {Physical Review Research}\ }\textbf {\bibinfo {volume} {7}},\
  \bibinfo {pages} {043314} (\bibinfo {year} {2025})}\BibitemShut {NoStop}%
\bibitem [{\citenamefont {Fraedrich}\ \emph {et~al.}(2005)\citenamefont
  {Fraedrich}, \citenamefont {Jansen}, \citenamefont {Kirk}, \citenamefont
  {Luksch},\ and\ \citenamefont {Lunkeit}}]{fraedrich_planet_2005}%
  \BibitemOpen
  \bibfield  {author} {\bibinfo {author} {\bibfnamefont {K.}~\bibnamefont
  {Fraedrich}}, \bibinfo {author} {\bibfnamefont {H.}~\bibnamefont {Jansen}},
  \bibinfo {author} {\bibfnamefont {E.}~\bibnamefont {Kirk}}, \bibinfo {author}
  {\bibfnamefont {U.}~\bibnamefont {Luksch}},\ and\ \bibinfo {author}
  {\bibfnamefont {F.}~\bibnamefont {Lunkeit}},\ }\bibfield  {title} {\bibinfo
  {title} {{The {{Planet Simulator}}: {{Towards}} a User Friendly Model}},\
  }\href {https://doi.org/10.1127/0941-2948/2005/0043} {\bibfield  {journal}
  {\bibinfo  {journal} {Meteorologische Zeitschrift}\ ,\ \bibinfo {pages}
  {299}} (\bibinfo {year} {2005})}\BibitemShut {NoStop}%
\bibitem [{\citenamefont {Zhao}\ \emph {et~al.}(2021)\citenamefont {Zhao},
  \citenamefont {Guo}, \citenamefont {Ye}, \citenamefont {Gasparrini},
  \citenamefont {Tong}, \citenamefont {Overcenco}, \citenamefont {Urban},
  \citenamefont {Schneider}, \citenamefont {Entezari}, \citenamefont
  {Vicedo-Cabrera} \emph {et~al.}}]{zhao2021global}%
  \BibitemOpen
  \bibfield  {author} {\bibinfo {author} {\bibfnamefont {Q.}~\bibnamefont
  {Zhao}}, \bibinfo {author} {\bibfnamefont {Y.}~\bibnamefont {Guo}}, \bibinfo
  {author} {\bibfnamefont {T.}~\bibnamefont {Ye}}, \bibinfo {author}
  {\bibfnamefont {A.}~\bibnamefont {Gasparrini}}, \bibinfo {author}
  {\bibfnamefont {S.}~\bibnamefont {Tong}}, \bibinfo {author} {\bibfnamefont
  {A.}~\bibnamefont {Overcenco}}, \bibinfo {author} {\bibfnamefont
  {A.}~\bibnamefont {Urban}}, \bibinfo {author} {\bibfnamefont
  {A.}~\bibnamefont {Schneider}}, \bibinfo {author} {\bibfnamefont
  {A.}~\bibnamefont {Entezari}}, \bibinfo {author} {\bibfnamefont {A.~M.}\
  \bibnamefont {Vicedo-Cabrera}}, \emph {et~al.},\ }\bibfield  {title}
  {\bibinfo {title} {{Global, regional, and national burden of mortality
  associated with non-optimal ambient temperatures from 2000 to 2019: a
  three-stage modelling study}},\ }\href@noop {} {\bibfield  {journal}
  {\bibinfo  {journal} {The Lancet Planetary Health}\ }\textbf {\bibinfo
  {volume} {5}},\ \bibinfo {pages} {e415} (\bibinfo {year} {2021})}\BibitemShut
  {NoStop}%
\bibitem [{\citenamefont {Newman}\ and\ \citenamefont
  {Noy}(2023)}]{newman2023global}%
  \BibitemOpen
  \bibfield  {author} {\bibinfo {author} {\bibfnamefont {R.}~\bibnamefont
  {Newman}}\ and\ \bibinfo {author} {\bibfnamefont {I.}~\bibnamefont {Noy}},\
  }\bibfield  {title} {\bibinfo {title} {{The global costs of extreme weather
  that are attributable to climate change}},\ }\href@noop {} {\bibfield
  {journal} {\bibinfo  {journal} {Nature Communications}\ }\textbf {\bibinfo
  {volume} {14}},\ \bibinfo {pages} {6103} (\bibinfo {year}
  {2023})}\BibitemShut {NoStop}%
\bibitem [{\citenamefont {Thompson}\ \emph {et~al.}(2023)\citenamefont
  {Thompson}, \citenamefont {Mitchell}, \citenamefont {Hegerl}, \citenamefont
  {Collins}, \citenamefont {Leach},\ and\ \citenamefont
  {Slingo}}]{thompson2023most}%
  \BibitemOpen
  \bibfield  {author} {\bibinfo {author} {\bibfnamefont {V.}~\bibnamefont
  {Thompson}}, \bibinfo {author} {\bibfnamefont {D.}~\bibnamefont {Mitchell}},
  \bibinfo {author} {\bibfnamefont {G.~C.}\ \bibnamefont {Hegerl}}, \bibinfo
  {author} {\bibfnamefont {M.}~\bibnamefont {Collins}}, \bibinfo {author}
  {\bibfnamefont {N.~J.}\ \bibnamefont {Leach}},\ and\ \bibinfo {author}
  {\bibfnamefont {J.~M.}\ \bibnamefont {Slingo}},\ }\bibfield  {title}
  {\bibinfo {title} {{The most at-risk regions in the world for high-impact
  heatwaves}},\ }\href@noop {} {\bibfield  {journal} {\bibinfo  {journal}
  {Nature Communications}\ }\textbf {\bibinfo {volume} {14}},\ \bibinfo {pages}
  {2152} (\bibinfo {year} {2023})}\BibitemShut {NoStop}%
\bibitem [{\citenamefont {Miloshevich}\ \emph {et~al.}(2023)\citenamefont
  {Miloshevich}, \citenamefont {Cozian}, \citenamefont {Abry}, \citenamefont
  {Borgnat},\ and\ \citenamefont {Bouchet}}]{miloshevich_probabilistic_2023-1}%
  \BibitemOpen
  \bibfield  {author} {\bibinfo {author} {\bibfnamefont {G.}~\bibnamefont
  {Miloshevich}}, \bibinfo {author} {\bibfnamefont {B.}~\bibnamefont {Cozian}},
  \bibinfo {author} {\bibfnamefont {P.}~\bibnamefont {Abry}}, \bibinfo {author}
  {\bibfnamefont {P.}~\bibnamefont {Borgnat}},\ and\ \bibinfo {author}
  {\bibfnamefont {F.}~\bibnamefont {Bouchet}},\ }\bibfield  {title} {\bibinfo
  {title} {{Probabilistic Forecasts of Extreme Heatwaves Using Convolutional
  Neural Networks in a Regime of Lack of Data}},\ }\href
  {https://doi.org/10.1103/PhysRevFluids.8.040501} {\bibfield  {journal}
  {\bibinfo  {journal} {Physical Review Fluids}\ }\textbf {\bibinfo {volume}
  {8}},\ \bibinfo {pages} {040501} (\bibinfo {year} {2023})}\BibitemShut
  {NoStop}%
\bibitem [{\citenamefont {Kantz}\ \emph {et~al.}(2006)\citenamefont {Kantz},
  \citenamefont {Altmann}, \citenamefont {Hallerberg}, \citenamefont
  {Holstein},\ and\ \citenamefont {Riegert}}]{kantz2006dynamical}%
  \BibitemOpen
  \bibfield  {author} {\bibinfo {author} {\bibfnamefont {H.}~\bibnamefont
  {Kantz}}, \bibinfo {author} {\bibfnamefont {E.~G.}\ \bibnamefont {Altmann}},
  \bibinfo {author} {\bibfnamefont {S.}~\bibnamefont {Hallerberg}}, \bibinfo
  {author} {\bibfnamefont {D.}~\bibnamefont {Holstein}},\ and\ \bibinfo
  {author} {\bibfnamefont {A.}~\bibnamefont {Riegert}},\ }\bibfield  {title}
  {\bibinfo {title} {Dynamical interpretation of extreme events: predictability
  and predictions},\ }in\ \href@noop {} {\emph {\bibinfo {booktitle} {Extreme
  events in nature and society}}}\ (\bibinfo  {publisher} {Springer},\ \bibinfo
  {year} {2006})\ pp.\ \bibinfo {pages} {69--93}\BibitemShut {NoStop}%
\bibitem [{\citenamefont {Ghil}\ \emph {et~al.}(2011)\citenamefont {Ghil},
  \citenamefont {Yiou}, \citenamefont {Hallegatte}, \citenamefont {Malamud},
  \citenamefont {Naveau}, \citenamefont {Soloviev}, \citenamefont
  {Friederichs}, \citenamefont {Keilis-Borok}, \citenamefont {Kondrashov},
  \citenamefont {Kossobokov}, \citenamefont {Mestre}, \citenamefont {Nicolis},
  \citenamefont {Rust}, \citenamefont {Shebalin}, \citenamefont {Vrac},
  \citenamefont {Witt},\ and\ \citenamefont
  {Zaliapin}}]{Ghil2011ExtremeEvents}%
  \BibitemOpen
  \bibfield  {author} {\bibinfo {author} {\bibfnamefont {M.}~\bibnamefont
  {Ghil}}, \bibinfo {author} {\bibfnamefont {P.}~\bibnamefont {Yiou}}, \bibinfo
  {author} {\bibfnamefont {S.}~\bibnamefont {Hallegatte}}, \bibinfo {author}
  {\bibfnamefont {B.~D.}\ \bibnamefont {Malamud}}, \bibinfo {author}
  {\bibfnamefont {P.}~\bibnamefont {Naveau}}, \bibinfo {author} {\bibfnamefont
  {A.}~\bibnamefont {Soloviev}}, \bibinfo {author} {\bibfnamefont
  {P.}~\bibnamefont {Friederichs}}, \bibinfo {author} {\bibfnamefont
  {V.}~\bibnamefont {Keilis-Borok}}, \bibinfo {author} {\bibfnamefont
  {D.}~\bibnamefont {Kondrashov}}, \bibinfo {author} {\bibfnamefont
  {V.}~\bibnamefont {Kossobokov}}, \bibinfo {author} {\bibfnamefont
  {O.}~\bibnamefont {Mestre}}, \bibinfo {author} {\bibfnamefont
  {C.}~\bibnamefont {Nicolis}}, \bibinfo {author} {\bibfnamefont {H.~W.}\
  \bibnamefont {Rust}}, \bibinfo {author} {\bibfnamefont {P.}~\bibnamefont
  {Shebalin}}, \bibinfo {author} {\bibfnamefont {M.}~\bibnamefont {Vrac}},
  \bibinfo {author} {\bibfnamefont {A.}~\bibnamefont {Witt}},\ and\ \bibinfo
  {author} {\bibfnamefont {I.}~\bibnamefont {Zaliapin}},\ }\bibfield  {title}
  {\bibinfo {title} {Extreme events: Dynamics, statistics and prediction},\
  }\href {https://doi.org/10.5194/npg-18-295-2011} {\bibfield  {journal}
  {\bibinfo  {journal} {Nonlinear Processes in Geophysics}\ }\textbf {\bibinfo
  {volume} {18}},\ \bibinfo {pages} {295} (\bibinfo {year} {2011})}\BibitemShut
  {NoStop}%
\bibitem [{sup(2026)}]{suppmat}%
  \BibitemOpen
  \href@noop {} {} (\bibinfo {year} {2026}),\ \bibinfo {note} {see Supplemental
  Material at \url{http://link.aps.org/supplemental/10.1103/b1gc-9c2q} for
  details on the PlaSim GCM configuration, AI emulator architecture and
  training, RES algorithm implementation and hyperparameters, additional
  baselines (including PFS+RES), speed-up metrics, and supplementary
  figures}\BibitemShut {NoStop}%
\bibitem [{\citenamefont {Alet}\ \emph {et~al.}(2025)\citenamefont {Alet},
  \citenamefont {Price}, \citenamefont {El-Kadi}, \citenamefont {Masters},
  \citenamefont {Markou}, \citenamefont {Andersson}, \citenamefont {Stott},
  \citenamefont {Lam}, \citenamefont {Willson}, \citenamefont
  {Sanchez-Gonzalez} \emph {et~al.}}]{alet2025skillful}%
  \BibitemOpen
  \bibfield  {author} {\bibinfo {author} {\bibfnamefont {F.}~\bibnamefont
  {Alet}}, \bibinfo {author} {\bibfnamefont {I.}~\bibnamefont {Price}},
  \bibinfo {author} {\bibfnamefont {A.}~\bibnamefont {El-Kadi}}, \bibinfo
  {author} {\bibfnamefont {D.}~\bibnamefont {Masters}}, \bibinfo {author}
  {\bibfnamefont {S.}~\bibnamefont {Markou}}, \bibinfo {author} {\bibfnamefont
  {T.~R.}\ \bibnamefont {Andersson}}, \bibinfo {author} {\bibfnamefont
  {J.}~\bibnamefont {Stott}}, \bibinfo {author} {\bibfnamefont
  {R.}~\bibnamefont {Lam}}, \bibinfo {author} {\bibfnamefont {M.}~\bibnamefont
  {Willson}}, \bibinfo {author} {\bibfnamefont {A.}~\bibnamefont
  {Sanchez-Gonzalez}}, \emph {et~al.},\ }\bibfield  {title} {\bibinfo {title}
  {{Skillful joint probabilistic weather forecasting from marginals}},\
  }\href@noop {} {\bibfield  {journal} {\bibinfo  {journal} {arXiv preprint
  arXiv:2506.10772}\ } (\bibinfo {year} {2025})}\BibitemShut {NoStop}%
\bibitem [{\citenamefont {Lang}\ \emph {et~al.}(2024)\citenamefont {Lang},
  \citenamefont {Alexe}, \citenamefont {Clare}, \citenamefont {Roberts},
  \citenamefont {Adewoyin}, \citenamefont {Bouall{\`e}gue}, \citenamefont
  {Chantry}, \citenamefont {Dramsch}, \citenamefont {Dueben}, \citenamefont
  {Hahner} \emph {et~al.}}]{lang2024aifs}%
  \BibitemOpen
  \bibfield  {author} {\bibinfo {author} {\bibfnamefont {S.}~\bibnamefont
  {Lang}}, \bibinfo {author} {\bibfnamefont {M.}~\bibnamefont {Alexe}},
  \bibinfo {author} {\bibfnamefont {M.~C.}\ \bibnamefont {Clare}}, \bibinfo
  {author} {\bibfnamefont {C.}~\bibnamefont {Roberts}}, \bibinfo {author}
  {\bibfnamefont {R.}~\bibnamefont {Adewoyin}}, \bibinfo {author}
  {\bibfnamefont {Z.~B.}\ \bibnamefont {Bouall{\`e}gue}}, \bibinfo {author}
  {\bibfnamefont {M.}~\bibnamefont {Chantry}}, \bibinfo {author} {\bibfnamefont
  {J.}~\bibnamefont {Dramsch}}, \bibinfo {author} {\bibfnamefont {P.~D.}\
  \bibnamefont {Dueben}}, \bibinfo {author} {\bibfnamefont {S.}~\bibnamefont
  {Hahner}}, \emph {et~al.},\ }\bibfield  {title} {\bibinfo {title}
  {{AIFS-CRPS: ensemble forecasting using a model trained with a loss function
  based on the continuous ranked probability score}},\ }\href@noop {}
  {\bibfield  {journal} {\bibinfo  {journal} {arXiv preprint arXiv:2412.15832}\
  } (\bibinfo {year} {2024})}\BibitemShut {NoStop}%
\bibitem [{\citenamefont {Couairon}\ \emph {et~al.}(2024)\citenamefont
  {Couairon}, \citenamefont {Singh}, \citenamefont {Charantonis}, \citenamefont
  {Lessig},\ and\ \citenamefont {Monteleoni}}]{couairon2024archesweather}%
  \BibitemOpen
  \bibfield  {author} {\bibinfo {author} {\bibfnamefont {G.}~\bibnamefont
  {Couairon}}, \bibinfo {author} {\bibfnamefont {R.}~\bibnamefont {Singh}},
  \bibinfo {author} {\bibfnamefont {A.}~\bibnamefont {Charantonis}}, \bibinfo
  {author} {\bibfnamefont {C.}~\bibnamefont {Lessig}},\ and\ \bibinfo {author}
  {\bibfnamefont {C.}~\bibnamefont {Monteleoni}},\ }\bibfield  {title}
  {\bibinfo {title} {{ArchesWeather \& ArchesWeatherGen: a deterministic and
  generative model for efficient ML weather forecasting}},\ }\href@noop {}
  {\bibfield  {journal} {\bibinfo  {journal} {arXiv preprint arXiv:2412.12971}\
  } (\bibinfo {year} {2024})}\BibitemShut {NoStop}%
\bibitem [{\citenamefont {Zhou}\ \emph {et~al.}(2025)\citenamefont {Zhou},
  \citenamefont {Wikner}, \citenamefont {Lancelin}, \citenamefont
  {Hassanzadeh},\ and\ \citenamefont {Farimani}}]{zhou2025reframing}%
  \BibitemOpen
  \bibfield  {author} {\bibinfo {author} {\bibfnamefont {A.}~\bibnamefont
  {Zhou}}, \bibinfo {author} {\bibfnamefont {A.}~\bibnamefont {Wikner}},
  \bibinfo {author} {\bibfnamefont {A.}~\bibnamefont {Lancelin}}, \bibinfo
  {author} {\bibfnamefont {P.}~\bibnamefont {Hassanzadeh}},\ and\ \bibinfo
  {author} {\bibfnamefont {A.~B.}\ \bibnamefont {Farimani}},\ }\bibfield
  {title} {\bibinfo {title} {{Reframing Generative Models for Physical Systems
  using Stochastic Interpolants}},\ }\href@noop {} {\bibfield  {journal}
  {\bibinfo  {journal} {arXiv preprint arXiv:2509.26282}\ } (\bibinfo {year}
  {2025})}\BibitemShut {NoStop}%
\bibitem [{\citenamefont {D'Andrea}\ \emph {et~al.}(2006)\citenamefont
  {D'Andrea}, \citenamefont {Provenzale}, \citenamefont {Vautard},\ and\
  \citenamefont {{De Noblet-Decoudr{\'e}}}}]{dandrea_hot_2006}%
  \BibitemOpen
  \bibfield  {author} {\bibinfo {author} {\bibfnamefont {F.}~\bibnamefont
  {D'Andrea}}, \bibinfo {author} {\bibfnamefont {A.}~\bibnamefont
  {Provenzale}}, \bibinfo {author} {\bibfnamefont {R.}~\bibnamefont
  {Vautard}},\ and\ \bibinfo {author} {\bibfnamefont {N.}~\bibnamefont {{De
  Noblet-Decoudr{\'e}}}},\ }\bibfield  {title} {\bibinfo {title} {{Hot and Cool
  Summers: {{Multiple}} Equilibria of the Continental Water Cycle}},\
  }\bibfield  {journal} {\bibinfo  {journal} {Geophysical Research Letters}\
  }\textbf {\bibinfo {volume} {33}},\ \href
  {https://doi.org/10.1029/2006GL027972} {10.1029/2006GL027972} (\bibinfo
  {year} {2006})\BibitemShut {NoStop}%
\bibitem [{\citenamefont {Fischer}\ \emph {et~al.}(2007)\citenamefont
  {Fischer}, \citenamefont {Seneviratne}, \citenamefont {Vidale}, \citenamefont
  {L{\"u}thi},\ and\ \citenamefont {Sch{\"a}r}}]{fischer_soil_2007}%
  \BibitemOpen
  \bibfield  {author} {\bibinfo {author} {\bibfnamefont {E.~M.}\ \bibnamefont
  {Fischer}}, \bibinfo {author} {\bibfnamefont {S.~I.}\ \bibnamefont
  {Seneviratne}}, \bibinfo {author} {\bibfnamefont {P.~L.}\ \bibnamefont
  {Vidale}}, \bibinfo {author} {\bibfnamefont {D.}~\bibnamefont {L{\"u}thi}},\
  and\ \bibinfo {author} {\bibfnamefont {C.}~\bibnamefont {Sch{\"a}r}},\
  }\bibfield  {title} {\bibinfo {title} {{Soil {{Moisture}}--{{Atmosphere
  Interactions}} during the 2003 {{European Summer Heat Wave}}}},\ }\href
  {https://doi.org/10.1175/JCLI4288.1} {\bibfield  {journal} {\bibinfo
  {journal} {Journal of Climate}\ }\textbf {\bibinfo {volume} {20}},\ \bibinfo
  {pages} {5081} (\bibinfo {year} {2007})}\BibitemShut {NoStop}%
\bibitem [{\citenamefont {Vautard}\ \emph {et~al.}(2007)\citenamefont
  {Vautard}, \citenamefont {Yiou}, \citenamefont {D'Andrea}, \citenamefont {{de
  Noblet}}, \citenamefont {Viovy}, \citenamefont {Cassou}, \citenamefont
  {Polcher}, \citenamefont {Ciais}, \citenamefont {Kageyama},\ and\
  \citenamefont {Fan}}]{vautard_summertime_2007}%
  \BibitemOpen
  \bibfield  {author} {\bibinfo {author} {\bibfnamefont {R.}~\bibnamefont
  {Vautard}}, \bibinfo {author} {\bibfnamefont {P.}~\bibnamefont {Yiou}},
  \bibinfo {author} {\bibfnamefont {F.}~\bibnamefont {D'Andrea}}, \bibinfo
  {author} {\bibfnamefont {N.}~\bibnamefont {{de Noblet}}}, \bibinfo {author}
  {\bibfnamefont {N.}~\bibnamefont {Viovy}}, \bibinfo {author} {\bibfnamefont
  {C.}~\bibnamefont {Cassou}}, \bibinfo {author} {\bibfnamefont
  {J.}~\bibnamefont {Polcher}}, \bibinfo {author} {\bibfnamefont
  {P.}~\bibnamefont {Ciais}}, \bibinfo {author} {\bibfnamefont
  {M.}~\bibnamefont {Kageyama}},\ and\ \bibinfo {author} {\bibfnamefont
  {Y.}~\bibnamefont {Fan}},\ }\bibfield  {title} {\bibinfo {title} {{Summertime
  {{European}} Heat and Drought Waves Induced by Wintertime {{Mediterranean}}
  Rainfall Deficit}},\ }\bibfield  {journal} {\bibinfo  {journal} {Geophysical
  Research Letters}\ }\textbf {\bibinfo {volume} {34}},\ \href
  {https://doi.org/10.1029/2006GL028001} {10.1029/2006GL028001} (\bibinfo
  {year} {2007})\BibitemShut {NoStop}%
\bibitem [{\citenamefont {Del~Moral}(2004)}]{moral2004feynman}%
  \BibitemOpen
  \bibfield  {author} {\bibinfo {author} {\bibfnamefont {P.}~\bibnamefont
  {Del~Moral}},\ }\href@noop {} {\emph {\bibinfo {title} {{Feynman-Kac
  formulae: genealogical and interacting particle systems with
  applications}}}}\ (\bibinfo  {publisher} {Springer},\ \bibinfo {year}
  {2004})\BibitemShut {NoStop}%
\bibitem [{\citenamefont {Deville}\ and\ \citenamefont
  {Tille}(1998)}]{deville1998unequal}%
  \BibitemOpen
  \bibfield  {author} {\bibinfo {author} {\bibfnamefont {J.-C.}\ \bibnamefont
  {Deville}}\ and\ \bibinfo {author} {\bibfnamefont {Y.}~\bibnamefont
  {Tille}},\ }\bibfield  {title} {\bibinfo {title} {{Unequal probability
  sampling without replacement through a splitting method}},\ }\href@noop {}
  {\bibfield  {journal} {\bibinfo  {journal} {Biometrika}\ }\textbf {\bibinfo
  {volume} {85}},\ \bibinfo {pages} {89} (\bibinfo {year} {1998})}\BibitemShut
  {NoStop}%
\bibitem [{\citenamefont {Chattopadhyay}\ \emph {et~al.}(2020)\citenamefont
  {Chattopadhyay}, \citenamefont {Nabizadeh},\ and\ \citenamefont
  {Hassanzadeh}}]{chattopadhyay2020analog}%
  \BibitemOpen
  \bibfield  {author} {\bibinfo {author} {\bibfnamefont {A.}~\bibnamefont
  {Chattopadhyay}}, \bibinfo {author} {\bibfnamefont {E.}~\bibnamefont
  {Nabizadeh}},\ and\ \bibinfo {author} {\bibfnamefont {P.}~\bibnamefont
  {Hassanzadeh}},\ }\bibfield  {title} {\bibinfo {title} {{Analog forecasting
  of extreme-causing weather patterns using deep learning}},\ }\href@noop {}
  {\bibfield  {journal} {\bibinfo  {journal} {Journal of Advances in Modeling
  Earth Systems}\ }\textbf {\bibinfo {volume} {12}},\ \bibinfo {pages}
  {e2019MS001958} (\bibinfo {year} {2020})}\BibitemShut {NoStop}%
\bibitem [{\citenamefont {Finkel}\ \emph {et~al.}(2021)\citenamefont {Finkel},
  \citenamefont {Webber}, \citenamefont {Gerber}, \citenamefont {Abbot},\ and\
  \citenamefont {Weare}}]{finkel2021learning}%
  \BibitemOpen
  \bibfield  {author} {\bibinfo {author} {\bibfnamefont {J.}~\bibnamefont
  {Finkel}}, \bibinfo {author} {\bibfnamefont {R.~J.}\ \bibnamefont {Webber}},
  \bibinfo {author} {\bibfnamefont {E.~P.}\ \bibnamefont {Gerber}}, \bibinfo
  {author} {\bibfnamefont {D.~S.}\ \bibnamefont {Abbot}},\ and\ \bibinfo
  {author} {\bibfnamefont {J.}~\bibnamefont {Weare}},\ }\bibfield  {title}
  {\bibinfo {title} {{Learning forecasts of rare stratospheric transitions from
  short simulations}},\ }\href@noop {} {\bibfield  {journal} {\bibinfo
  {journal} {Monthly Weather Review}\ }\textbf {\bibinfo {volume} {149}},\
  \bibinfo {pages} {3647} (\bibinfo {year} {2021})}\BibitemShut {NoStop}%
\bibitem [{\citenamefont {Finkel}\ \emph {et~al.}(2023)\citenamefont {Finkel},
  \citenamefont {Gerber}, \citenamefont {Abbot},\ and\ \citenamefont
  {Weare}}]{finkel2023revealing}%
  \BibitemOpen
  \bibfield  {author} {\bibinfo {author} {\bibfnamefont {J.}~\bibnamefont
  {Finkel}}, \bibinfo {author} {\bibfnamefont {E.~P.}\ \bibnamefont {Gerber}},
  \bibinfo {author} {\bibfnamefont {D.~S.}\ \bibnamefont {Abbot}},\ and\
  \bibinfo {author} {\bibfnamefont {J.}~\bibnamefont {Weare}},\ }\bibfield
  {title} {\bibinfo {title} {{Revealing the statistics of extreme events hidden
  in short weather forecast data}},\ }\href@noop {} {\bibfield  {journal}
  {\bibinfo  {journal} {AGU Advances}\ }\textbf {\bibinfo {volume} {4}},\
  \bibinfo {pages} {e2023AV000881} (\bibinfo {year} {2023})}\BibitemShut
  {NoStop}%
\bibitem [{\citenamefont {Mascolo}\ \emph {et~al.}(2025)\citenamefont
  {Mascolo}, \citenamefont {Lovo}, \citenamefont {Herbert},\ and\ \citenamefont
  {Bouchet}}]{mascolo2025gaussian}%
  \BibitemOpen
  \bibfield  {author} {\bibinfo {author} {\bibfnamefont {V.}~\bibnamefont
  {Mascolo}}, \bibinfo {author} {\bibfnamefont {A.}~\bibnamefont {Lovo}},
  \bibinfo {author} {\bibfnamefont {C.}~\bibnamefont {Herbert}},\ and\ \bibinfo
  {author} {\bibfnamefont {F.}~\bibnamefont {Bouchet}},\ }\bibfield  {title}
  {\bibinfo {title} {{Gaussian framework and optimal projection of weather
  fields for prediction of extreme events}},\ }\href@noop {} {\bibfield
  {journal} {\bibinfo  {journal} {Journal of Advances in Modeling Earth
  Systems}\ }\textbf {\bibinfo {volume} {17}},\ \bibinfo {pages}
  {e2024MS004487} (\bibinfo {year} {2025})}\BibitemShut {NoStop}%
\bibitem [{\citenamefont {Lovo}\ \emph {et~al.}(2025)\citenamefont {Lovo},
  \citenamefont {Lancelin}, \citenamefont {Herbert},\ and\ \citenamefont
  {Bouchet}}]{lovo2025tackling}%
  \BibitemOpen
  \bibfield  {author} {\bibinfo {author} {\bibfnamefont {A.}~\bibnamefont
  {Lovo}}, \bibinfo {author} {\bibfnamefont {A.}~\bibnamefont {Lancelin}},
  \bibinfo {author} {\bibfnamefont {C.}~\bibnamefont {Herbert}},\ and\ \bibinfo
  {author} {\bibfnamefont {F.}~\bibnamefont {Bouchet}},\ }\bibfield  {title}
  {\bibinfo {title} {{Tackling the Accuracy--Interpretability Trade-Off in a
  Hierarchy of Machine Learning Models for the Prediction of Extreme
  Heatwaves}},\ }\href {https://doi.org/10.1175/AIES-D-24-0094.1} {\bibfield
  {journal} {\bibinfo  {journal} {Artificial Intelligence for the Earth
  Systems}\ }\textbf {\bibinfo {volume} {4}},\ \bibinfo {pages} {240094}
  (\bibinfo {year} {2025})}\BibitemShut {NoStop}%
\bibitem [{\citenamefont {Finkel}\ \emph {et~al.}(2020)\citenamefont {Finkel},
  \citenamefont {Abbot},\ and\ \citenamefont {Weare}}]{finkel2020path}%
  \BibitemOpen
  \bibfield  {author} {\bibinfo {author} {\bibfnamefont {J.}~\bibnamefont
  {Finkel}}, \bibinfo {author} {\bibfnamefont {D.~S.}\ \bibnamefont {Abbot}},\
  and\ \bibinfo {author} {\bibfnamefont {J.}~\bibnamefont {Weare}},\ }\bibfield
   {title} {\bibinfo {title} {{Path properties of atmospheric transitions:
  {I}llustration with a low-order sudden stratospheric warming model}},\
  }\href@noop {} {\bibfield  {journal} {\bibinfo  {journal} {Journal of the
  Atmospheric Sciences}\ }\textbf {\bibinfo {volume} {77}},\ \bibinfo {pages}
  {2327} (\bibinfo {year} {2020})}\BibitemShut {NoStop}%
\bibitem [{\citenamefont {C{\'e}rou}(2006)}]{cerou2006genetic}%
  \BibitemOpen
  \bibfield  {author} {\bibinfo {author} {\bibfnamefont {F.}~\bibnamefont
  {C{\'e}rou}},\ }\emph {\bibinfo {title} {{Genetic genealogical models in rare
  event analysis}}},\ \href@noop {} {Ph.D. thesis},\ \bibinfo  {school} {INRIA}
  (\bibinfo {year} {2006})\BibitemShut {NoStop}%
\bibitem [{\citenamefont {Miloshevich}\ \emph {et~al.}(2024)\citenamefont
  {Miloshevich}, \citenamefont {Lucente}, \citenamefont {Yiou},\ and\
  \citenamefont {Bouchet}}]{miloshevich_extreme_2024-3}%
  \BibitemOpen
  \bibfield  {author} {\bibinfo {author} {\bibfnamefont {G.}~\bibnamefont
  {Miloshevich}}, \bibinfo {author} {\bibfnamefont {D.}~\bibnamefont
  {Lucente}}, \bibinfo {author} {\bibfnamefont {P.}~\bibnamefont {Yiou}},\ and\
  \bibinfo {author} {\bibfnamefont {F.}~\bibnamefont {Bouchet}},\ }\bibfield
  {title} {\bibinfo {title} {{Extreme Heat Wave Sampling and Prediction with
  Analog {{Markov}} Chain and Comparisons with Deep Learning}},\ }\href
  {https://doi.org/10.1017/eds.2024.7} {\bibfield  {journal} {\bibinfo
  {journal} {Environmental Data Science}\ }\textbf {\bibinfo {volume} {3}},\
  \bibinfo {pages} {e9} (\bibinfo {year} {2024})}\BibitemShut {NoStop}%
\bibitem [{\citenamefont {Taylor}\ \emph {et~al.}(2001)\citenamefont {Taylor},
  \citenamefont {Williamson},\ and\ \citenamefont {Zwiers}}]{taylor_sea_2001}%
  \BibitemOpen
  \bibfield  {author} {\bibinfo {author} {\bibfnamefont {K.}~\bibnamefont
  {Taylor}}, \bibinfo {author} {\bibfnamefont {D.}~\bibnamefont {Williamson}},\
  and\ \bibinfo {author} {\bibfnamefont {F.}~\bibnamefont {Zwiers}},\
  }\href@noop {} {\emph {\bibinfo {title} {{The Sea Surface Temperature and Sea
  Ice Concentration Boundary Conditions for {{AMIP II}} Simulations}}}},\
  \bibinfo {type} {{{PCDMI Report}}}\ \bibinfo {number} {60}\ (\bibinfo
  {institution} {{Program for Climate Model Diagnosis and Intercomparison}},\
  \bibinfo {address} {Lawrence Livermore National Laboratory},\ \bibinfo {year}
  {2001})\BibitemShut {NoStop}%
\bibitem [{\citenamefont {Lunkeit}\ \emph {et~al.}(2011)\citenamefont
  {Lunkeit}, \citenamefont {Kirk}, \citenamefont {Borth}, \citenamefont
  {Kleidon}, \citenamefont {Böttinger}, \citenamefont {Luksch}, \citenamefont
  {Fraedrich}, \citenamefont {Jansen}, \citenamefont {Paiewonsky},
  \citenamefont {Schubert}, \citenamefont {Sielmann},\ and\ \citenamefont
  {Wan}}]{Lunkeit2011PlanetS}%
  \BibitemOpen
  \bibfield  {author} {\bibinfo {author} {\bibfnamefont {F.}~\bibnamefont
  {Lunkeit}}, \bibinfo {author} {\bibfnamefont {E.}~\bibnamefont {Kirk}},
  \bibinfo {author} {\bibfnamefont {H.}~\bibnamefont {Borth}}, \bibinfo
  {author} {\bibfnamefont {A.}~\bibnamefont {Kleidon}}, \bibinfo {author}
  {\bibfnamefont {M.}~\bibnamefont {Böttinger}}, \bibinfo {author}
  {\bibfnamefont {U.}~\bibnamefont {Luksch}}, \bibinfo {author} {\bibfnamefont
  {K.}~\bibnamefont {Fraedrich}}, \bibinfo {author} {\bibfnamefont
  {H.}~\bibnamefont {Jansen}}, \bibinfo {author} {\bibfnamefont
  {P.}~\bibnamefont {Paiewonsky}}, \bibinfo {author} {\bibfnamefont
  {S.}~\bibnamefont {Schubert}}, \bibinfo {author} {\bibfnamefont
  {F.}~\bibnamefont {Sielmann}},\ and\ \bibinfo {author} {\bibfnamefont
  {H.}~\bibnamefont {Wan}},\ }\bibfield  {title} {\bibinfo {title} {{Planet
  Simulator - Reference Manual, Version 16.0}}\ }(\bibinfo {year}
  {2011})\BibitemShut {NoStop}%
\bibitem [{\citenamefont {Chattopadhyay}\ \emph {et~al.}(2024)\citenamefont
  {Chattopadhyay}, \citenamefont {Sun},\ and\ \citenamefont
  {Hassanzadeh}}]{chattopadhyay_challenges_2024}%
  \BibitemOpen
  \bibfield  {author} {\bibinfo {author} {\bibfnamefont {A.}~\bibnamefont
  {Chattopadhyay}}, \bibinfo {author} {\bibfnamefont {Y.~Q.}\ \bibnamefont
  {Sun}},\ and\ \bibinfo {author} {\bibfnamefont {P.}~\bibnamefont
  {Hassanzadeh}},\ }\href {https://doi.org/10.48550/arXiv.2304.07029} {\bibinfo
  {title} {{Challenges of Learning Multi-Scale Dynamics with {{AI}} Weather
  Models: {{Implications}} for Stability and One Solution}}} (\bibinfo {year}
  {2024})\BibitemShut {NoStop}%
\bibitem [{\citenamefont {Toth}\ and\ \citenamefont
  {Kalnay}(1993)}]{toth1993ensemble}%
  \BibitemOpen
  \bibfield  {author} {\bibinfo {author} {\bibfnamefont {Z.}~\bibnamefont
  {Toth}}\ and\ \bibinfo {author} {\bibfnamefont {E.}~\bibnamefont {Kalnay}},\
  }\bibfield  {title} {\bibinfo {title} {{Ensemble forecasting at NMC: The
  generation of perturbations}},\ }\href@noop {} {\bibfield  {journal}
  {\bibinfo  {journal} {Bulletin of the american meteorological society}\
  }\textbf {\bibinfo {volume} {74}},\ \bibinfo {pages} {2317} (\bibinfo {year}
  {1993})}\BibitemShut {NoStop}%
\bibitem [{\citenamefont {Lee}\ and\ \citenamefont
  {Whiteley}(2018)}]{lee2018variance}%
  \BibitemOpen
  \bibfield  {author} {\bibinfo {author} {\bibfnamefont {A.}~\bibnamefont
  {Lee}}\ and\ \bibinfo {author} {\bibfnamefont {N.}~\bibnamefont {Whiteley}},\
  }\bibfield  {title} {\bibinfo {title} {{Variance estimation in the particle
  filter}},\ }\href@noop {} {\bibfield  {journal} {\bibinfo  {journal}
  {Biometrika}\ }\textbf {\bibinfo {volume} {105}},\ \bibinfo {pages} {609}
  (\bibinfo {year} {2018})}\BibitemShut {NoStop}%
\bibitem [{\citenamefont {Liu}\ and\ \citenamefont {Liu}(2001)}]{liu2001monte}%
  \BibitemOpen
  \bibfield  {author} {\bibinfo {author} {\bibfnamefont {J.~S.}\ \bibnamefont
  {Liu}}\ and\ \bibinfo {author} {\bibfnamefont {J.~S.}\ \bibnamefont {Liu}},\
  }\href@noop {} {\emph {\bibinfo {title} {{Monte Carlo strategies in
  scientific computing}}}},\ Vol.~\bibinfo {volume} {10}\ (\bibinfo
  {publisher} {Springer},\ \bibinfo {year} {2001})\BibitemShut {NoStop}%
\bibitem [{\citenamefont {Coles}(2001)}]{coles2001introduction}%
  \BibitemOpen
  \bibfield  {author} {\bibinfo {author} {\bibfnamefont {S.}~\bibnamefont
  {Coles}},\ }\href@noop {} {\emph {\bibinfo {title} {{An introduction to
  statistical modeling of extreme values}}}},\ Vol.\ \bibinfo {volume} {208}\
  (\bibinfo  {publisher} {Springer},\ \bibinfo {year} {2001})\BibitemShut
  {NoStop}%
\bibitem [{\citenamefont {Virtanen}\ \emph {et~al.}(2020)\citenamefont
  {Virtanen}, \citenamefont {Gommers}, \citenamefont {Oliphant}, \citenamefont
  {Haberland}, \citenamefont {Reddy}, \citenamefont {Cournapeau}, \citenamefont
  {Burovski}, \citenamefont {Peterson}, \citenamefont {Weckesser},
  \citenamefont {Bright}, \citenamefont {{van der Walt}}, \citenamefont
  {Brett}, \citenamefont {Wilson}, \citenamefont {Millman}, \citenamefont
  {Mayorov}, \citenamefont {Nelson}, \citenamefont {Jones}, \citenamefont
  {Kern}, \citenamefont {Larson}, \citenamefont {Carey}, \citenamefont {Polat},
  \citenamefont {Feng}, \citenamefont {Moore}, \citenamefont {{VanderPlas}},
  \citenamefont {Laxalde}, \citenamefont {Perktold}, \citenamefont {Cimrman},
  \citenamefont {Henriksen}, \citenamefont {Quintero}, \citenamefont {Harris},
  \citenamefont {Archibald}, \citenamefont {Ribeiro}, \citenamefont
  {Pedregosa}, \citenamefont {{van Mulbregt}},\ and\ \citenamefont {{SciPy 1.0
  Contributors}}}]{2020SciPy-NMeth}%
  \BibitemOpen
  \bibfield  {author} {\bibinfo {author} {\bibfnamefont {P.}~\bibnamefont
  {Virtanen}}, \bibinfo {author} {\bibfnamefont {R.}~\bibnamefont {Gommers}},
  \bibinfo {author} {\bibfnamefont {T.~E.}\ \bibnamefont {Oliphant}}, \bibinfo
  {author} {\bibfnamefont {M.}~\bibnamefont {Haberland}}, \bibinfo {author}
  {\bibfnamefont {T.}~\bibnamefont {Reddy}}, \bibinfo {author} {\bibfnamefont
  {D.}~\bibnamefont {Cournapeau}}, \bibinfo {author} {\bibfnamefont
  {E.}~\bibnamefont {Burovski}}, \bibinfo {author} {\bibfnamefont
  {P.}~\bibnamefont {Peterson}}, \bibinfo {author} {\bibfnamefont
  {W.}~\bibnamefont {Weckesser}}, \bibinfo {author} {\bibfnamefont
  {J.}~\bibnamefont {Bright}}, \bibinfo {author} {\bibfnamefont {S.~J.}\
  \bibnamefont {{van der Walt}}}, \bibinfo {author} {\bibfnamefont
  {M.}~\bibnamefont {Brett}}, \bibinfo {author} {\bibfnamefont
  {J.}~\bibnamefont {Wilson}}, \bibinfo {author} {\bibfnamefont {K.~J.}\
  \bibnamefont {Millman}}, \bibinfo {author} {\bibfnamefont {N.}~\bibnamefont
  {Mayorov}}, \bibinfo {author} {\bibfnamefont {A.~R.~J.}\ \bibnamefont
  {Nelson}}, \bibinfo {author} {\bibfnamefont {E.}~\bibnamefont {Jones}},
  \bibinfo {author} {\bibfnamefont {R.}~\bibnamefont {Kern}}, \bibinfo {author}
  {\bibfnamefont {E.}~\bibnamefont {Larson}}, \bibinfo {author} {\bibfnamefont
  {C.~J.}\ \bibnamefont {Carey}}, \bibinfo {author} {\bibfnamefont
  {{\.I}.}~\bibnamefont {Polat}}, \bibinfo {author} {\bibfnamefont
  {Y.}~\bibnamefont {Feng}}, \bibinfo {author} {\bibfnamefont {E.~W.}\
  \bibnamefont {Moore}}, \bibinfo {author} {\bibfnamefont {J.}~\bibnamefont
  {{VanderPlas}}}, \bibinfo {author} {\bibfnamefont {D.}~\bibnamefont
  {Laxalde}}, \bibinfo {author} {\bibfnamefont {J.}~\bibnamefont {Perktold}},
  \bibinfo {author} {\bibfnamefont {R.}~\bibnamefont {Cimrman}}, \bibinfo
  {author} {\bibfnamefont {I.}~\bibnamefont {Henriksen}}, \bibinfo {author}
  {\bibfnamefont {E.~A.}\ \bibnamefont {Quintero}}, \bibinfo {author}
  {\bibfnamefont {C.~R.}\ \bibnamefont {Harris}}, \bibinfo {author}
  {\bibfnamefont {A.~M.}\ \bibnamefont {Archibald}}, \bibinfo {author}
  {\bibfnamefont {A.~H.}\ \bibnamefont {Ribeiro}}, \bibinfo {author}
  {\bibfnamefont {F.}~\bibnamefont {Pedregosa}}, \bibinfo {author}
  {\bibfnamefont {P.}~\bibnamefont {{van Mulbregt}}},\ and\ \bibinfo {author}
  {\bibnamefont {{SciPy 1.0 Contributors}}},\ }\bibfield  {title} {\bibinfo
  {title} {{{SciPy} 1.0: Fundamental Algorithms for Scientific Computing in
  Python}},\ }\href {https://doi.org/10.1038/s41592-019-0686-2} {\bibfield
  {journal} {\bibinfo  {journal} {Nature Methods}\ }\textbf {\bibinfo {volume}
  {17}},\ \bibinfo {pages} {261} (\bibinfo {year} {2020})}\BibitemShut
  {NoStop}%
\end{thebibliography}%
